\pdfoutput=1
\documentclass[
11pt, 
english, 
singlespacing, 
hyperref, 
headsepline, 
consistentlayout, 
]{BScThesis} 

\usepackage[utf8]{inputenc} 
\usepackage[T1]{fontenc} 

\usepackage{palatino} 

\usepackage[autostyle=true]{csquotes} 

\usepackage{etoolbox} 

\usepackage{lipsum}

\usepackage{caption}

\usepackage{titlesec} 

\usepackage{listings}
\usepackage{bera}

\usepackage{bigfoot}

\usepackage[labelfont=bf,margin=0.2in,labelsep=endash,justification=raggedright]{caption}

\usepackage[utf8]{inputenc}
\usepackage{subcaption}
\usepackage{color}

\usepackage{glossaries}

\usepackage{fixltx2e}

\usepackage[bottom]{footmisc}

\usepackage{tabto}

\thesistitle{IOMMU Support for Virtual-Address Remote DMA in an ARMv8 environment} 
\supervisor{Manolis G.H. \textsc{Katevenis}} 
\examiner{} 
\degree{B.Sc.} 
\author{Antonios \textsc{Psistakis}} 
\addresses{csd3196@csd.uoc.gr} 

\subject{Remote DMA acceses} 
\university{\href{http://www.uoc.gr}{University of Crete}} 
\department{\href{http://www.csd.uoc.gr}{Computer Science Department}} 
\group{\href{http://www.ics.forth.gr/carv/}{Computer Architecture and VLSI Systems (CARV) Laboratory}} 
\faculty{\href{http://www.ics.forth.gr}{Institute of Computer Science (ICS), Foundation of Research and Technology - Hellas (FORTH)}} 

\AtBeginDocument{
	\hypersetup{pdftitle=\ttitle} 
	\hypersetup{pdfauthor=\authorname} 
	\hypersetup{pdfkeywords=\keywordnames} 
}

\definecolor{codegreen}{rgb}{0,0.6,0}
\definecolor{codegray}{rgb}{0.5,0.5,0.5}
\definecolor{codepurple}{rgb}{0.58,0,0.82}
\definecolor{backcolour}{rgb}{0,0,0}
\definecolor{whitecolour}{rgb}{1,1,1}
\lstdefinestyle{mystyle}{
	backgroundcolor=\color{backcolour},   
	basicstyle=\ttfamily\color{white},
	commentstyle=\color{codegreen},
	keywordstyle=\color{magenta},
	numberstyle=\tiny\color{codegray},
	stringstyle=\color{codepurple},
	breakatwhitespace=false,         
	breaklines=true,                 
	captionpos=b,                    
	numbers=left,                    
	numbersep=5pt,                  
	showspaces=false,                
	showstringspaces=false,
	showtabs=false,                  
	tabsize=2,
	escapeinside={~~},
	mathescape=true,
	language=bash,
	xleftmargin=0.025\textwidth
}

\begin{document}
	
	\hypersetup{%
		colorlinks = true,
		linkcolor  = black
	}

	\newcommand{\source}[1]{\caption*{Source: {#1}} }

	\makeatletter
	\let\latex@footnoterule\footnoterule
	\patchcmd{\footnoterule}{\latex@footnoterule}{}{}
	\renewcommand{\footnoterule}{%
		\setlength{\parindent}{0ex} 
		\vtop to 0pt{
			\newgeometry{ textheight=1cm}
			\vspace{\textheight}
			
			\hrule \@width .4\columnwidth
			\vskip 4\p@
			\hbox{\footnotesize\itshape\fixedfootnotetext}
			\vskip 4\p@
			\hrule \@height \z@

			
			\begin{minipage}[t]{\textwidth}
				\begin{flushleft} \large
					\emph{CARV, ICS, FORTH {  }{  }{  }{  }{  }{  }{  }{  }{  }{  }{  }{  } B.Sc. Thesis by A. Psistakis {  }{  }{  }{  }{  }{  }{  }{  }{  }{  }{  }{  }   CSD, UoC}\\
				\end{flushleft}
			\end{minipage}

		}
	}
	\setlength{\skip\footins}{4ex}
	\makeatother
	\newcommand\fixedfootnotetext

	
	\pagestyle{plain} 
	
	
	\begin{titlepage}
		\begin{center}
			
			\vspace*{.06\textheight}
			
			{\scshape\LARGE \deptname\par}\vspace{0.2cm}
			{\scshape\LARGE \univname\par}\vspace{1.5cm} 
			\textsc{\Large B.Sc. Thesis}\\[0.5cm] 
			
			\HRule \\[0.4cm] 
			{\huge \bfseries \ttitle\par}\vspace{0.4cm} 
			\HRule \\[1.5cm] 
			
			\begin{minipage}[t]{0.4\textwidth}
				\begin{flushleft} \large
					\emph{Author:}\\
					{\authorname} 
				\end{flushleft}
			\end{minipage}
			\begin{minipage}[t]{0.5\textwidth}
				\begin{flushright} \large
					\emph{Supervisor:} \\
					Prof. \href{http://users.ics.forth.gr/~kateveni/}{\supname} 
				\end{flushright}
			\end{minipage}\\[3cm]
			
			\vfill
			
			
			
		\vfill\vfill\vfill\vfill\vfill\vfill\vfill\vfill\vfill\vfill\vfill\vfill\vfill
		
		
		{\large March 20, 2017}\\[1cm] 

			\renewcommand{\fixedfootnotetext}{%
		
					\vtop to 0pt{
						\hrule
						
					}
		
		
		%

			}
		
			\newcommand\blfootnote[1]{%
				\begingroup
				\renewcommand\thefootnote{}\footnote{#1}%
				\addtocounter{footnote}{-1}%
				\endgroup
			}
			
			\blfootnote{The thesis is submitted in fulfillment of the requirements for the degree of \degreename\ in the \deptname~(CSD), \univname~(UoC). It took place in the \groupname\ of the \facname, from June 2016 to December 2016 and was funded by the \href{https://www.forth.gr/index_main.php?l=e&c=56&i=9}{distinguished Undergraduate Scholarship "Stelios Orphanoudakis"}.}
			\flushbottom
			
			
		\end{center}
	\end{titlepage}

	\begin{abstract}
	\addchaptertocentry{\abstractname} 
	
	\setlength{\parindent}{0ex} 
	In complex systems with many compute nodes consisting of many CPUs, that are coherent inside each node, one issue is to achieve coherence between them in an efficient and valid way. Unimem system \cite{Reference12} addresses this issue and in order to support it, it suggests a virtualized system (virtual global address space) that is able to handle this. This approach requires the use of the I/O Memory Management Unit (IOMMU) in each node.
	\setlength{\parindent}{3ex} 
	
	The goal of this thesis is to support this approach, by testing and using successfully the IOMMU of one node. In order to do that, we used ARM's IOMMU, which is called System Memory Management Unit (SMMU), that, in general, enables the translation of the virtual addresses to their corresponding physical addresses. Using the SMMU is complex - the documentation when it comes to Linux support is not very clear, so we had to implement our own modules, in order to test and use the SMMU.
	
	First, we tested the SMMU in the Processing System (PS) of the Xilinx Zynq UltraScale+ MPSoC \cite{Reference2}, by writing a module that adds a virtual address and its corresponding translation to a physical address as an entry to the SMMU. Then, we triggered a DMA transaction sending data to the virtual address that we added to the SMMU and we noticed that our DMA transaction passed through the SMMU in order to translate the virtual address to the corresponding physical address. Later, we tested the SMMU, by doing the same but this time, by triggering the transaction from the Programmable Logic (PL). We noticed, again, that the transaction passed through the SMMU in order to translate the virtual address of the destination. Finally, we created a module that makes possible to trigger transactions from the PL, without the need of mapping the entries of virtual addresses and physical addresses before -- we managed to have the SMMU being able to handle and translate all virtual addresses related to a user process, by setting the SMMU with the pointer of the page table of the user process.
	
	In general, the result for all scenarios and modules mentioned above is that we managed to have the SMMU working. Due to the limited time of this thesis, future work for understanding and using more features of the SMMU is expected.
	

	\end{abstract}

	
	\dedicatory{Dedicated to the memory of a great, kind, smart and charming person and loyal friend to me, my grandmother Maria A. Psistaki (1925 - 2016).} 
	
	\label{ack}
	\begin{acknowledgements}
	\addchaptertocentry{\acknowledgementname} 
	The work for this thesis took place in the \groupname\ of the \facname, from June 2016 to December 2016 and was funded by the \href{https://www.forth.gr/index_main.php?l=e&c=56&i=9}{distinguished Undergraduate Scholarship "Stelios Orphanoudakis"}.

	\setlength{\parindent}{3ex}
	
	I would like to thank and give credit to a lot of people who helped me in many ways throughout the thesis.
	
	First of all, I would like to start by thanking my supervisor Professor \supname, who trusted me with such new, exciting and perhaps challenging task and advised me many times during the time I was working on my thesis. \par
	
	Then, I would like to thank Dr. Fabien Chaix.~He was very supportive from the beginning of my thesis until the last day. We worked together in many parts of this thesis, like: Petalinux and booting Linux from our Trenz Boards, the theory of the IOMMU and the implementation of it (the kernel modules - Chapter \ref{Chapter4}). It was his suggestion, that parts (bits) of the StreamID are the AXI ID and the Master Device (in our tests, the HPC0 - Chapter \ref{Chapter5}). He also helped me with the understanding of how to use Vivado and SDK, in order to create our own Linux images. Dr. Chaix was actually involved in everything I was involved. I thank him for being patient with me and also for his advices and support. \par
	
	Also, I would like to thank Dr. Vassilis Papaefstathiou. We met almost at the end of my thesis, but we worked together in the understanding of the IOMMU (devices, groups, domains, SMR$n$, S2CR$n$ etc), how it works and how we could "pass" the page table of a process to the SMMU. The focus of the work we did together was on the last kernel module (Chapter \ref{Chapter4}) -- he explained me how the ioctl works and we investigated together the translation/context faults we had at first and we wanted to fix, before we manage to have a translation table of a user process in a context bank of the SMMU working. I thank him for his support and persistence in guiding me to become better. \par
	
	For the support in the first three kernel modules (Chapter \ref{Chapter4}), I would like to thank my colleague John Vardas. He helped me with debugging and also he did some changes/additions to some of the modules, that made them work more efficiently. In both SMMU modules (with transactions from the PS and the PL), he run many tests and helped us understand many things about the SMMU. \par
	
	I would like to thank Dr. Nikolaos Chrysos, for helping John Vardas and me, with the discussions we had in order to make the third kernel "module" (the module that allowed us to have a single-page transaction from the PL, using the SMMU) work.
	
	For the system software support, I would like to thank Dimitris Poulios (M.Sc.) and Charalampos Aronis. Dimitris helped us boot Linux from the QSPI and the SD Card of the Trenz board and he was always helping us, when we needed Linux support. Charalampos helped us understand the kernel environment and he was the person who gave me guidance, in order to use the IOMMU API and build our first SMMU module for testing purposes. \par
	\setlength{\parindent}{3ex} 
	\noindent
	
	For the hardware support, I would like to thank Konstantinos Harteros (M.Sc.), for the discussions we had because of his experience working with an IOMMU in the past, Antonis Psathakis (M.Sc.), for his support with Xilinx Vivado and SDK and Aggelos Ioannou (M.Sc.), for providing us the hardware files that we needed, in order to create our own Linux images and use them to boot from a Trenz board. \par
	
	I greatly appreciate the feedback that Dr. Fabien Chaix and Panos Peristerakis gave me for my thesis before I submit it.
	
	Last but not least and while I am hoping I did not forget to thank personally anyone that I should have, I would like to thank everyone in the CARV Laboratory of the Institute of Computer Science, FORTH, for all the support throughout my thesis. I could not be more thankful. \par

	\end{acknowledgements}
	
	\tableofcontents 

	\pagestyle{thesis} 
	


\chapter{Introduction} 

\label{Chapter1} 


In this chapter, we will describe the motivation behind this thesis, the contributions and the context/background that was needed for the purposes of this thesis, in terms of the Hardware and the Software.
\setlength{\parindent}{3ex}

\section{Motivation}


\label{unimem}

As the years go by, when it comes to the Computer Science and the industry, it seems there is a direction in creating more complex sets of computers. Looking the image below, we can see in a very simplistic way that many nodes, or "Coherent Islands", as they are described in Unimem \cite{Reference12}, are trying to communicate. It is obvious that the coherence between all CPUs in one node is somehow granted. The problem begins when we have many nodes and we want to achieve coherence between them -- the more nodes the bigger the problem. This is the issue that the Unimem is trying to address, in order for big and complex systems, as we described, to be able to communicate in a valid and efficient way.

\begin{figure}[h]
	
	\centering
	
	
	
	\captionbox[Text]{A system with many nodes that are connected \label{fig:3.1}}{%
		\includegraphics[width=0.7\textwidth]{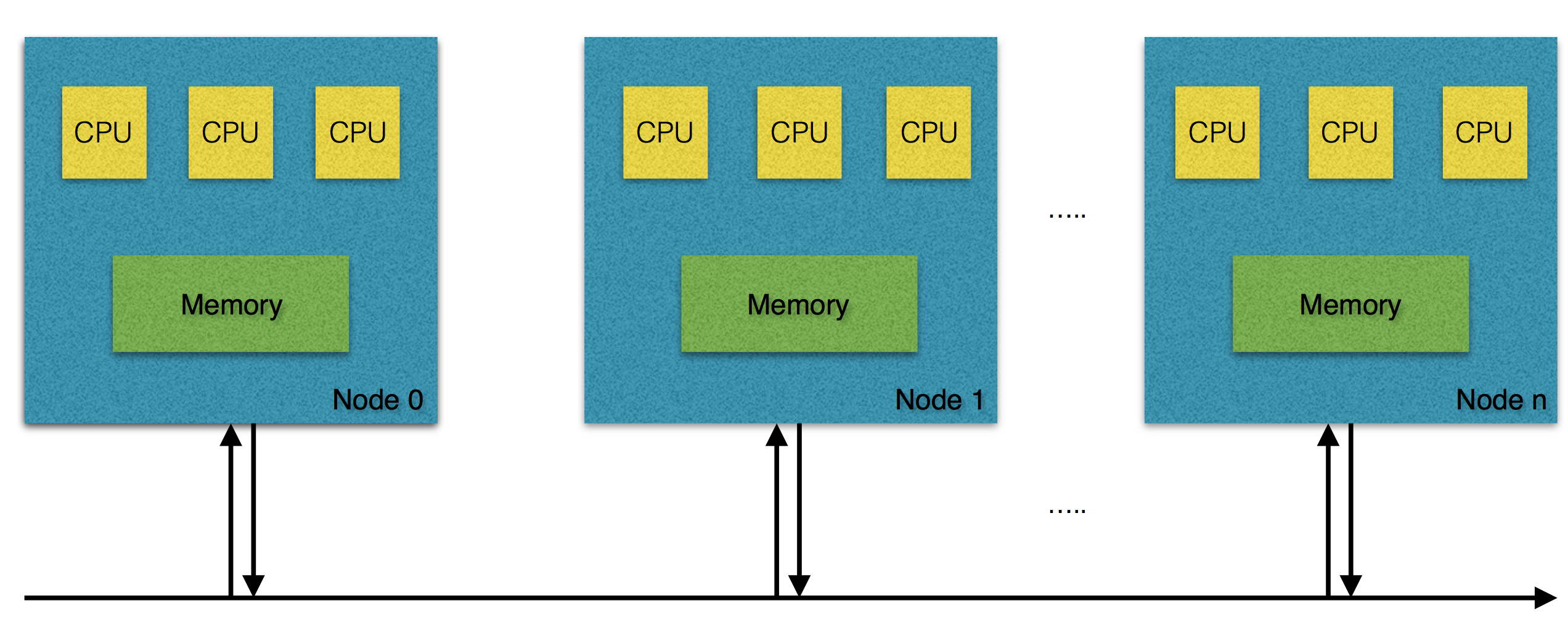}  
	}
\end{figure}

The Unimem suggests that if we want such system to be coherent, each node that needs to access information that belongs to the memory of another node, it should request the information from the other node, without keeping a copy to its memory. In order to achieve this goal, Unimem proposed a virtualized system that distinguishes all nodes and more specifically, a virtual global address space. This has as a result the necessity of an "arbiter" in each node, that will handle all incoming transactions in order to access their memory. This arbiter will be no other than the I/O Memory Management Unit (IOMMU).




Part of this thesis would, in a way, support the implementation of the Unimem.

\vfill
\footnoterule
\pagebreak

\section{Contributions}
\setlength{\parindent}{3ex}

This thesis is trying to support the Unimem, as described in Section \ref{unimem}, by understanding how the SMMU (ARM's IOMMU) works and how we could use it in order to achieve virtual-address remote Direct Memory Access (DMA) transactions in one or between multiple nodes -- for the purposes of this thesis only one node was considered in the experiments, but the information and results generated by this thesis could also support future implementations for multiple nodes. Having the understanding and information needed, we could proceed to the implementation of the kernel modules that would test and use the SMMU. In order to achieve these goals, the work for this thesis consisted of three different parts.

The first part is the background/theory of ARM's SMMU, which includes all the information and knowledge that was collected and used in order to proceed with the implementation of the kernel modules. Chapter \ref{Chapter3} has a detailed description of all this information.

The second part is the kernel modules that were implemented in order to test and use the SMMU of the Zynq UltraScale+ MPSoC. Chapter \ref{Chapter4} has a detailed description of all the different modules that were implemented during this thesis. 

The third part is the experiments that took place for all the kernel modules that are described in Chapter \ref{Chapter4} and the notes we used, in order to reach our big goal that was having the SMMU ready (initialized) to support the virtual-address remote DMA transactions. Detailed information about the experiments can be found in Chapter \ref{Chapter5}.

\vfill
\footnoterule
\pagebreak

\section{Context}
\label{context}
\subsection{Hardware}
\setlength{\parindent}{3ex}

\subsubsection{Trenz board}

The hardware that was used in order to achieve our goal was basically a \href{https://wiki.trenz-electronic.de/pages/viewpage.action?pageId=24153560}{TET0808 Trenz board} with the FPGA: XCZU9EG-FFVC900-1 \cite{Reference9}. It is a MPSoC module integrating a Xilinx Zynq UltraScale+, with 2 Giga Byte DDR4 SDRAM with 64-Bit width, 64 MByte Flash memory for configuration and operation, 20 Gigabit tranceivers, and switch-mode power supplies for all on-board voltages.\par

\subsubsection{Zynq UltraScale+}
\setlength{\parindent}{3ex}


Zynq UltraScale+ MPSoC is the Xilinx second-generation Zynq platform, combining a processing system (PS) and user-programmable logic (PL) into the same device. 
The Zynq UltraScale+ MPSoC has four different power domains.

\begin{itemize}
	
	\item Low-power domain (LPD).
	\item Full-power domain (FPD).
	
\end{itemize}
\setlength{\parindent}{0ex}
It also has the PL power domain (PLPD) and the Battery power domain (BPD), that we will not use for the purposes of this thesis.

\setlength{\parindent}{0ex}
The Zynq UltraScale+ MPSoC PS block has three major processing units:

\begin{itemize}
	
	\item Cortex-A53 application processing unit (APU)—ARM v8 architecture-based 64-bit quad-core or dual-core multiprocessing CPU
	\item Cortex-R5 real-time processing unit (RPU)—ARM v7 architecture-based 32-bit dual real-time processing unit with dedicated tightly coupled memory (TCM)
\end{itemize}
\setlength{\parindent}{0ex}

The third processing unit is the Mali-400 graphics processing unit (GPU) with pixel and geometry processor and 64KB L2 cache. For the purposes of this thesis we will not use it.

\setlength{\parindent}{3ex}
\noindent

As we can see below (Figure \ref{fig:3.2}), in the top-level block diagram of the Zynq UltraScale+ MPSoC, part of the Processing System (PS) is the APU that consists of four Cortex-A53 processors that will run the system and more specifically they will run the Linux that we will use in order to achieve our goals, using the IOMMU. Also, part of the PS is the RPU that consists of two Cortex-R5 real-time processors, that we will not use for the purposes of this thesis. As we can see, the SMMU and the CCI (Cache Coherent Interconnect) are in the same block in the top-level block diagram -- that of course does not mean that they indeed are part of the same block, but that they collaborate in order to have coherent accesses to the memory. For the purposes of the thesis and because of the limited time, we will not focus on coherence issues. Also, as we will see later, we will use both the LPD-DMA (Low-Power Domain DMA) and the FPD-DMA (FPD-DMA) that are parts of the PS, in order to trigger transactions that we want to go through the SMMU, when we have virtual addresses. 
Last but not least, part of the top-level diagram is the Programmable Logic (PL), where a user can program the FPGA, 
\vfill
\footnoterule
\pagebreak
add blocks that use the PS etc. In our case, we will use the PL to trigger transactions, that we want to go through the SMMU, from there.

\setlength{\parindent}{0ex}

\begin{figure}[h!]
	
	\centering
	\captionbox[Text]{Zynq UltraScale+ MPSoC Top-Level Block Diagram \label{fig:3.2}\addtocounter{figure}{-1}\subcaption*{Source: Xilinx \cite{Reference2}}}{%
		\includegraphics[width=\textwidth]{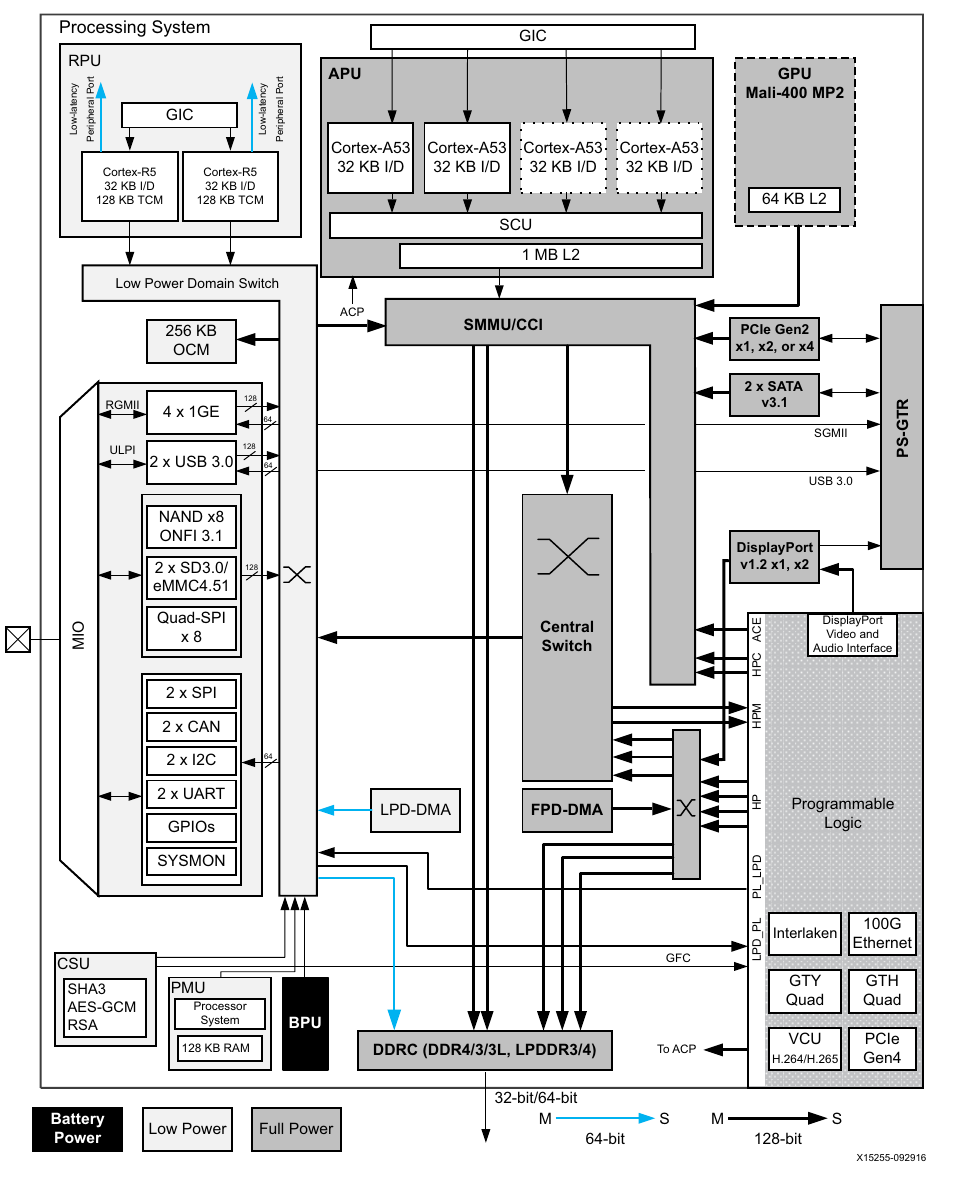}  
	}
\end{figure}

\vfill
\footnoterule
\pagebreak

\setlength{\parindent}{3ex}

\subsubsection{Memory Management}

For the purposes of this thesis, the most interesting part of the Zynq UltraScale+ MPSoC is the System Memory Management Unit (SMMU) block. Before we continue with the remaining of the thesis, we should explain the basic idea of the Memory Management Unit (MMU).

\paragraph{The Memory Management Unit (MMU)}\mbox{}\\

\setlength{\parindent}{0ex} 

The Memory Management Unit (MMU), in general, is a computer hardware unit having all memory references passed through itself, primarily performing the translation of virtual memory addresses to physical addresses. It is usually implemented as part of the CPU, but it also can be in the form of a separate integrated circuit. 

\setlength{\parindent}{3ex} 

A MMU can effectively perform virtual memory management, handling at the same time memory protection, cache control, bus arbitration and, in simpler computer architectures (especially 8-bit systems), bank switching. 

\begin{figure}[h!]
	
	\centering
	\captionbox[Text]{A simple use of the MMU \label{fig:3.3}}{%
		\includegraphics[width=0.6\textwidth]{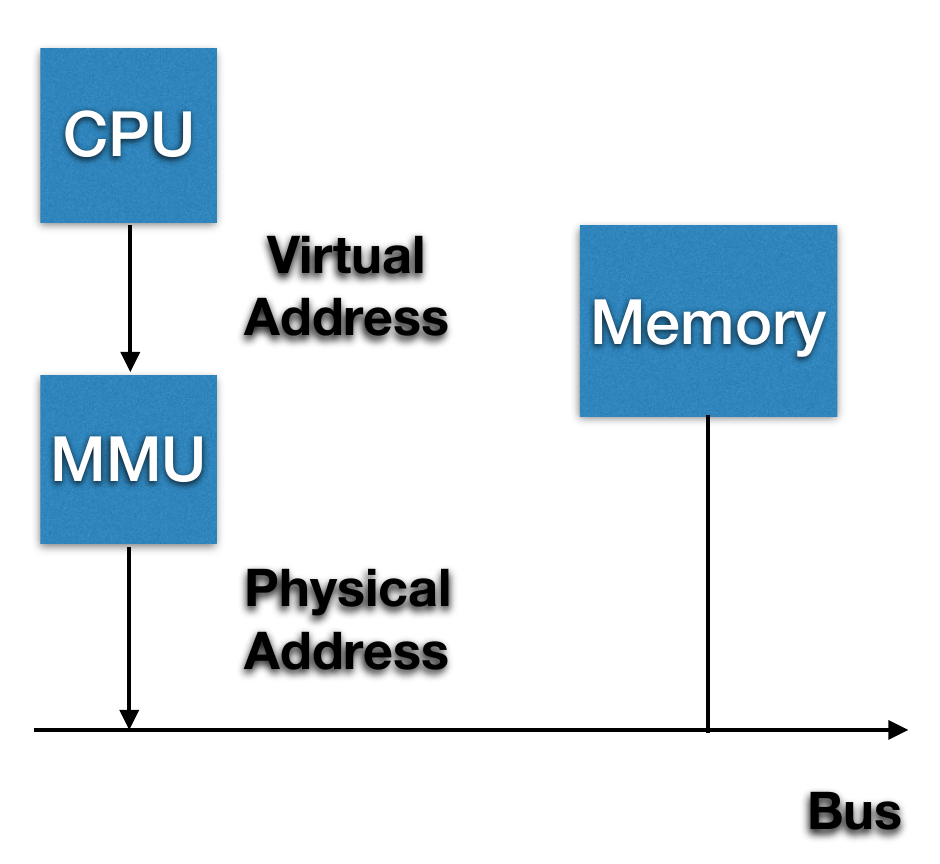}  
	}
\end{figure}

\paragraph{The I/O Memory Management Unit (IOMMU)}\mbox{}\\

\setlength{\parindent}{0ex} 

I/O Memory Management Unit (IOMMU) is a MMU that connects a Direct Memory Access-capable (DMA-capable) I/O bus to the main memory. IOMMU is basically responsible to map device-visible virtual addresses (also called device addresses or I/O addresses in this context) to physical addresses.

\setlength{\parindent}{3ex} 

\begin{figure}[h]
	
	\centering
	\captionbox[Text]{Comparison of the IOMMU to the MMU \label{fig:3.4}\addtocounter{figure}{-1}\subcaption*{Source: \href{https://en.wikipedia.org/wiki/Input-output_memory_management_unit}{Wikipedia}}}{%
		\includegraphics[width=0.4\textwidth]{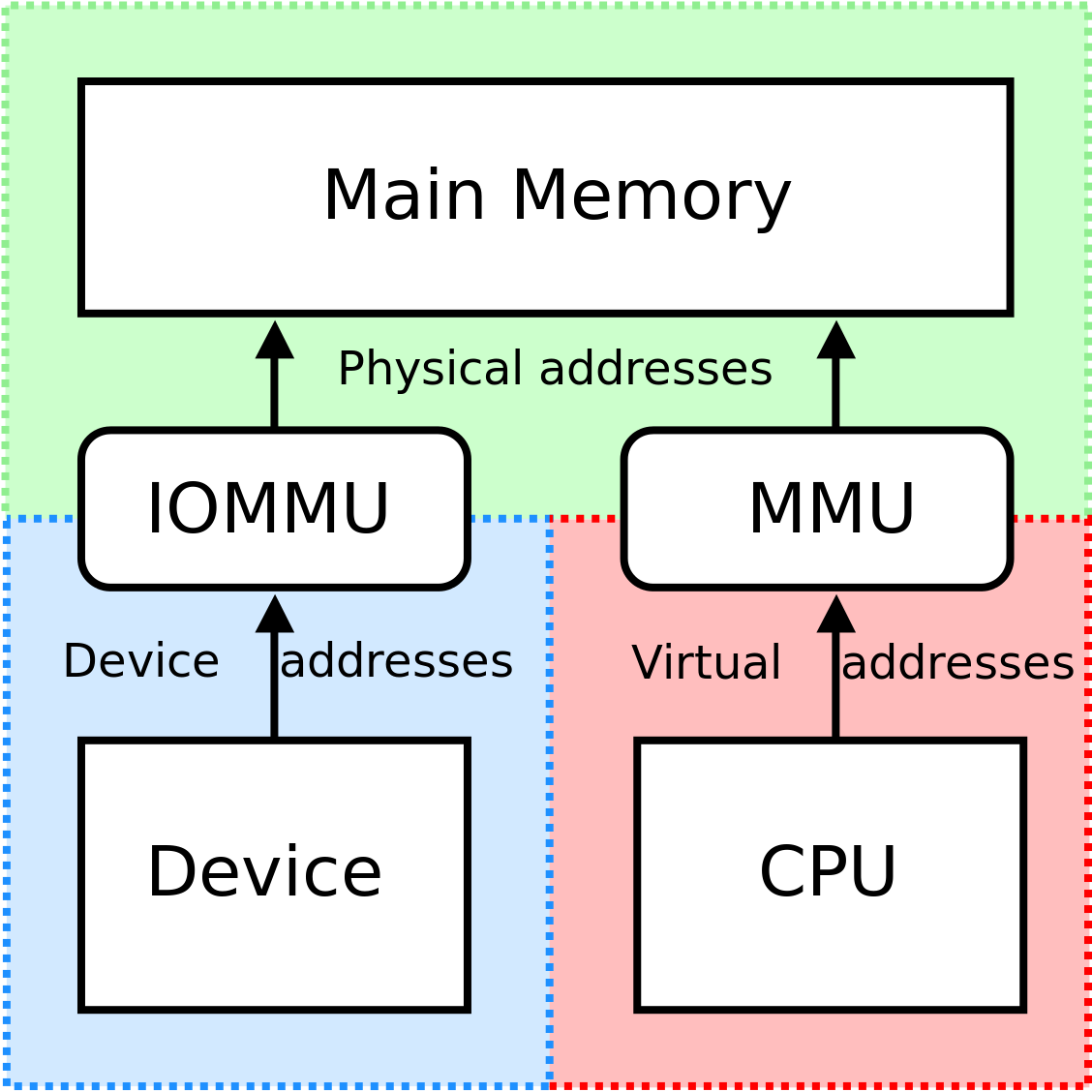}  
	}
\end{figure}

\vfill
\footnoterule
\pagebreak
\paragraph{The System Memory Management Unit (SMMU)}\mbox{}\\

\setlength{\parindent}{0ex} 

In order to address some issues (e.g.: memory fragmentation, multiple DMA-capable masters, system integrity guarantee) and limitations, the ARM Architecture Virtualization Extensions introduced the System Memory Management Unit (SMMU) concept to the ARM Architecture. In other words, as we mentioned before, ARM's IOMMU is called SMMU. 

\setlength{\parindent}{3ex} 

The SMMU performs address translation of an incoming AXI (Advanced eXtensible Interface) \cite{Reference10} address and AXI ID (mapped to context) to an outgoing address (physical address - PA), based on address mapping and memory attribute information held in translation tables. 

\begin{figure}[h]
	
	\centering
	
	\captionbox[Text]{Examples of where a SMMU could be located in a system. Coherent interconnects ensure cache coherence between masters. \label{fig:3.5}\addtocounter{figure}{-1}\subcaption*{Source: \href{https://www.arm.com/files/pdf/System-MMU-Whitepaper-v8.0.pdf}{ARM} }}{%
		\includegraphics[width=0.6\textwidth]{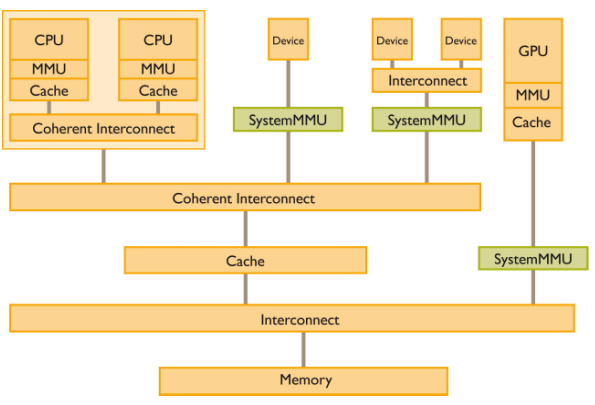}  
	}
	
\end{figure}

\vfill
\footnoterule
\pagebreak

\subsection{Software}

\setlength{\parindent}{3ex}

Most of the effort for this thesis was actually spent on Software aspects. That is because, in general, the configuration of the SMMU is difficult to understand and handle, without proper documentation and experience. 

\subsubsection{Linux}

\setlength{\parindent}{3ex} We chose to use the Linux OS as the Operating System for the purposes of this thesis - the Linux version was 4.4.0.

At first, we used the pre-built images provided by Trenz in order to flash the QSPI (Queued Serial Peripheral Interface -- a type of SPI controller that uses a data queue to transfer data across the SPI bus) memory with Linux. Although the pre-built images worked, it was not enough for our purposes because those pre-built images had a very minimal version of Linux. When we tried to find another way to create our own Linux images, the Petalinux method was the suggested one by Trenz and Xilinx. Although, again, there was a document with the process of creating images via Petalinux environment, the steps and the details were not clear, and after spending much time trying to make them work, we decided to use part of the working files (or images) like the device tree (.dtb) and the first-stage-boot-loader (FSBL) and the rest of the images, that we needed, from other projects that were working already.

As mentioned before, we used to boot from the QSPI (that we were flashing using the JTAG), but currently, since we now have more extensive base boards that support the SD card, we are booting from there. Appendix \ref{AppendixA} describes the process of booting from the SD Card. 

\subsubsection{Xilinx}

\setlength{\parindent}{3ex} 

We used Vivado, a software suite produced by Xilinx for synthesis and analysis of HDL designs and SDK (Software Development Kit), that is the Integrated Design Environment for creating embedded applications on any of Xilinx's microprocessors like the Zynq UltraScale+ MPSoC. 

We are using Xilinx Vivado to export the bitstream (.bit) and the hardware design/description file (.hdf), and the Xilinx SDK to create the device tree (.dtb) and the first-stage-boot-loader (.fsbl) from the exported hardware design/description file (.hdf). In Appendix \ref{AppendixA} (part of Section \ref{fsbl_generation}), we can see how to generate the FSBL and in Appendix \ref{AppendixC}, we can see how to generate the device tree. We used the 2016.2 versions of the Xilinx Vivado and SDK. Xilinx tools also helped us create the BOOT.bin image, that we are using to boot the Linux. In Appendix \ref{AppendixA} (Section \ref{bootbin}) we see how we create the BOOT.bin, with commands like "bootgen" being part of the Xilinx tools.

\subsubsection{Modules/Drivers}

\setlength{\parindent}{3ex} At this point, since we mentioned the "modules" and "drivers" before and they are a big part of this thesis, it is good to define them before we continue.

A device driver (commonly referred to simply as a driver) is a computer program that operates or controls a particular type of device that is attached to a 
\vfill
\footnoterule
\pagebreak
computer. A driver provides a software interface to hardware devices, enabling operating systems and other computer programs to access hardware functions without needing to know precise details of the hardware being used.

\setlength{\parindent}{3ex}
\noindent

A module is a piece of code that can be loaded and unloaded into the kernel upon demand. Modules extend the functionality of the kernel without the need to reboot the system. For example, one type of module is the device driver, which allows the kernel to access hardware connected to the system. Without modules, we would have to build monolithic kernels and add new functionality directly into the kernel image. Besides having larger kernels, this has the disadvantage of requiring us to rebuild and reboot the kernel every time we want new functionality.

In order to test and use the SMMU, we had to write or modify some drivers and modules, which will be described in Chapter \ref{modules} and Appendix \ref{AppendixD}.\\

\setlength{\parindent}{0ex}

Below, in Figure \ref{fig:3.6}, we can see how a system can use the drivers and the modules.


\begin{figure}[h]
	
	\centering
	
	\captionbox[Text]{The Kernel, The Processes And The Hardware. \label{fig:3.6}\addtocounter{figure}{-1}\subcaption*{Source: \href{http://haifux.org/lectures/86-sil/kernel-modules-drivers/kernel-modules-drivers.html}{haifux.org} }}{%
		\includegraphics[width=0.8\textwidth]{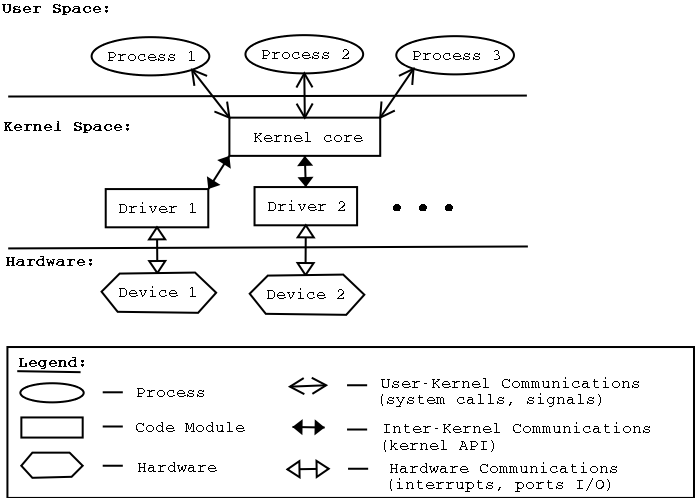}  
	}
	
\end{figure}


\noindent

\vfill
\footnoterule

\chapter{Related work} 

\label{Chapter2} 

\setlength{\parindent}{3ex} 

The truth is that when we were looking for information or previous work about anything related to the implementations and uses of the IOMMU (or the SMMU) with Linux support, we could not find many things. 

There is a book, called \textbf{"Linux Device Drivers"}, which introduces the idea of the Input/Output Memory Management Unit (IOMMU) and the virtualization. There are, also, some examples of information coming from companies, who implement and make use of the IOMMU. \par 
In this chapter, we will give a brief description of all the references/information coming from the following sources:

\begin{enumerate}
	\item Linux Device Drivers
	\item Intel IOMMU
	\item AMD IOMMU
	\item ARM IOMMU
\end{enumerate}

\section{Linux Device Drivers, Third Edition}
\setlength{\parindent}{0ex}
In the Linux Device Drivers \cite{Reference6}, and more particularly in Chapter 15 "Memory Mapping and DMA" it is mentioned that some architectures can provide an IOMMU that remaps addresses between a bus and main memory. An IOMMU can be helpful in many ways (i.e.: making a buffer scattered in memory
appear contiguous to the device), but programming the IOMMU is an extra step that must be performed when setting up DMA operations.
\setlength{\parindent}{3ex} 

Later, it suggests that a DMA mapping is a combination of allocating a DMA buffer and generating an address for that buffer that is accessible by the device. One approach is to get that address with a simple call to $virt{\_}to{\_}bus$, but there are many good reasons for avoiding this approach. The first of those is that reasonable hardware comes with an IOMMU that provides a set of mapping registers for the bus. The IOMMU can arrange for any physical memory to appear within the address range accessible by the device, and it can cause physically scattered buffers to look contiguous to the device. Making use of the IOMMU requires using the generic DMA layer and the $virt{\_}to{\_}bus$ is not able to handle this.

According to the book not all architectures have an IOMMU; in particular, by that time, the popular x86 platform had no IOMMU support -- a properly written driver did not need to 
be aware of the I/O support hardware it is running over.

\vfill
\footnoterule
\pagebreak

\setlength{\parindent}{3ex} 
\noindent

\section{Intel IOMMU}

\setlength{\parindent}{3ex}

In Intel systems, the use of the IOMMU is part of the "Intel VT" (Virtualization Technology). More specifically, Intel's "Intel VT-d" (Intel Virtualization Technology for Directed I/O), makes direct access to a PCI device possible for guest systems with the help of the Input/Output Memory Management Unit (IOMMU) provided. This allows a LAN card to be dedicated to a guest system, which makes possible to allow better attainment of increased network performance beyond that of an emulated LAN card. Of course, once such a direct access system has been implemented, live migration of the guest operating system is no longer possible. For instance, VMware can be configured for use with an activated Intel VT-d system using VMware VMDirectPath for direct access to PCI cards.\par

Another reference on the Intel IOMMU, when it comes to Linux, is that the Intel IOMMU Support is related to the IOVA (IO Virtual Address) generation. 
"Well behaved" drivers call $pci{\_}map{\_}*$ function before sending command to the device that needs to do a DMA transaction. Once the DMA transaction is completed and the mapping is no longer required, device performs a $pci{\_}unmap{\_}*$ call to unmap the region that was mapped before. The Intel IOMMU driver allocates a virtual address per domain. Each PCI (PCI Express) device has its own domain, which is translated to protection. Devices under p2p (peer to peer) bridges share the virtual address with all devices under the p2p bridge due to transaction id aliasing for the p2p bridges. The IOVA generation is pretty generic - Intel used the same technique as used in $vmalloc$ function, but these are not global address spaces, but are separate for each domain. Different DMA engines may support different number of domains. 

Intel, also, allocates guard pages with each mapping, so it can be attempted to catch any overflow that might happen. 

More information about the Intel IOMMU can be found in the Intel Virtualization Technology for Directed I/O specifications \cite{Reference8}. \par

\section{AMD IOMMU}

In the case of the AMD processors, the I/O Memory Management Unit (IOMMU) extends the AMD64 system architecture by adding support for address translation and system memory access protection on DMA transfers from peripheral devices. The AMD IOMMU also helps filter and remap interrupts from peripheral devices.

\setlength{\parindent}{3ex}
The AMD IOMMU enables several significant system-level enhancements:
\begin{itemize}
		\addtolength{\itemindent}{1cm}
		\item Legacy 32-bit I/O device support on 64-bit systems (generally without 
		\tabto{1cm}	requiring bounce buffers and expensive memory copies)
		\item More secure user-level application access to selected I/O devices
		\item More secure virtual machine guest operating system access to 
		\tabto{1cm} selected I/O devices
\end{itemize}

\setlength{\parindent}{0ex}

and can be used to replace the existing Graphics Address Remapping Table (GART) mechanism. It can also be used to remap addresses above 4 GB for I/O 
\vfill
\footnoterule
\pagebreak

\setlength{\parindent}{0ex}
devices that do not support 64-bit addressing and allow a guest OS running on a virtual machine to have direct control of a device. It can provide page granularity control of device access to system memory, a device direct access to user space I/O and direct delivery of interrupts to a guest Operating System. Also, it can filter and remap interrupts and share process virtual address space with selected peripheral devices.

\setlength{\parindent}{3ex} The IOMMU can be thought of as a generalization of two facilities included in the AMD64 architecture: the GART and the Device Exclusion Vector (DEV). The GART provides address translation of I/O device accesses to a small range of the system physical address space, and the DEV provides a limited degree of I/O device classification and memory protection. With appropriate software support, the IOMMU can emulate the capabilities of the GART or DEV.

\setlength{\parindent}{3ex} The IOMMU extends the concept of protection domains (or domains) first introduced with the AMD64 DEV. The IOMMU allows each I/O device in the system to be assigned to a specific domain and a distinct set of I/O page tables. When an I/O device attempts to read or write system memory, the IOMMU intercepts the access, determines the domain to which the device has been assigned, and uses the TLB entries associated with that domain or the I/O page tables associated with that I/O device to determine whether the access is to be permitted as well as the actual location in system memory that is to be accessed. \par

The IOMMU may include optional support for remote IOTLBs (input/output translation look-aside buffers, that speed-up the address resolution). A trusted I/O device with IOTLB support can cooperate with the IOMMU to maintain its own cache of address translations. This creates a framework for creating scalable systems with an IOMMU in which I/O devices may have different usage models and working set sizes. IOTLB-capable I/O devices contain private TLBs tailored for their own needs, creating a scalable distributed system of TLBs. The performance of IOTLB-capable I/O devices is not limited by the number of TLB entries implemented in the IOMMU. A peripheral with an IOTLB may issue untranslated addresses or pre-translated addresses that are determined from IOTLB entries. Pre-translated addresses are not checked by the IOMMU except to validate that the peripheral has the IOTLB enable bit set in the corresponding Device Table Entry. 

\setlength{\parindent}{3ex}
Optionally, the IOMMU may include support for Peripheral Page Requests (PPR) for peripherals that use Address Translation Services (ATS). This creates a mechanism for peripherals and software to reduce the need for pinned pages during I/O. The IOMMU may include optional support for interrupt virtualization. This uses a virtualized guest APIC (Advanced Programmable Interrupt Controller), which is a family of interrupt controllers, (one implementation of a guest APIC is the Advanced Virtual Interrupt Controller) with memory tables to deliver interrupts to guest VMs. 

More information about the AMD IOMMU can be found in the AMD I/O Virtualization Technology (IOMMU) specification \cite{Reference7}.

\vfill
\footnoterule
\pagebreak

\section{ARM IOMMU}

\setlength{\parindent}{3ex}

ARM's I/O Memory Management Unit (IOMMU) is called SMMU (System Memory Management Unit) and is the IOMMU that was described in Section \ref{context}. A more detailed information about the SMMU Architecture and Operation can be found in Chapter \ref{Chapter3}.

Most of the information related to the SMMU is coming from the corresponding manual of the ARM IOMMU \cite{Reference1} and the Xilinx UltraScale+ TRM (Technical Reference Manual) \cite{Reference2}.

\vfill
\footnoterule


\chapter{Overview of the ARM SMMU} 

\label{Chapter3} 


\setlength{\parindent}{3ex} 

In this chapter, we will describe the first part of the contributions of this thesis, which is the background/theory related to the SMMU. More specifically, we will present an overview of the information related to the Architecture and the Operation of the SMMU, that we found reading the manuals of the ARM SMMU version 2 \cite{Reference1} and the Xilinx Zynq UltraScale+ MPSoC \cite{Reference2}.


\section{Architecture}

The ARM SMMU architecture supports:

\begin{itemize}
	
	\item Aarch32 short descriptor (up to a 32-bit virtual-address and up to 32-bit physical-address)
	\item Aarch32 long descriptor (up to a 32-bit virtual-address and up to 40-bit physical-address)
	\item Aarch64 descriptor (up to a 49-bit virtual-address and up to 48-bit physical-address)
	
\end{itemize}

\setlength{\parindent}{3ex}
\noindent

An implementation of the ARM SMMU architecture can provide multiple transaction contexts, that apply to specific streams of transactions and single or two stage translation. For any stage of translation, it can provide multiple levels of address lookup for fine-grained memory control. Also, it can provide fault handling, logging, signaling and debug and optional performance monitoring functionality.

%
%

\setlength{\parindent}{3ex} 
\noindent

\begin{figure}[h!]
	
	\centering
	
	\captionbox[Text]{An ARM SMMU in the memory system \label{fig:4.1}\addtocounter{figure}{-1}\subcaption*{Source: ARM \cite{Reference1} }}{%
		\includegraphics[width=0.6\textwidth]{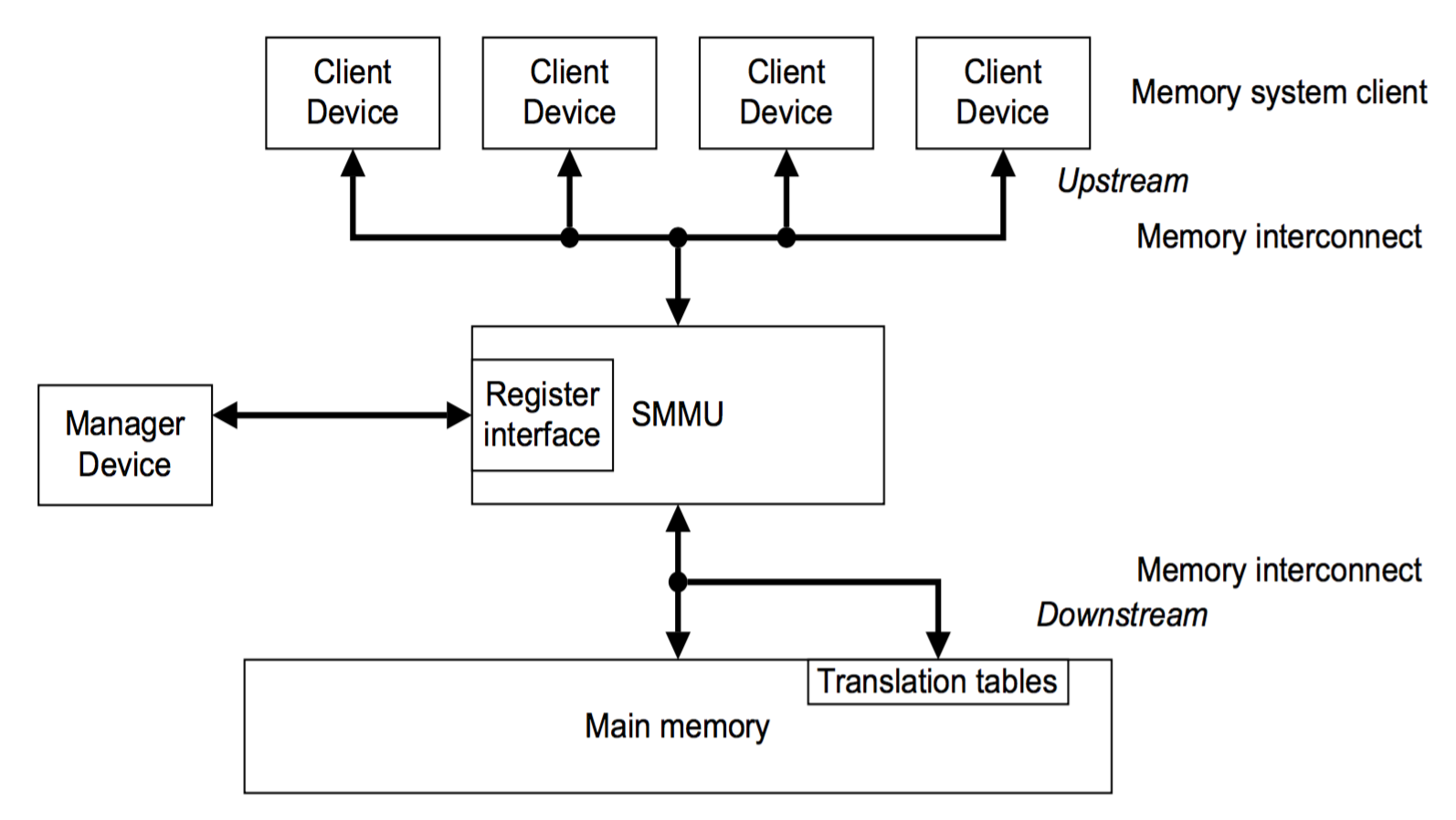}  
	}
	
\end{figure}

\vfill
\footnoterule
\pagebreak

\setlength{\parindent}{3ex} 

As we can see in Figure \ref{fig:4.1}, which shows an implementation of an ARM SMMU in the memory system, one or more Client devices are connected to the SMMU through the memory. Client devices are described as being the "upstream" of the SMMU. The connection between the SMMU and the client devices is the upstream bus. The SMMU is connected to the rest of the memory system, through the main memory. The rest of the memory system is described as being the "downstream" of the SMMU. The connection between the SMMU and the rest of the memory system is the downstream bus. A client device issues a transaction request to the SMMU. The SMMU processes that transaction and returns a response to the client.

\subsection{ARM Exception levels and Execution states}
\setlength{\parindent}{3ex} 

The ARMv8 exception model defines the Exception Level EL$n$, where $n$ can have the values 0-3. Software execution privilege increases with the increase in the value of $n$, with EL$0$ software having the lowest level of privilege. Execution at EL$0$ is described as unprivileged execution. It is implementation defined whether a processor implementation includes EL$2$ or EL$3$. \\

\setlength{\parindent}{0ex} 
The typical use of the different Exception levels is:

\begin{itemize}
	
	\item EL$0$  Application software $\rightarrow$ Secure state (i.e.: The processor can access both the Secure and the Non-secure memory address space. When executing at EL3, the processor can access all the system control resources.) or Non-secure state (i.e.: The processor can access only the Non-secure memory address space and cannot access the Secure system control resources.)
	\item EL$1$  Operating system $\rightarrow$ Secure or Non-secure state
	\item EL$2$  Hypervisor $\rightarrow$ Non-secure state only
	\item EL$3$  Secure monitor
	
\end{itemize}

\subsection{ARM translation regimes}
\setlength{\parindent}{3ex} 

The ARM architecture supports different translation regimes and stages for the AArch32 (32-bits) and the AArch64 (64-bits). Figure \ref{fig:4.2} (next page) shows the ARM architecture translation regimes for memory accesses made from AArch64 state, where:

\begin{itemize}
	\item VA is the Virtual Address
	\item PA is the Physical Address
	\item IPA is the Intermediate Physical Address (used in virtualization context)
\end{itemize}

\begin{figure}[h!]
	
	\centering
	
	\captionbox[Text]{ARM processor architecture AArch64 translation regimes and stages \label{fig:4.2}\addtocounter{figure}{-1}\subcaption*{Source: ARM \cite{Reference1} }}{%
		\includegraphics[width=0.9\textwidth]{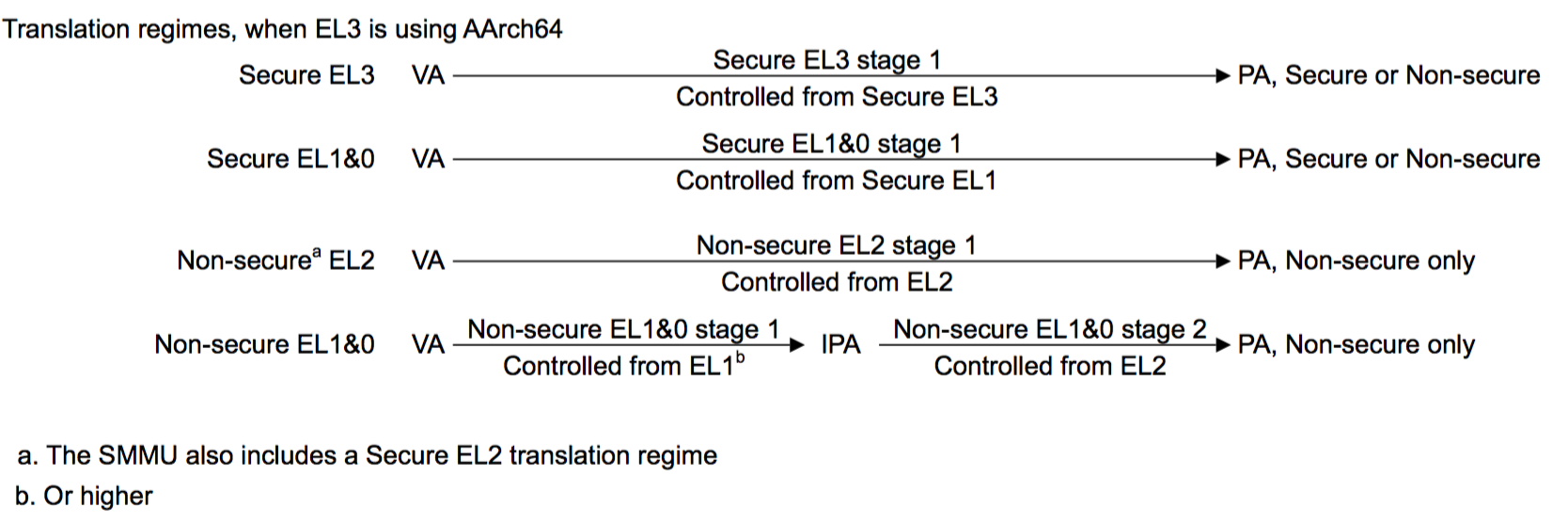}  
	}
\end{figure}

\vfill
\footnoterule
\pagebreak

\setlength{\parindent}{3ex} 

In the ARM processor architecture, the behavior of a PE (Processing Element) that is executing in AArch32 state is defined in relation to different PE modes. Most of these modes have the level of execution privilege that is appropriate to an operating system. In a processor implementation that includes EL3, the Exception level of these modes can depend on whether EL3 is using AArch32 or AArch64, something we can also confirm from the Figure \ref{fig:4.3} below.

\begin{figure}[h!]
	
	\centering
	
	\captionbox[Text]{Mapping of AArch32 PE modes to Exception levels \label{fig:4.3}\addtocounter{figure}{-1}\subcaption*{Source: ARM \cite{Reference1} }}{%
		\includegraphics[width=\textwidth]{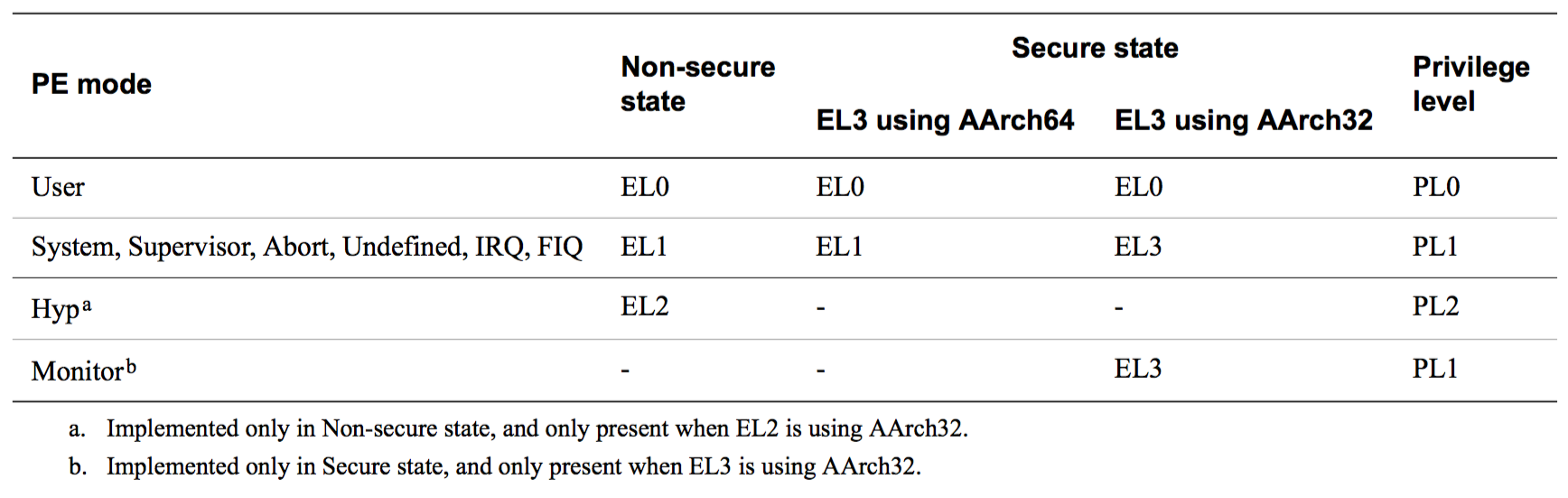}  
	}
	
\end{figure}

Figure \ref{fig:4.4} shows the ARM architecture translation regimes for memory accesses made from AArch32 state.

\begin{figure}[h!]
	
	\centering
	
	\captionbox[Text]{ARM processor architecture AArch32 translation regimes and stages \label{fig:4.4}\addtocounter{figure}{-1}\subcaption*{Source: ARM \cite{Reference1} }}{%
		\includegraphics[width=0.8\textwidth]{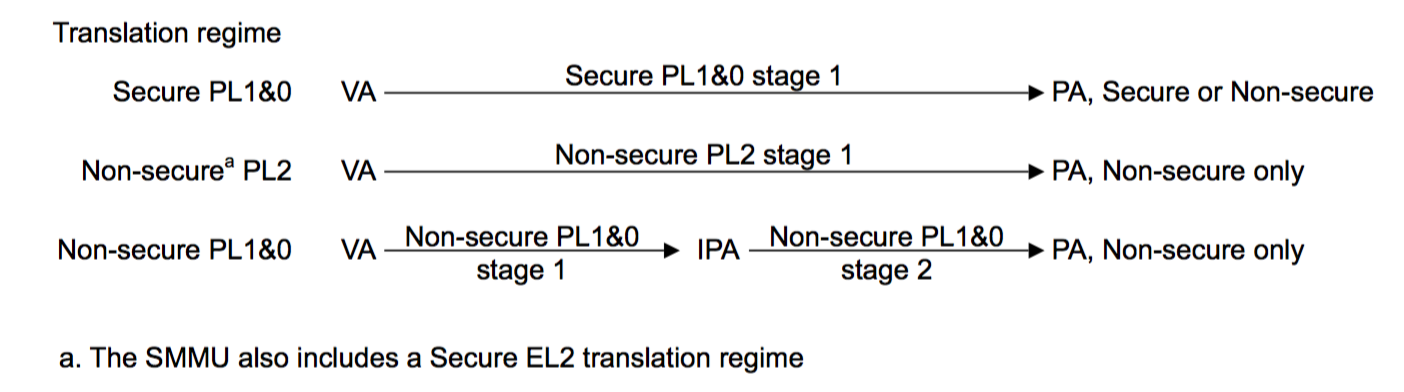}  
	}
	
\end{figure}

\vfill
\footnoterule
\pagebreak

\subsection{SMMU Translation Schemes}
\setlength{\parindent}{3ex} 

The SMMUv2 translation schemes support address translations required by both the AArch32 and AArch64 Execution states, for each translation regime, as follows:

The Aarch32 short descriptor uses 32-bit entries, or descriptors, in its translation tables. Also, this descriptor uses up to a 32-bit VA and 32-bit (or 40 bits with loss of granularity) PA and is compatible with the ARMv7 architecture.

The Aarch32 long descriptor uses 64-bit entries in its translation tables. Also, this descriptor uses up to a 32-bit VA and 40-bit IPA or PA and is compatible with the ARMv7 architecture.

The Aarch64 descriptor uses 64-bit entries in its translation tables. Also the Aarch64 descriptor uses up to a 49-bit VA and 48-bit, as two independent address ranges and is defined by the ARMv8 architecture.

\textbf{Translation granule} is the smallest region of input address space that an SMMU implementation can be configured to support. It defines both the maximum size of a single translation table and the memory page size, that is, the granularity at which attributes can be assigned to memory regions.
%

\setlength{\parindent}{0ex} 
\textbf{Aarch32 translation scheme}: supports a translation granule of 4 KB.

\begin{figure}[h!]
	
	\centering
	
	\captionbox[Text]{AArch32 translation regimes \label{fig:4.5}\addtocounter{figure}{-1}\subcaption*{Source: ARM \cite{Reference1} }}{%
		\includegraphics[width=0.8\textwidth]{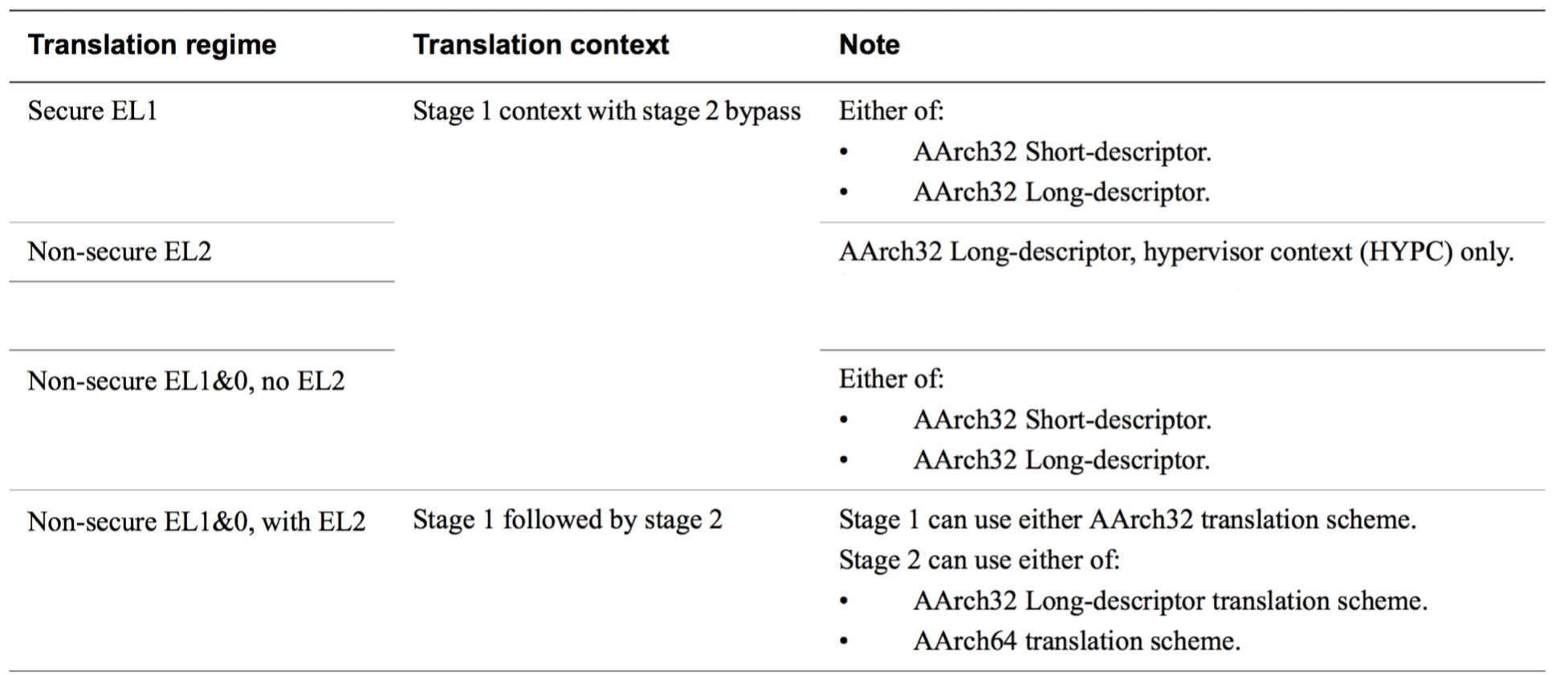}  
	}
	
\end{figure}

\textbf{Aarch64 translation scheme}: supports a translation granule of 4 KB, 16 KB, or 64 KB.\\

\begin{figure}[h!]
	
	\centering
	
	\captionbox[Text]{AArch64 translation regimes \label{fig:4.6}\addtocounter{figure}{-1}\subcaption*{Source: ARM \cite{Reference1} }}{%
		\includegraphics[width=0.8\textwidth]{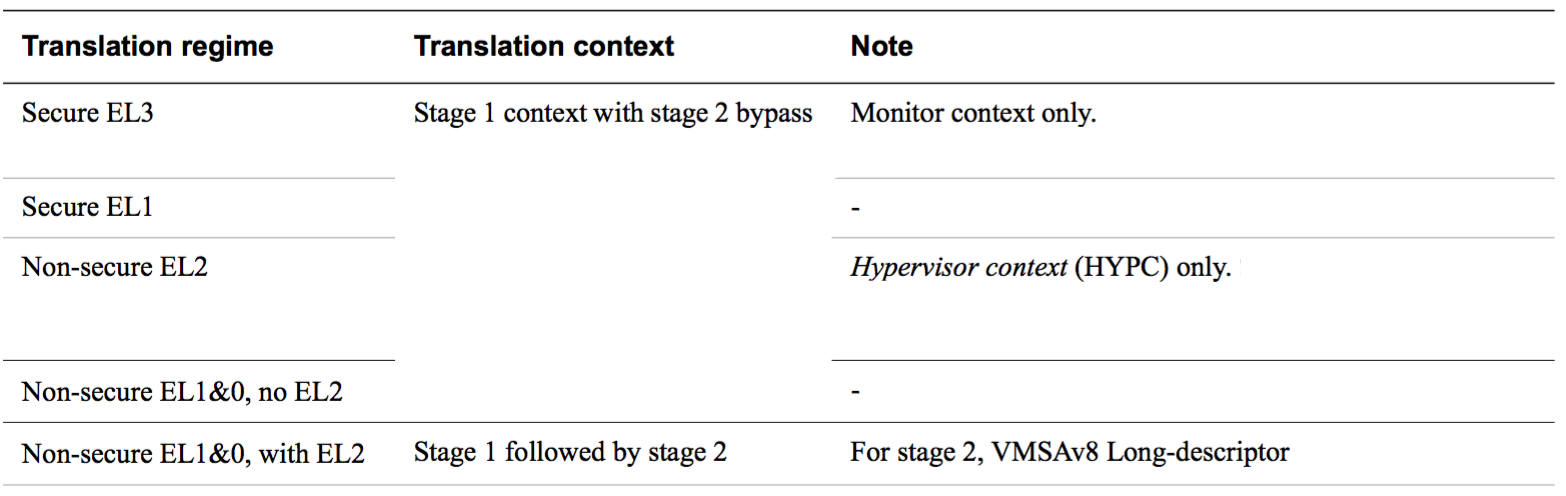}  
	}

\end{figure}

\vfill
\footnoterule
\pagebreak

\subsection{SMMU address support}
\setlength{\parindent}{3ex} 

The SMMUv2 architecture supports virtual addresses up to 32 bits when held in 32-bit registers, up to 49 bits when held in 64-bit registers (when bit[49] is valid, it determines the Translation Table Base Register (TTBR)).
It also supports Translation granule sizes of 4 KB, 16 KB, and 64 KB - it is implementation defined which of these granule sizes are supported. 

%

\subsection{SMMU address handling sequence}
\setlength{\parindent}{3ex}

A transaction bypasses SMMU translation or it does not. In certain cases, the input address is sign-extended. For each stage of translation, the SMMU performs checks on the input address and the output address (i.e.: whether a stage 1 or stage 2 translation context bank is used, the value of SMMU{\_}CB$n${\_}SCTLR.M for the context bank, whether a 32-bit or 64-bit descriptor format is used).

\section{SMMU Operation}
\setlength{\parindent}{3ex}

In this section we will describe the steps that the SMMU performs on receiving a memory access request. At the memory system level, in performing address translation, an SMMU controls: 
\begin{itemize}
	
	\item Security state determination, which identifies whether a transaction is from a Secure or Non-secure device
	\item Context determination, which identifies the stage 1 or stage 2 context resources the SMMU uses to process a transaction. In some cases, the configuration settings for a transaction mean that transaction bypasses the translation process, or is faulted
	\item Memory access permissions and determination of memory attributes
	\item Memory attribute checks
	\item TLB operation
	
\end{itemize}
\vspace{0.05cm}
Figure \ref{fig:4.7} shows the SMMU partitioning and the different SMMU transactions. 
\begin{figure}[h!]
	
	\centering

	\captionbox[Text]{SMMU partitioning \label{fig:4.7}\addtocounter{figure}{-1}\subcaption*{Source: ARM \cite{Reference1} }}{%
		\includegraphics[width=0.55\textwidth]{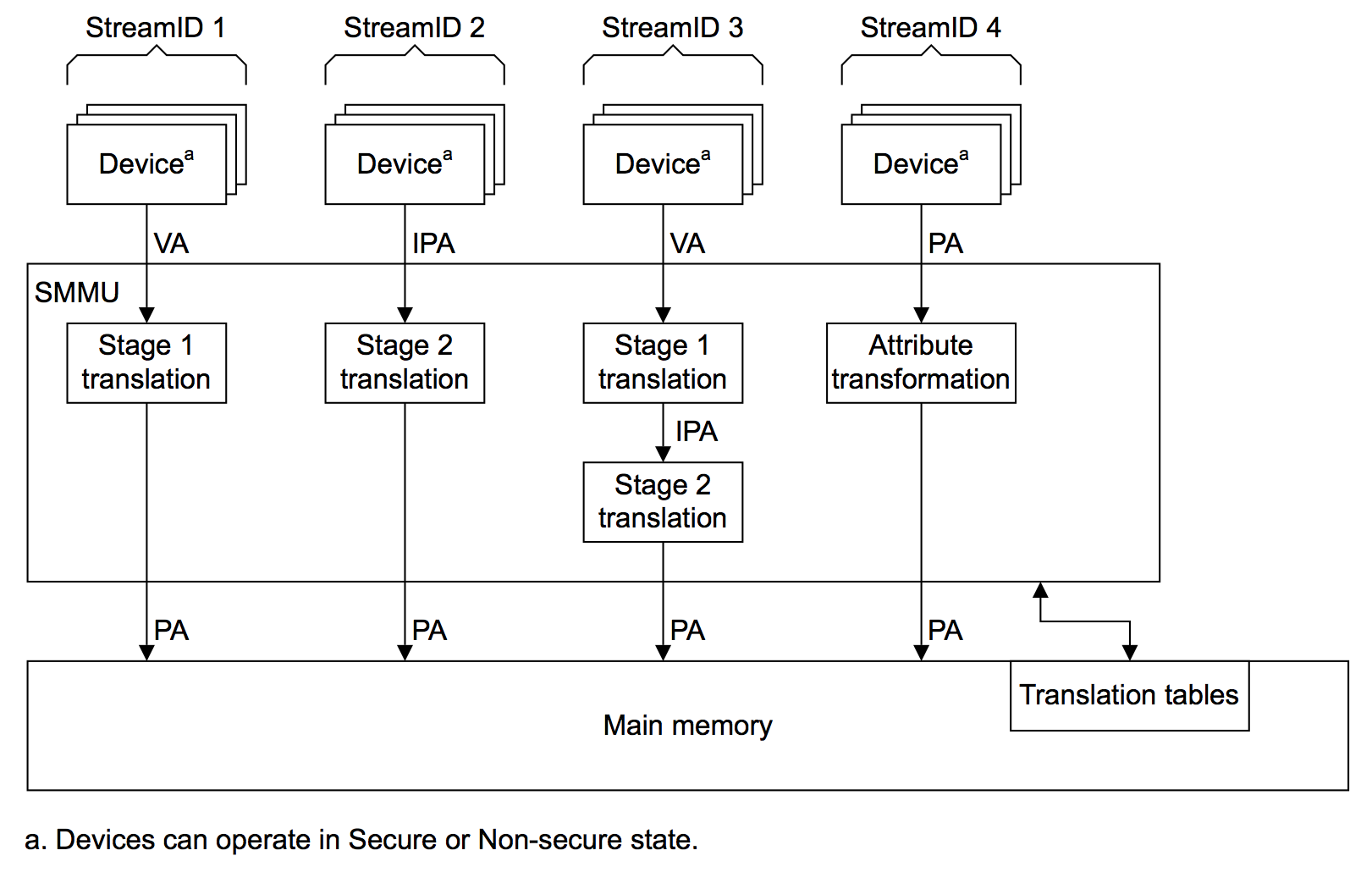}  
	}
	
\end{figure}

\vfill
\footnoterule
\pagebreak

\setlength{\parindent}{3ex} 

The \textbf{Stream Identifier (StreamID)} uniquely identifies a stream of transactions that can originate from different devices, but are associated with the same context, and are therefore subject to the same type of SMMU processing. The mechanism by which the StreamID is obtained from the incoming bus transaction is implementation defined.

The \textbf{Memory attributes} can be modified by SMMU. In some cases, SMMU performs attribute transformation only, with no address translation. In this case, the output physical address is the same with the input virtual address.

\setlength{\parindent}{3ex} 
\noindent

An access to the SMMU is referred to as a \textbf{transaction}. A client transaction is an access by a client device, that the SMMU is to process. A configuration transaction is a device access to a register in the SMMU configuration address space.

\setlength{\parindent}{0ex} 

\vspace{0.05cm}

Figure \ref{fig:4.8} below shows the generic SMMU process flow.


\begin{figure}[h!]
	
	\centering
	
	\captionbox[Text]{SMMU process flow \label{fig:4.8}\addtocounter{figure}{-1}\subcaption*{Source: ARM \cite{Reference1} }}{%
		\includegraphics[width=0.77\textwidth]{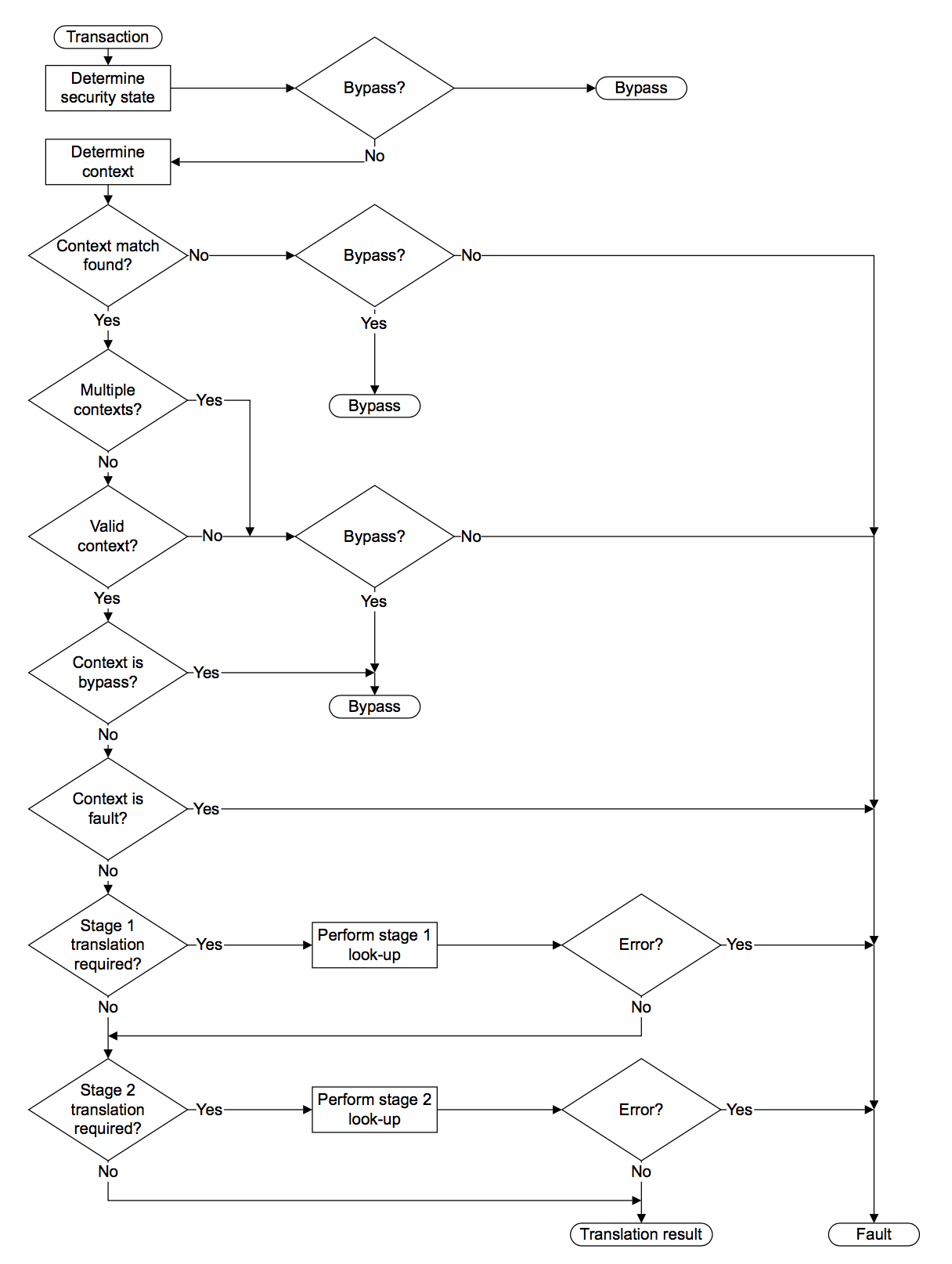}  
	}
	
\end{figure}

\vfill
\footnoterule
\pagebreak

\subsection{Context determination}

The SMMU processes a transaction in one of the following ways:
\begin{itemize}
	\item Bypass address translation but transform attributes
	\item Fault the transaction
	\item Require the resources of either one or two translation context banks, each having its own set of translations, attributes and permissions, that apply to a transaction being processed by that context bank
\end{itemize}
\setlength{\parindent}{0ex} 

The context determination actually determines the resources the SMMU uses to process a transaction. The SMMU can process transactions from multiple sources, potentially using a different context for each transaction. 
\setlength{\parindent}{3ex} 
\noindent

The \textbf{Transaction stream} is a sequence of transactions associated with a particular thread of activity in the system. The \textbf{StreamID} is derived from transaction identification information, such as:
\begin{itemize}
	\item The transaction ID
	\item The read and write status
	\item The Security state of the upstream bus transaction
\end{itemize}
\setlength{\parindent}{0ex} 
The number of implemented StreamID bits in SMMUv2 lies in the range: 0-16 and in SMMUv1 lies in the range: 0-15. A zero-bit StreamID might exist if the source of a single StreamID has a dedicated SMMU.
\setlength{\parindent}{3ex} 
\noindent

\textbf{Stream mapping} is the process of mapping a StreamID to the associated Stream-to-Context register (SMMU{\_}S2CR$n$), that defines the processing required for the stream. There are three (3) Stream mapping mechanisms defined by SMMUv2:
\begin{itemize}
	\item \textbf{Stream matching} \newline
	The StreamID is looked up in the set of the Stream Match registers, as known as the SMMU{\_}SMR$n$. When a unique match is found, the corresponding SMMU{\_}S2CR$n$ holds the context for the stream.
	SMMUv1 provides up to 128 Stream Match registers. An SMMUv2 implementation can provide up to 128 Stream Match Registers or can implement the optional Extended Stream Matching Extension, that provides up to 1024 Stream Match Registers and the associated Stream-to-Context registers.
	
	\textbf{Note}: The implementation/architecture of our Zynq UltraScale+ MPSoC provides 48 Stream Match registers and Stream-to-Context registers.
	
	\item \textbf{Stream indexing} \newline
	When Stream indexing is used, the StreamID is a direct index to the required SMMU{\_}S2CR$n$ (register). That means, if the StreamID is $m$, the required Stream-to-Context register is the SMMU{\_}S2CR$m$. The maximum StreamID is determined by the number of implemented SMMU{\_}S2CR$n$ registers.
	\vfill
	\footnoterule
	\pagebreak
	\item \textbf{Compressed StreamID indexing} (available only in an SMMUv2 implementation that includes the optional StreamID Compressed Indexing extension.) \newline
	When Compressed StreamID indexing is used, the StreamID is an indirect index to the required SMMU{\_}S2CR$n$, as follows:
	\begin{enumerate}
		\item The StreamID, $m$, indexes a single-byte S2CRIndex$i$ field in the array of SMMU{\_}COMPINDEX$n$ registers provided by the StreamID Compressed Indexing extension.
		\item The S2CRIndex$i$ field holds the value of the SMMU{\_}S2CR$n$ for the stream. So, if the value of the S2CRIndex$i$ field is $x$, the required Stream-to-Context register is SMMU{\_}S2CR$x$.
	\end{enumerate}
	Only a single Security state can use Compressed StreamID indexing. In an SMMUv2 implementation that supports two Security states, if one Security state is using Compressed StreamID indexing then the other Security state must use Stream matching.
	
\end{itemize}

\setlength{\parindent}{0ex} 
\textbf{Note}: In our case, we used the Stream Matching mechanism -- it is considered the default mechanism for the remaining of this thesis.

\setlength{\parindent}{3ex} 
\noindent

The \textbf{Stream mapping table} (Figure \ref{fig:4.19}) maps a transaction stream to a context and consists of a number of entries, where each entry is a Stream mapping register group containing the following registers, with $n$ defining the Stream mapping register group number. Each entry has:
\begin{enumerate}
	\item a SMMU{\_}SMR$n$, which determines whether a transaction matches the group.
	During configuration, the Stream Match Register table can have multiple entries that match the same Stream Identifier value, possibly resulting in "unpredictable" behavior. To prevent multiple matches, software must ensure that no transactions that might match the StreamID are received, by:
	\begin{itemize}
		\item Disabling any source client devices that might match
		\item Ensuring that no outstanding transactions from these client devices are in progress
	\end{itemize}
	As an extra precaution, software can first disable all affected SMMU{\_}SMR$n$ table entries by setting the SMMU{\_}SMR$n$.VALID bit to 0, then reprogramming the entries as appropriate.
	
	\textbf{Format of the SMMU{\_}SMR$n$ register}:\\
	When the SMMU{\_}sCR0.EXIDENABLE (Extended Stream Matching Extension) is: 
	\begin{itemize}
		\item 0 (zero):
		\begin{figure}[h!]
			
			\centering
			
			\captionbox[Text]{SMR$n$ \label{fig:4.9}\addtocounter{figure}{-1}\subcaption*{Source: ARM \cite{Reference1} }}{%
				\includegraphics[width=0.6\textwidth]{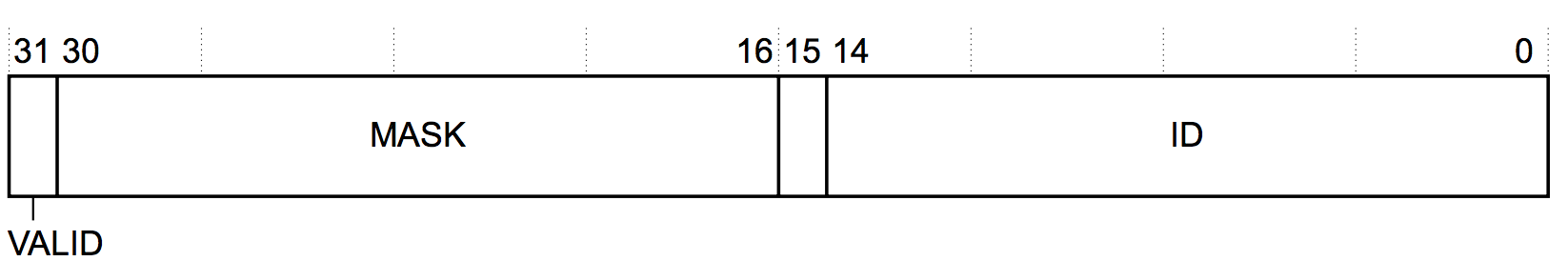}  
			}

		\end{figure}
		\vfill
		\footnoterule
		\pagebreak
		\item 1 (one): 
		\begin{figure}[h!]
			
			\centering
			
			\captionbox[Text]{SMR$n$ "extended" \label{fig:4.10}\addtocounter{figure}{-1}\subcaption*{Source: ARM \cite{Reference1} }}{%
				\includegraphics[width=0.6\textwidth]{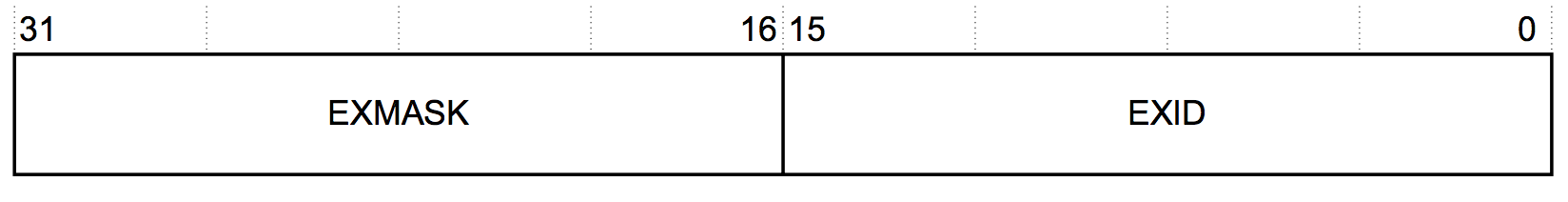}  
			}
		\end{figure}
	\end{itemize}
	Some explanations:
	When MASK[i] == 1, then the ID[i] is ignored. Same with EXMASK and EXID. ID/EXID is actually the StreamID to be matched. When VALID==1, the entry is included in the Stream mapping table search.
	
	
	\item and a SMMU{\_}S2CR$n$, which specifies the initial context for the translation process.\\
	\textbf{Format of the SMMU{\_}S2CR$n$ registers}:\\
	When the bits[17:16], the "Type" field of these registers is:

	\begin{itemize}
		\item 00, the type means Translation Context:
		\begin{figure}[h!]
			
			\centering
			
			\captionbox[Text]{S2CR$n$ Type: 00 \label{fig:4.11}\addtocounter{figure}{-1}\subcaption*{Source: ARM \cite{Reference1} }}{%
				\includegraphics[width=0.7\textwidth]{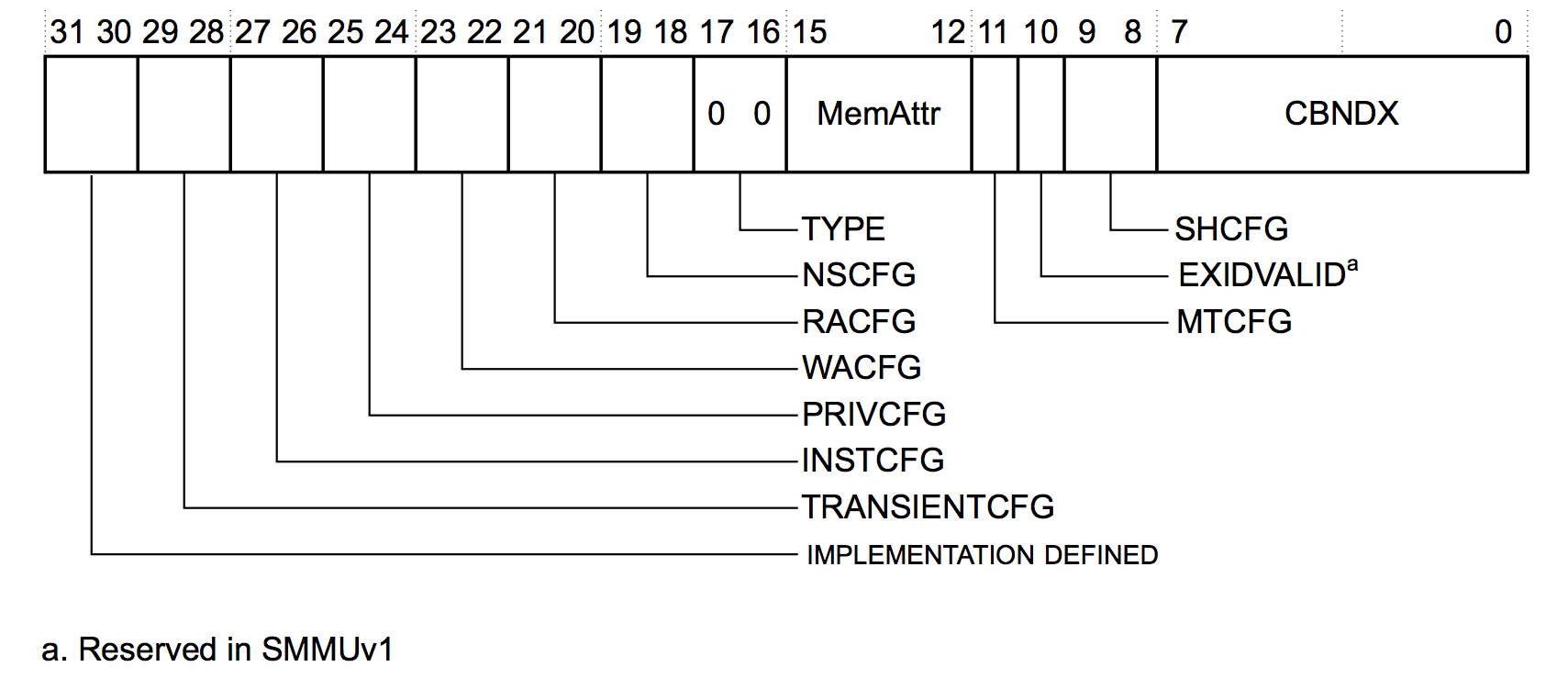}  
			}
			
		\end{figure}
		\item 01, the type means Bypass Mode:
		\begin{figure}[h!]
			
			\centering
			
			\captionbox[Text]{S2CR$n$ Type: 01 \label{fig:4.12}\addtocounter{figure}{-1}\subcaption*{Source: ARM \cite{Reference1} }}{%
				\includegraphics[width=0.7\textwidth]{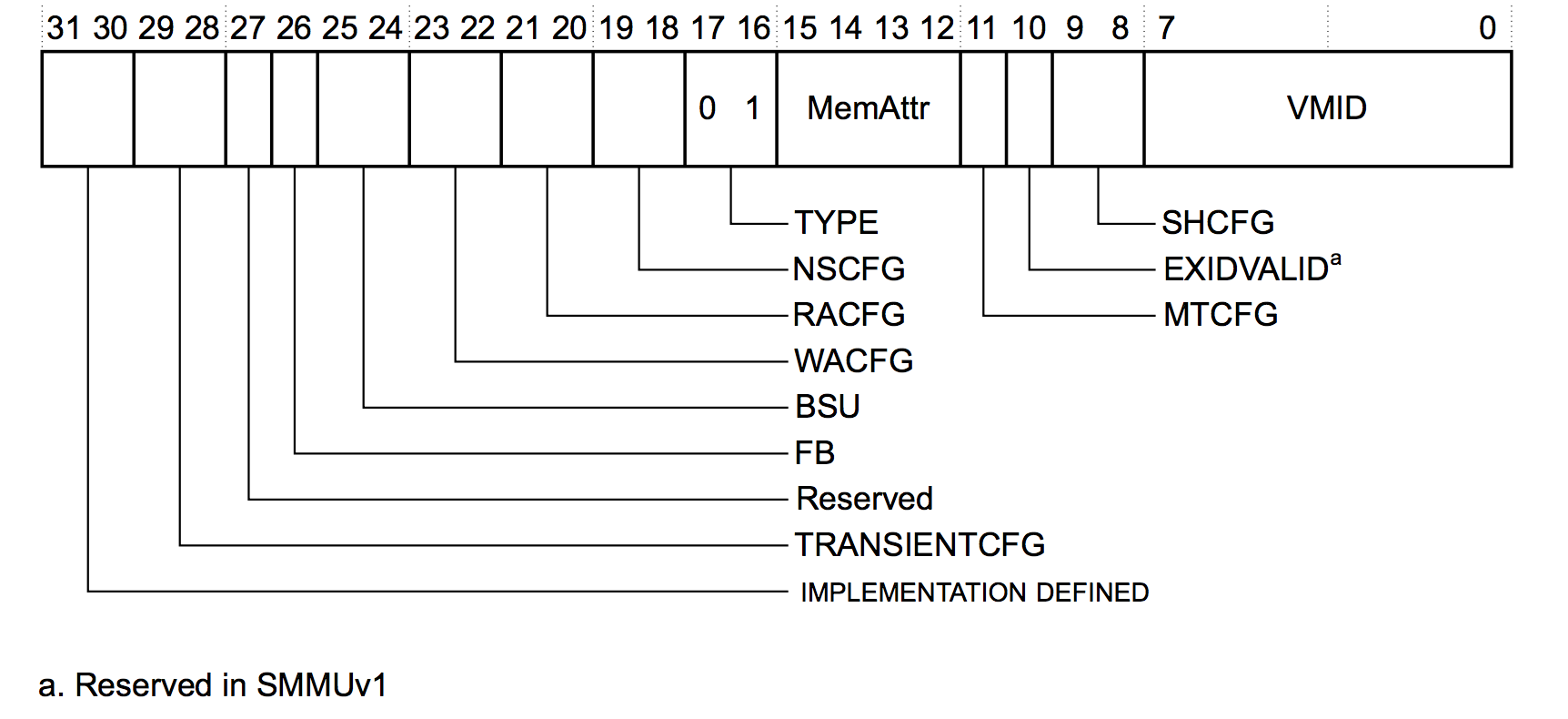}  
			}
		\end{figure}
		
		\vfill
		\footnoterule
		\pagebreak
		
		\item 10, the type means Fault Context:
	
		\begin{figure}[h!]
			
			\centering
			
			\captionbox[Text]{S2CR$n$ Type: 10 \label{fig:4.13}\addtocounter{figure}{-1}\subcaption*{Source: ARM \cite{Reference1} }}{%
				\includegraphics[width=0.7\textwidth]{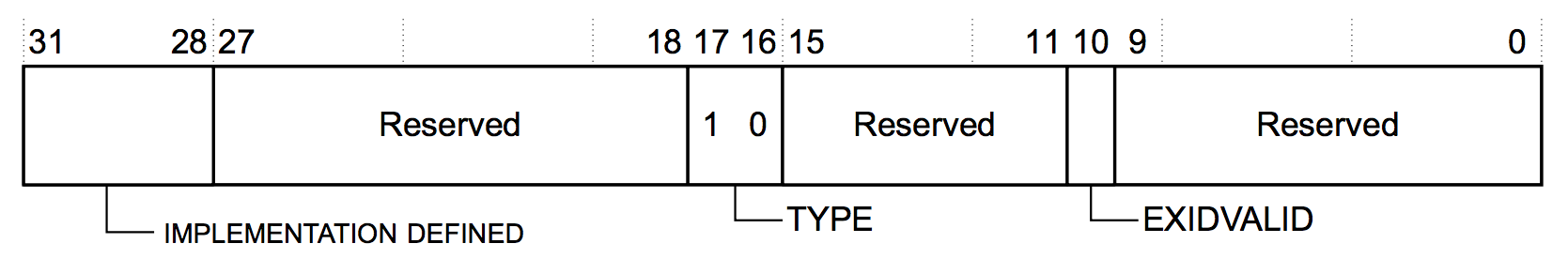}  
			}
		\end{figure}
		\item 11 is Reserved and treated as a fault context
	\end{itemize}
\end{enumerate}

\setlength{\parindent}{3ex} 
\noindent

At this point, it is not important to describe the meaning of all fields of the SMMU{\_}S2CR$n$ register, since someone can easily find it on the related manual. For the purposes of this thesis, it is important to mention that when the "Type" field is 00, meaning Translation Context, after a match from the S2CR$n$ register, the CBNDX field of the corresponding SMMU{\_}S2CR$n$ is the index of the Context Bank that is "responsible" for the matched transaction.

\setlength{\parindent}{3ex} 
\noindent

When the SMMU{\_}S2CR$n$ register specifies that the initial context for a transaction is a translation context bank, or as we said the "Type" field is 00, then the \textbf{SMMU{\_}CBAR$n$} register specifies the configuration attributes for the translation context bank $n$.

\setlength{\parindent}{0ex}
\textbf{Format of the SMMU{\_}CBAR$n$ registers}:\\
When the bits[17:16], the "Type" field of these registers is:
\begin{itemize}
	
	\item 00, the type means Stage 2 context. The context bank provides stage 1 translation, and stage 2 memory attribute-only transformations:
	\begin{figure}[h!]
		
		\centering
		
		\captionbox[Text]{CBAR$n$ Type: 00 \label{fig:4.14}\addtocounter{figure}{-1}\subcaption*{Source: ARM \cite{Reference1} }}{%
			\includegraphics[width=0.7\textwidth]{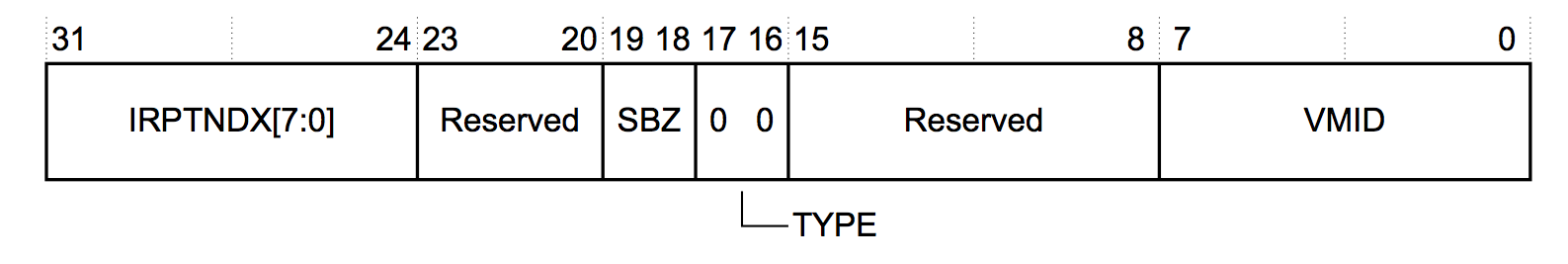}
		}
	\end{figure}
	\item 01, the type means Stage 1 context with stage 2 bypass:
	In SMMUv2, if the value of the SMMU{\_}IDR2.E2HS bit is 1, then the implementation supports an additional context type, E2HC, that is intended to be used by an OS, that is also acting as a Hypervisor.\\
	When: 
	\begin{itemize}
		\item no support for E2HC , or SMMU{\_}CR0.HYPMODE == 0:
		
		\begin{figure}[h!]
			
			\centering
			\captionbox[Text]{CBAR$n$ Type: 01 with no E2HC \label{fig:4.15}\addtocounter{figure}{-1}\subcaption*{Source: ARM \cite{Reference1} }}{%
				\includegraphics[width=0.7\textwidth]{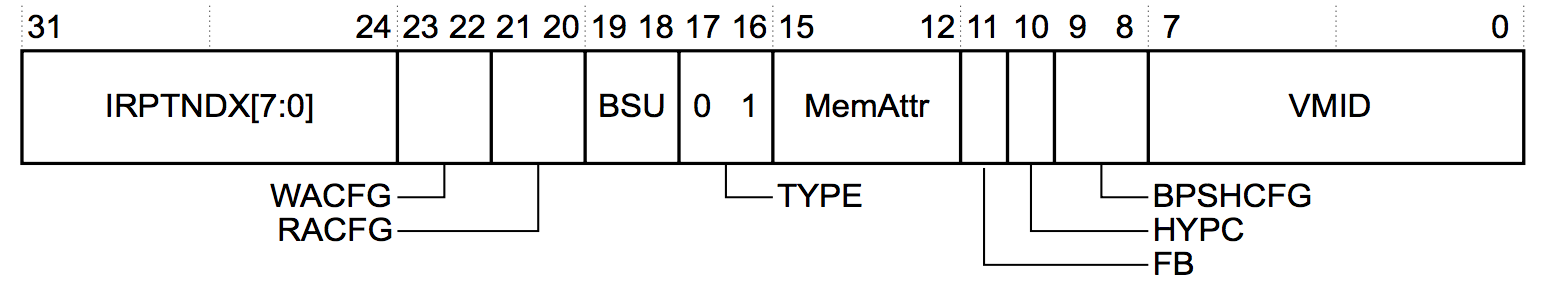}
			}
		\end{figure}
		\vfill
		\footnoterule
		\pagebreak
		\item SMMU{\_}CR0.HYPMODE == 1:
		\begin{figure}[h!]
			
			\centering
			
			\captionbox[Text]{CBAR$n$ Type: 01 with E2HC \label{fig:4.16}\addtocounter{figure}{-1}\subcaption*{Source: ARM \cite{Reference1} }}{%
				\includegraphics[width=0.7\textwidth]{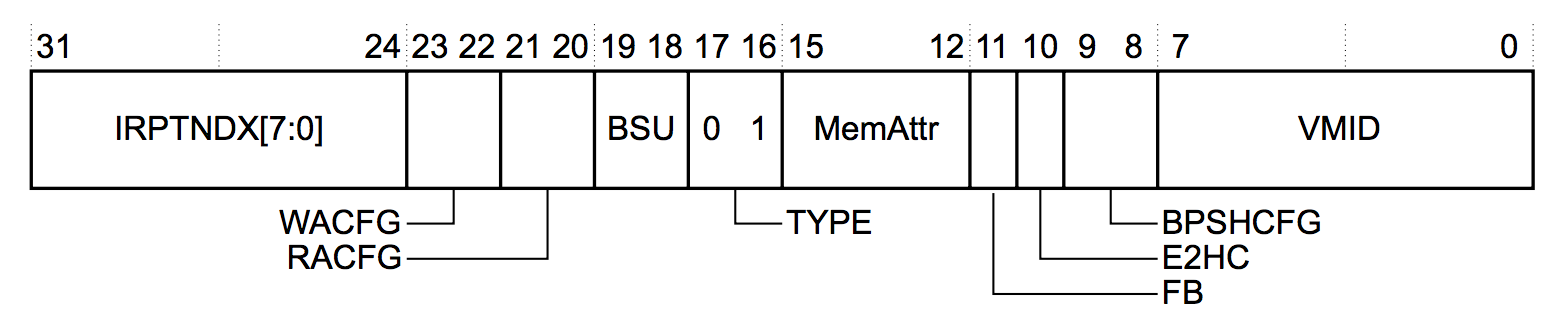}
			}
			
		\end{figure}
		
	\end{itemize}	
	
	\item 10, the type means Stage 1 context with stage 2 fault. The context bank generates an invalid context fault:
	\begin{figure}[h!]
		
		\centering
		
		\captionbox[Text]{CBAR$n$ Type: 10 \label{fig:4.17}\addtocounter{figure}{-1}\subcaption*{Source: ARM \cite{Reference1} }}{%
			\includegraphics[width=0.7\textwidth]{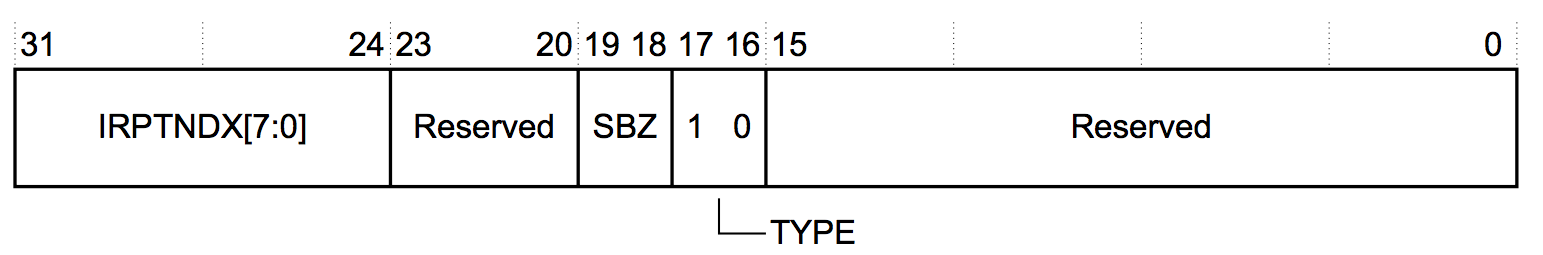}
		}
		
	\end{figure}

	
	\item 11, the type means Reserved and Stage 1 followed by stage 2 translation context. The context bank provides stage 1 translation followed by stage 2 translation, and specifies an additional translation context bank for the stage 2 translation.:
	\begin{figure}[h!]
		
		\centering
		
		\captionbox[Text]{CBAR$n$ Type: 11 \label{fig:4.18}\addtocounter{figure}{-1}\subcaption*{Source: ARM \cite{Reference1} }}{%
			\includegraphics[width=0.7\textwidth]{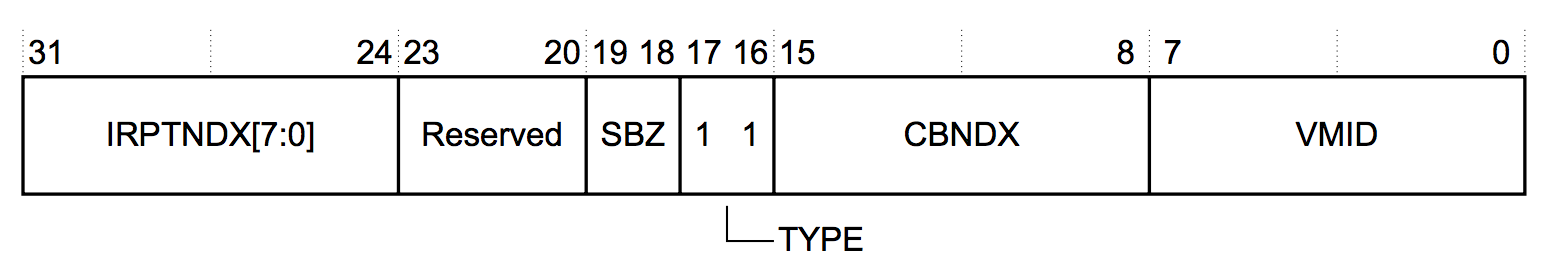}
		}
		
	\end{figure}
	
\end{itemize}

\setlength{\parindent}{3ex} 
\noindent

From the information of the SMMU{\_}CBAR$n$ registers, it is important to mention that each stage of translation (1 and 2) has its own context bank. For instance, if we have a stage 1 translation followed by a stage 2 translation, the "Type" for the initial context bank related to the SMMU{\_}CBAR$n$ will be "11" and this register will also give us the index of the context bank for the stage 2 translation (the CBNDX field of the SMMU{\_}CBAR$n$ register). The context bank (index) for the stage 1, as we mentioned before, can be found on the S2CR$n$ register shown in Figure \ref{fig:4.11}.\\
\setlength{\parindent}{0ex}

\begin{figure}[h!]
	\centering
	\captionbox[Text]{Stream Mapping Table \label{fig:4.19}}[0.6\textwidth][c]{
		\scalebox{1.3}{
			\begin{tabular}{ |c|c| } 
				\hline
				SMR$0$ & S2CR$0$ \\ 
				SMR$1$ & S2CR$1$ \\ 
				SMR$2$ & S2CR$2$ \\ 
				... & ... \\ 
				... & ... \\ 
				... & ... \\ 
				SMR$n$ & S2CR$n$ \\ 
				\hline
			\end{tabular}
		}
	}
\end{figure}

\vfill
\footnoterule
\pagebreak

\textbf{Note}: In our Zynq UltraScale+ MPSoC we have only 48 groups, which means $n$ lies in the range: 0-47.\\


\subsection{Translation Context}
\setlength{\parindent}{3ex}

The translation context (as known as the context bank) provides information and resources required by the SMMU to process a transaction. SMMU can process multiple transaction streams from different threads of execution and supports multiple live translation contexts.\\

\setlength{\parindent}{0ex} 
A translation context bank includes:
\begin{itemize}
	\item state for configuring the translation process 
	\item capturing fault status, and operations for maintaining cached translations
\end{itemize}

\setlength{\parindent}{0ex} 
In implementations that support stage 1 followed by stage 2 translations: 
\begin{itemize}
	\item one translation context bank is specified for single-stage translation.
	\item two translation context banks are specified for two-stage translation.
\end{itemize}

\setlength{\parindent}{3ex} 
\noindent

A translation context bank is arranged as a table in the SMMU configuration address map. Each entry in the table occupies a 4 KB or 64 KB address space.

In theory, SMMU architecture provides space for up to 128 translation context banks.\\

\setlength{\parindent}{0ex} 
\textbf{Note}: In our Zynq UltraScale+ MPSoC, we have only 16 translation context banks.

\subsection{Page Tables}
\label{PGD}
\setlength{\parindent}{3ex} 

A page table is the data structure used by a virtual memory system in a computer Operating System to store the mapping between virtual addresses and physical addresses.\\

\vfill
\footnoterule
\pagebreak

\setlength{\parindent}{0ex} 
\textbf{Linux}\\

\setlength{\parindent}{0ex} 

Architectures that manage their Memory Management Unit (MMU) differently are expected to emulate the three-level page tables or four-level page tables.

\setlength{\parindent}{0ex} 

Logically, Linux has four levels of page tables:
\begin{itemize}
	\item Page Global Directory (PGD)
	\item Page Upper Directory (PUD)
	\item Page Middle Directory (PMD)
	\item Page Table Entry (PTE)
\end{itemize}

\setlength{\parindent}{0ex} 
\textbf{Note}: On architectures that do not require all four levels, inner levels may be collapsed.

\setlength{\parindent}{3ex} 
\noindent

Each process has a pointer (mm{\_}struct$\rightarrow$pgd) to its own Page Global Directory (PGD), which is a physical page frame. This frame contains an array of type pgd{\_}t which is an architecture specific type defined in <asm/page.h>. Each active entry in the PGD table points to a page frame containing an array of Page Upper Directory (PUD) entries of type pud{\_}t. Each active entry in the PUD table points to a page frame containing an array of Page Middle Directory (PMD) entries of type pmd{\_}t, which in turn points to page frames containing Page Table Entries (PTE) of type pte{\_}t, which finally points to page frames containing the actual user data.

\begin{figure}[h!]
	
	\centering
	\captionbox[Text]{Linux 4-Levels Page Tables \label{fig:4.20}}{%
		\includegraphics[width=0.7\textwidth]{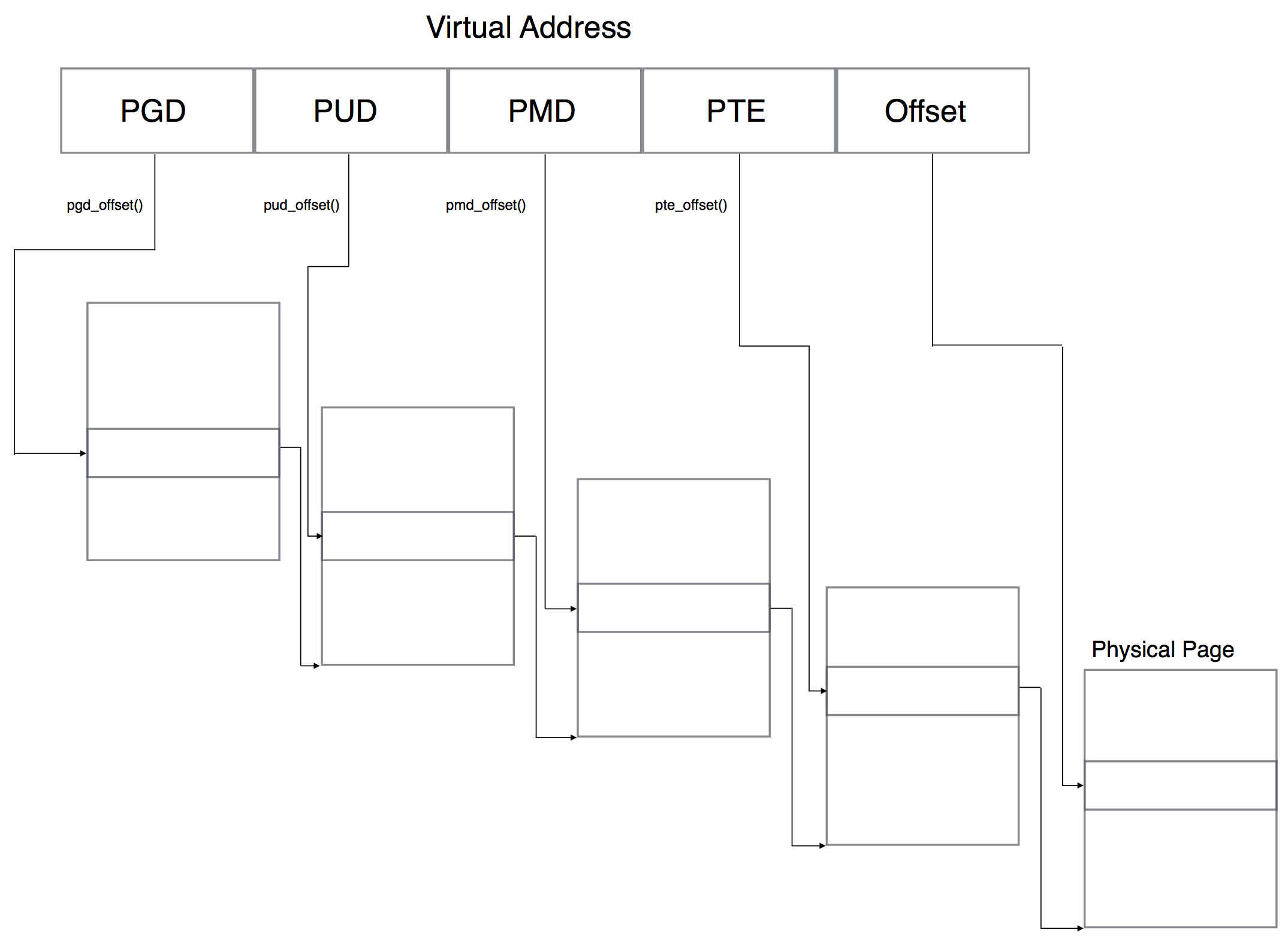}
	}
	
\end{figure}

\setlength{\parindent}{0ex} 
\textbf{SMMU}

\setlength{\parindent}{0ex}

Each context bank of the SMMU can be considered as one page table -- to be more accurate, each context bank has a field that points to a unique page table for this context.

\setlength{\parindent}{3ex}

Each context bank has the \textbf{SMMU{\_}CB$n${\_}TTBR$m$}, where $m$ can be 0 or 1. TTBR0, as known as the Translation Table Base Register 0 holds the base address of translation table 0. Respectively, the TTBR1 holds the base address of transla-
\vfill
\footnoterule
\pagebreak
tion table 0.

\setlength{\parindent}{3ex} 
\noindent

It is recommended by ARM that TTBR0 should be used to store the offset to the page tables used by user processes and TTBR1 should be used to store the offset to the page tables used by the kernel. It seems that most Linux implementations for ARM have decided to just basically eliminate the use of TTBR1 and stick to using TTBR0 for everything -- this is also happening with the ARM SMMU driver (arm-smmu.c), we worked with.

For each context bank there is also the Translation Control Register, called \textbf{SMMU{\_}CB$n${\_}TCR}, that determines translation properties, including which one of the Translation Table Base Registers,
SMMU{\_}CB$n${\_}TTBR$m$, defines the base address for the translation table walk required when an input address is not found in the TLB. An extension of the SMMU{\_}CB$n${\_}TCR exists with the name \textbf{SMMU{\_}CB$n${\_}TCR2}, that basically extends the SMMU{\_}CB$n${\_}TCR by adding control information about the translation granule size and the size of the intermediate physical address.

\noindent

\setlength{\parindent}{10ex}

\vfill
\footnoterule

\chapter{Kernel modules for SMMU support}
\label{Chapter4} 

\setlength{\parindent}{3ex} 


In this chapter, we will describe the second part of the contributions of this thesis, which was the kernel modules we implemented in order to test and use the SMMU of the Zynq UltraScale+ MPSoC. In total we wrote four kernel modules. Below we can see the description of each module:
\label{modules}

\begin{enumerate}
	\item The first module is triggering DMA transactions inside the Processing System using the LPD-DMA or the FPD-DMA.
	\item The second module is testing the ARM SMMU from the Processing System (PS) by triggering a transaction from the PS. In Figure \ref{fig:4.21}, we can see a simplified path for this.
	\begin{figure}[h!]
		
		\centering
		
		\captionbox[Text]{DMAs in the Processing System using the SMMU to access the DRAM \label{fig:4.21}}{%
			\includegraphics[width=0.49\textwidth]{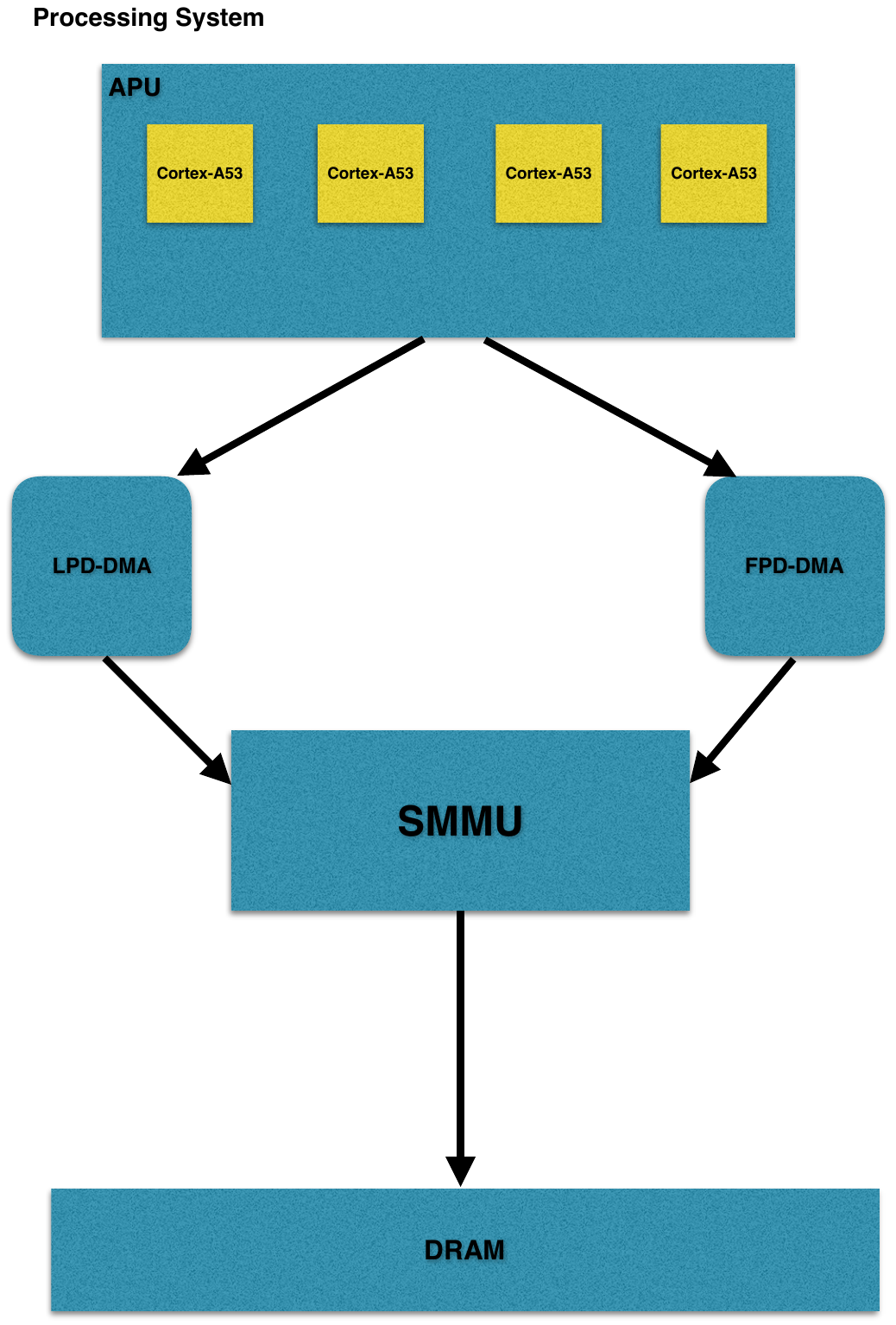}
		}
		
	\end{figure}
	\vfill
	\footnoterule
	\pagebreak
	\item The third module is testing the ARM SMMU from the Programmable Logic (PL) by triggering a transaction from the PL, while the SMMU is having a singe-page entry "pre-mapped". In Figure \ref{fig:4.22}, we can see a simplified path for this.	
	\item The fourth module is using the ARM SMMU from the Programmable Logic, by triggering a transaction from the PL, while the SMMU is having all pages of a user process not "pre-mapped" (from the translation table of the process).
	
	\begin{figure}[h!]
		
		\centering
		
		\captionbox[Text]{DMAs in the Programmable Logic using the SMMU to access the DRAM \label{fig:4.22}}{%
			\includegraphics[width=0.5\textwidth]{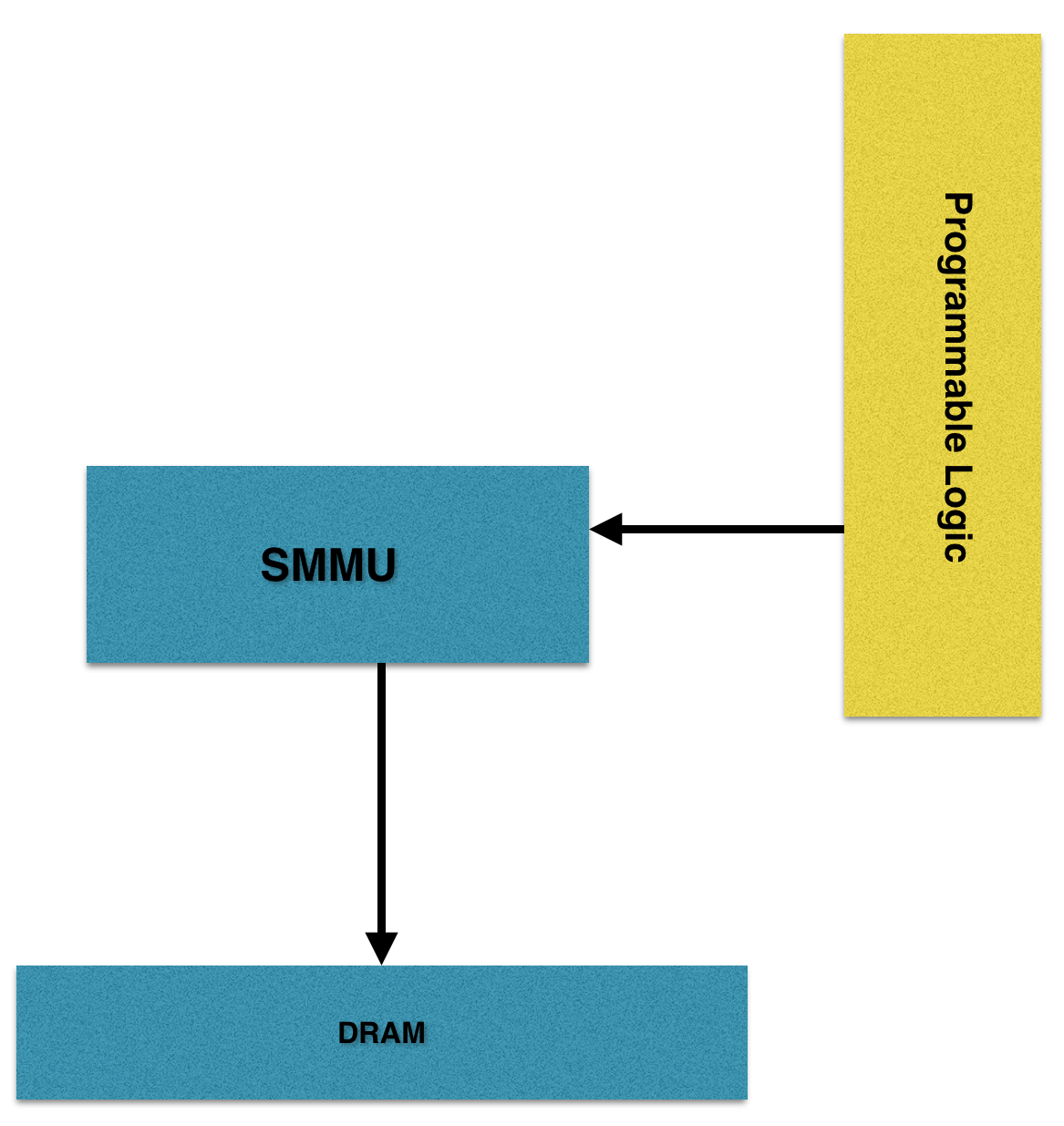}
		}
		
	\end{figure}
\end{enumerate}

\section{DMA test in PS}
\setlength{\parindent}{3ex}


For this module, we actually used a  module/driver, that already exists in Linux 4.4.0 with the name: \textbf{dmatest.c}. This driver tests all the DMA channels in the Processing System (more information about the DMA in the PS is provided in the Appendix \ref{AppendixB}) by triggering a DMA transaction of, in a way, random bytes from a random source address to a random destination address. It also has an option to run many threads per channel.

Although we liked its usage for testing our DMA from the user-space, we wanted something more accurate, meaning that we could give our own source and destination physical addresses, the length of the data and of course, a specific DMA channel of the Processing System (PS).

At this point, it is good to mention that Linux sees each DMA channel as a unique device, so that's why we could easily modify the module in order to "test" the DMA, or better, do a DMA transaction, using a specific DMA channel.

We faced a problem at first, because by default the LPD-DMA channels were disabled in the device-tree generated by Xilinx tools (Appendix \ref{AppendixC}), but we easily 
\vfill
\footnoterule
\pagebreak
solved it by following the steps mentioned in the same Appendix. Eventually, we could test 16 different DMA channels (8 for the FPD-DMA and 8 for the LPD-DMA).

\setlength{\parindent}{3ex} 

The most important function of this module is the \textbf{dmatransfer()}, that actually triggers a DMA transaction and has as arguments, the source address, the destination address, the size/length of the data and the name of the DMA channel.

\section{Transaction from PS with single-page entry}
\label{Transaction-from-the-PS}
\setlength{\parindent}{3ex} 

This module was created in order to test if a transaction with a virtual source and/or destination address can be translated to their physical address, successfully, using the SMMU.

In order to achieve this, we had to deeply understand part of the Linux kernel and the IOMMU API. In general, an application program interface (API), is a set of procedures, protocols, and tools for building software applications. IOMMU API was needed in our case because we wanted to access and use the SMMU. We can see the IOMMU API as the more generic driver that actually uses the more specific IOMMU driver and in our case ARM's SMMU driver (the arm-smmu.c).

First of all, before we continue with this module, it would be good to explain the relation between the devices, the groups and the domains, when it comes to the IOMMU API and drivers.

\begin{figure}[h!]
	
	\centering
	
	\captionbox[Text]{The tree of "hierarchy" of the devices, groups and domains in IOMMU API and drivers. \label{fig:4.23}}{%
		\includegraphics[width=\textwidth]{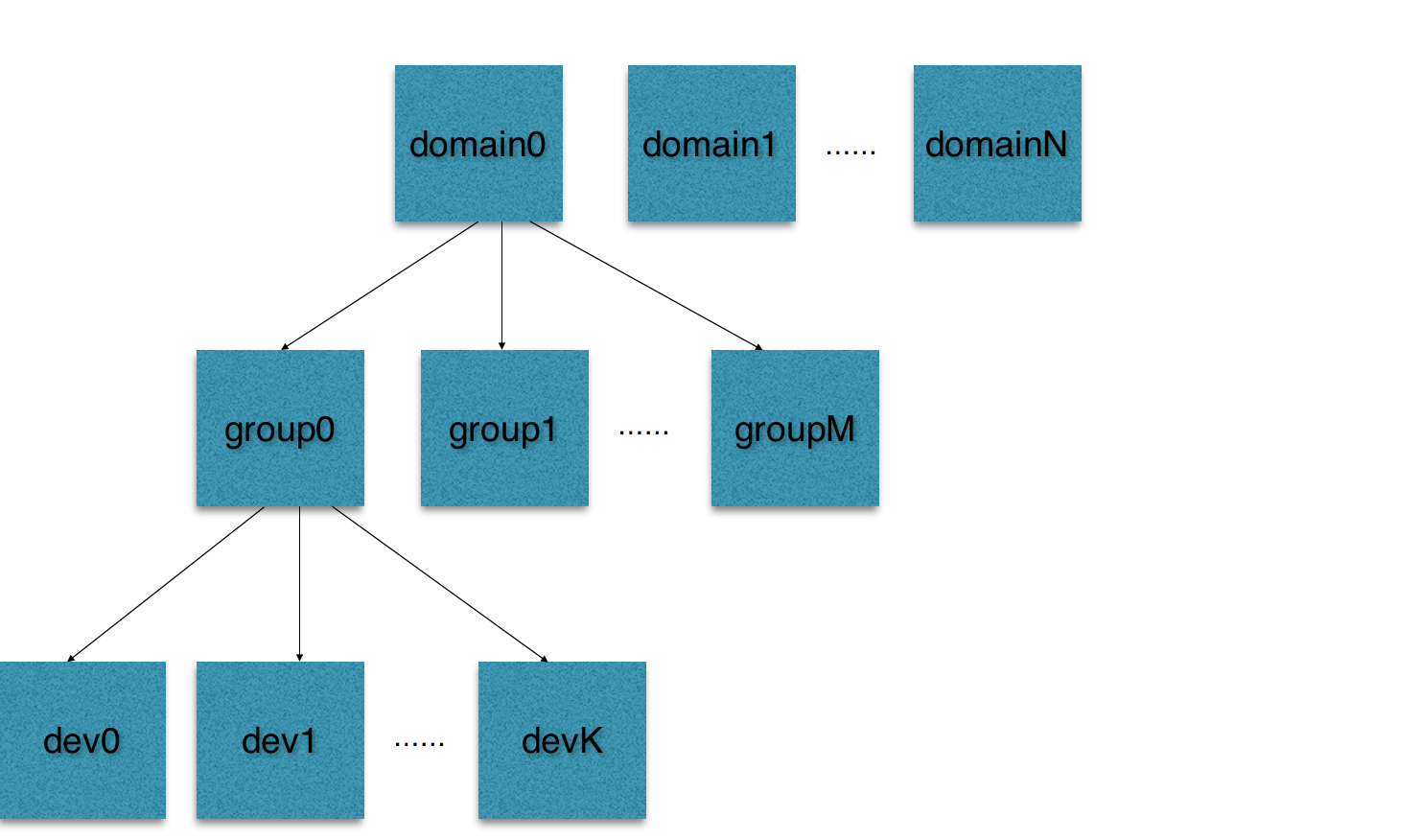}
	}
	
\end{figure}

\textbf{IOMMU groups} are the natural working unit of the IOMMU, but the IOMMU API works on \textbf{domains} and \textbf{devices}. That gap is bridged by iterating over the devices in a group. Ideally, we would have a single device which represents the requestor ID of the group, but it is also allowed to the IOMMU drivers to create policy-defined minimum sets, where the physical hardware may be able to disti-
\vfill
\footnoterule
\pagebreak
nguish members, but wishing to group them at a higher level (ex. untrusted multi-function PCI devices).

In practice (i.e.: when the system was booting), we noticed that each device is added by default to a different group. So, in simple words, if someone wants to use those devices as IOMMU (or SMMU) masters, they should first attach them to one or more domains.

\setlength{\parindent}{3ex}

That is exactly what we did with this module. But before starting the implementation of the SMMU module and because we needed to trigger a DMA transaction, we modified the previous module of the Processing System DMA test, in order for other modules to be able to call only the dmatransfer() function and trigger the DMA transaction. So, before inserting the SMMU test kernel module to the system, we make sure we insert first the modified version of the dmatest module -- otherwise, our SMMU test module, which is calling the dmatransfer() from the other module, will not work.

Back to the implementation of the SMMU test module, in order to initialize the SMMU and create a context bank, we called (from the IOMMU API):

\begin{enumerate}
	\item the \textbf{iommu{\_}domain{\_}alloc()}, that allocates a new IOMMU domain, using the IOMMU-master device
	\item \textbf{iommu{\_}group{\_}get()}, that returns the iommu group for our device (which is a DMA channel -- one of the 16 DMA channels in the Processing System)
	\item \textbf{iommu{\_}attach{\_}group()}, that attaches the IOMMU group to the iommu domain
\end{enumerate}

\setlength{\parindent}{0ex}
Those functions were enough to create one context bank that we would use for the translation.

\setlength{\parindent}{3ex}
\noindent

The only thing that we had to do then was to add the entry/pair of the virtual and physical address (the physical would be the translation of the virtual) to the translation table of the context bank. So, we called, again from the IOMMU API, the \textbf{iommu{\_}map()}, that takes as arguments the domain, that has been allocated (that also has the context bank), the virtual address, the physical address, the size and some flags.

\section{Transaction from PL with single-page entry}
\label{Transaction-from-the-PL}
\setlength{\parindent}{3ex} 

We followed the same procedure when it comes to the IOMMU groups and domains, with the SMMU test module that helped us trigger a DMA transaction from the PS, using the SMMU as we described before, but with one difference: For this module, we do not use the \textbf{dmatranfer()} function - this call was for the DMA transactions inside the PS. This time we have a new Intellectual Property (IP) block in the Programmable Logic (PL), that works like a DMA. This IP has these fields: the destination address, the data that they have to be sent to the destination (i.e.: no source address), some bits for the cacheability and the AXI ID of the transaction.

First we initialize the SMMU in the exact way we did before in the Section \ref{Transaction-from-the-PS} and then we do one mapping (virtual address, physical address) for the destina-
\vfill
\footnoterule
\pagebreak
tion (address).

\setlength{\parindent}{3ex} 

Someone might think that before, we used a device from the PS as an SMMU master in order to make it work - what about now that the IP block is in the PL and also is not an IOMMU master by default? About that, we figured out that the \textbf{StreamID} that is given for each device by default has indeed some meaning. In theory, the StreamID of each device in the device-tree has information of the AXI ID and the Master Device, meaning the port that the transaction will go to or is coming from.

So what we noticed was that in the Zynq UltraScale+ MPSoC, if the StreamID is up to 15 or 16 bits (in our case it is up to 15 bits), then the six (6) least significant bits would be about the AXI ID and the next four (4) bits, would be about the Master Device (the port). The rest of the most significant bits were tested and used with the value of zero (0).

Below we see an example of the field of a device, that is a SMMU master, in the source of the device-tree. In this example we see the first channel of the FPD-DMA.

\vspace{1cm}

\lstset{
	language=C,
	basicstyle=\footnotesize\color{white}
}

\begin{lstlisting}
	dma@fd500000 {
		status = "okay";
		compatible = "xlnx,zynqmp-dma-1.0";
		reg = <0x0 0xfd500000 0x1000>;
		interrupt-parent = <0x1>;
		interrupts = <0x0 0x7c 0x4>;
		clock-names = "clk_main", "clk_apb";
		xlnx,id = <0x0>;
		xlnx,bus-width = <0x80>;
		#stream-id-cells = <0x1>;
		~\textbf{iommus = <0x5 0x14e8>;}~
		power-domains = <0x6>;
		clocks = <0x7 0x3>;
		linux,phandle = <0x29>;
		phandle = <0x29>;
	};
\end{lstlisting}

\vspace{1cm}

In the field "iommus = <0x5 0x14e8>", we see the label (phandle) of the SMMU device, which is 0x5 and the StreamID of our device (first channel of the FPD-DMA) for this SMMU, which is 0x14e8 and is the default StreamID.

\setlength{\parindent}{3ex} 

Below (in the next page) we see the SMMU field in the source of the device-tree. In the \textbf{mmu-masters} field, we see all the devices that are "allowed" to use the SMMU. For example, the entry "0x29 0x14e8" means that the device with the label (or phandle) 0x29, which is the first channel of the FPD-DMA, is an SMMU master and the corresponding and default StreamID for this device is the 0x14e8, as we already saw before.

\vfill
\footnoterule
\pagebreak

\begin{lstlisting}
	smmu@fd800000 {
		compatible = "arm,mmu-500";
		reg = <0x0 0xfd800000 0x20000>;
		#iommu-cells = <0x1>;
		#global-interrupts = <0x1>;
		interrupt-parent = <0x1>;
		interrupts = <0x0 0x9b 0x4 0x0 0x9b 0x4 0x0 0x9b 0x4 0x0 0x9b 0x4 
		0x0 0x9b 0x4 0x0 0x9b 0x4 0x0 0x9b 0x4 0x0 0x9b 0x4 0x0 0x9b 0x4 
		0x0 0x9b 0x4 0x0 0x9b 0x4 0x0 0x9b 0x4 0x0 0x9b 0x4 0x0 0x9b 0x4 
		0x0 0x9b 0x4 0x0 0x9b 0x4 0x0 0x9b 0x4>;
		mmu-masters = < 0x1a 0x874
									0x1b 0x875
									0x1c 0x876
									0x1d 0x877
									0x1e 0x860
									0x1f 0x861
									0x20 0x873
									
									/*LPD-DMA */
									0x21 0x868
									0x22 0x869
									0x23 0x86a
									0x24 0x86b
									0x25 0x86c
									0x26 0x86d
									0x27 0x86e
									0x28 0x86f
									
									/* FPD-DMA */
									~\textbf{0x29 0x14e8}~
									0x2a 0x14e9
									0x2b 0x14ea
									0x2c 0x14eb
									0x2d 0x14ec 
									0x2e 0x14ed 
									0x2f 0x14ee 
									0x30 0x14ef  
									
									0x31 0x870
									0x32 0x871
									0x33 0x872>;
		linux,phandle = <0x5>;
		phandle = <0x5>;
	};
\end{lstlisting}

\setlength{\parindent}{3ex}

\lstset{style=mystyle,
	basicstyle=\ttfamily\color{white}}

\vfill
\footnoterule
\pagebreak

\section{Transaction from PL using process page table}
\label{Process-Page-Table}
\setlength{\parindent}{3ex} 

One of the biggest goals during the thesis was to achieve the SMMU having the translation table of a process, that belongs to a user, in a context bank. 

Once again, in the environment we have, we are using the same IP block in the PL (that was similar to a DMA). It is important here to mention that, again, we edit the device-tree accordingly, so the IP block can be treated as an IOMMU-master too (because as we mentioned before, by default it is not). We will try to trigger a transaction from the PL, but this time the destination address will not be a "random" virtual address that corresponds to a physical address, but the virtual address of the buffer that a user has allocated. So, we need one context bank or in other words a translation table of the SMMU to be able to translate the virtual address of the buffer. \par

In order to achieve this, we used a more clear way than the utility programs, that were somehow opening the /dev/mem and we used in previous modules - we used the ioctl (Input Output ConTroL). In Unix, every device can have its own ioctl commands, which can be: read ioctls (to send information from a process to the kernel), write ioctls (to return information to a process), both or neither. Basically we used ioctl to write to and read from the drivers/modules of the system. \par

The basic function of this module is the \textbf{exanest{\_}virt{\_}ioctl()}, that at first takes the device information we need in order to initialize the SMMU from a "misc device", which in general is a "small" device driver that supports custom "hacks", calling the to{\_}exanest{\_}virt{\_}local() function and then uses a switch() with two (2) cases:
\begin{enumerate}
	\item \textbf{DMA{\_}OPEN{\_}IOCTL}, that basically does the same things we did in the previous SMMU modules in order to initialize the SMMU and create a context bank
		\begin{enumerate}
			\item \textbf{iommu{\_}domain{\_}alloc()}, that allocates a new IOMMU domain
			\item calls the \textbf{virt{\_}to{\_}phys()}, that takes as argument the pointer of the current process page table, the pgd (\ref{PGD}) and by doing a traverse in all 4 levels of the page-table, it finds its physical translation and returns it
			\item \textbf{iommu{\_}group{\_}get()}, that returns the IOMMU group for our device (which is the IP, similar to a DMA, in the PL)
			\item \textbf{iommu{\_}attach{\_}group()}, that attaches the IOMMU group to the IOMMU domain
		\end{enumerate}
	\item \textbf{DMA{\_}START{\_}IOCTL}, that basically:
		\begin{enumerate}
			\item takes a structure, the \textbf{dmaCMD}, from the user that contains all the information that the IP in the PL needs, in order to trigger the transaction:
				\begin{itemize}
					 \item the virtual address of the destination (the buffer that the user allocated)
					 \item the data that we will "send"/write to the destination
					 \item the cacheability bits (which are fixed but still the user send them)
					 \item the AXI ID of the transaction 
				\end{itemize}
				\vfill
				\footnoterule
				\pagebreak
			\item flushes the 4-level translation table for the buffer
			\item writes all the information that came from the user to the addresses of the PL and triggers the transaction
		\end{enumerate}
\end{enumerate}

\par

\setlength{\parindent}{3ex} The program that the user runs in order to use the module we described before does the following:
\begin{enumerate}
	\item allocates a buffer, that we will use as a (virtual) destination address for the transaction
	\item initializes this buffer, so we know its value before the transaction
	\item opens the exanest{\_}virt module that we described before and that when we insert it in the kernel, it appears in this path: /dev/exanest{\_}virt
	\item with ioctl() function, it starts the \textbf{DMA{\_}OPEN{\_}IOCTL}, that we described before
	\item then, initializes the \textbf{dmaCMD} structure, with the values the user wants for the IP in the PL, that will trigger the transaction -- as it was mentioned before
	\item finally, it reads the buffer to see if the IP in the PL did the transaction successfully - the expected result would be the data he wrote in the \textbf{dmaCMD} structure and not the values that the buffer had after the initialization, at the beginning
\end{enumerate}

\setlength{\parindent}{0ex} 
At first, we noticed the following:

\setlength{\parindent}{3ex} 
If the \textbf{StreamID} in the device-tree was not written like we described in the Section \ref{Transaction-from-the-PL}, then we had no fault and the transaction was not happening (at the end of the user program the buffer had the same values when it was initialized). If the \textbf{StreamID} in the device-tree was written like we described in the Section \ref{Transaction-from-the-PL}, then by default we had two faults: a global fault and a context fault.

In effect, the value of the \textbf{SMMU{\_}S2CR$n$.INSTCFG} register was 0b11, meaning it was "expecting" an instruction and not data. So, in order to fix that, we overwrote this register and said to the arm-smmu driver, that when we pass our pgd (page table directory), it should also overwrite this value, so the SMMU can treat it as data.

Also, the the \textbf{SMMU{\_}CB$n${\_}TCR} (Translation Control Register) of the SMMU and especially the six (6) least significant bits, the field T0SZ, where the $2^{64 - T0SZ}$ gives the address space, had the value 0b01{\_}0000, meaning the input address space (ias) would expect 64-16 = 48 bits (for the incoming virtual address), while the corresponding register in the \textbf{MMU} of the processor had the value 0b01{\_}1001, meaning the corresponding address would be 64-24 = 40 bits (incoming virtual address). So, we changed the \textbf{T0SZ} field of the SMMU to expect 40 bits too.

We noticed, also, a similar effect of the SMMU and MMU, respectively, TCR register in the \textbf{SMMU{\_}CB$n${\_}TCR2} register in which, the three (3) least significant bits (the field PASize  - Physical Address Size), had the value 0xb101, meaning the expected physical address would have to be 48 bits (the out-coming physical address), while in the MMU of the processor, the corresponding field had
\vfill
\footnoterule
\pagebreak  
the value 0b010, meaning the expected physical address would have to be 40 bits (the out-coming physical address). So, again, we changed the field of the SMMU to expect 40 bits too.

\setlength{\parindent}{3ex} 
The major changes that we did in order to make this task work, were the changes that we mentioned above and mostly in the \textbf{arm-smmu.c} driver. A more detailed description of those changes can be found in the Appendix \ref{AppendixD}. 



\noindent

\setlength{\parindent}{3ex}

\lstset{style=mystyle,
	basicstyle=\ttfamily\color{white}}

\vfill
\footnoterule

\chapter{Kernel Experiments}
\label{Chapter5} 

\noindent

\setlength{\parindent}{3ex}

In this chapter, we will describe the third part of the contributions of this thesis, which is the experiments we did for all the kernel modules and the notes we used, in order to achieve our goals.

\section{DMA test in PS}


\setlength{\parindent}{3ex} 

In order to test the DMA module, we were following these steps: 
\begin{enumerate}
	\item Insert kernel module with the \textbf{insmod} command
	\begin{lstlisting}
		# mydmatest.ko is the kernel object of the module 
		# mydmatest.c
		insmod mydmatest.ko
	\end{lstlisting}	
	\item Use the utility programs (utils) of read/write from/to physical addresses, that existed in the Laboratory, in order to write to a physical address which would be the source of the DMA transaction and then we were reading from an address, that we would use for the destination address, just to make sure it has different data than the source address. It is important to mention that opening part of the /dev/mem in order to read/write from/to physical addresses is far from an acceptable way to access the memory, but at this point we were just testing all the DMA channels in the Processing System (PS). The read/write utils work like this: \\
	\begin{lstlisting}
		# ./read_physical32 <address in hex>
		./read_physical32 0x60000000
		# ./write_physical32 <address in hex> <data in hex>
		./write_physical32 0x60000000 0xcafebeef 
	\end{lstlisting}
	\item Set the proper parameters of the modules, like the source and the destination (physical) addresses and the name of the DMA channel we wanted to test. In order to set the value of the parameters we were calling these commands from a terminal:\\ 
	\begin{lstlisting}
		echo 0x60000000 > /sys/module/mydmatest/parameters/src
		echo 0x60002000 > /sys/module/mydmatest/parameters/dst
		echo dma1chan0 > /sys/module/mydmatest/parameters/channel
	\end{lstlisting}
	\vfill
	\footnoterule
	\pagebreak
	\item Set "1" to the parameter with the name "run", in order to trigger the transaction
	\begin{lstlisting}
		echo 1 > /sys/module/mydmatest/parameters/run
	\end{lstlisting}
	\item Read the data of the destination address to make sure they are the same with the source address
	\begin{lstlisting}
		./read_physical32 0x60002000
	\end{lstlisting}
\end{enumerate}

In our tests we used to test the physical address 0x60000000 as the source and the physical address 0x60002000 as the destination, but we also tested many other addresses and the DMA transactions were always successful.\\

\setlength{\parindent}{0ex} 

Below we can see the names of the DMA channels in the Processing System as they were detected in Linux.\\

\setlength{\parindent}{0ex}

\textbf{FPD-DMA} \\

\setlength{\parindent}{3ex}
\noindent

Channel 0 : dma1chan0
\noindent

Channel 1 : dma2chan0
\noindent

Channel 2 : dma3chan0
\noindent

Channel 3 : dma4chan0
\noindent

Channel 4 : dma5chan0
\noindent

Channel 5 : dma6chan0
\noindent

Channel 6 : dma7chan0
\noindent

Channel 7 : dma8chan0\\

\setlength{\parindent}{0ex}

\textbf{LPD-DMA} \\
\setlength{\parindent}{3ex}
\noindent

Channel 0 : dma9chan0
\noindent

Channel 1 : dma10chan0
\noindent

Channel 2 : dma11chan0
\noindent

Channel 3 : dma12chan0
\noindent

Channel 4 : dma13chan0
\noindent

Channel 5 : dma14chan0
\noindent

Channel 6 : dma15chan0
\noindent

Channel 7 : dma16chan0
\noindent

 \vfill
 \footnoterule
 \pagebreak

 \section{Transaction from PS with single-page entry}
 
\setlength{\parindent}{3ex}

 In order to test the SMMU from the Processing System, after the initialization of the SMMU and the creation/allocation of the context bank, we added two (2) entries in the context bank (meaning two mappings) because in the DMA transaction we wanted to use for both source and destination, a virtual address. So for the source the physical address was: 0x60000000 and the virtual address was: 0x70000000. For the destination the physical address was: 0x60002000 and the virtual address was: 0x70002000.
 
 For testing purposes we used the same read/write utility programs, that we mentioned before. Using those programs, we could write something at the physical address of the source (the 0x60000000 address) and we could expect after running the module (that at the end of it was also triggering a DMA transaction), that we would see the same data at the physical address of the destination (0x60002000).
 
The following lines are the lines we wrote to a terminal in order to run and test this SMMU module.

\begin{lstlisting}
		insmod mydmatest.ko
   ./read_physical32 0x60000000		 # to see out of curiosity
   ./write_physical32 0x60000000 0xcafebeef
   ./read_physical32 0x60000000		 # to be sure our data were 
							                     # written
   ./read_physical32 0x60002000    # to make sure the destination 
					                         # does not have the same data 
					                         # with the source
   insmod test_iommu.ko            # the kernel object of our
                                   # module 
   ./read_physical32 0x60002000    # checking if it has same data
	                                 # with the source (0x60000000)
\end{lstlisting}

At the last run of the read{\_}physical32 program, we saw as the result the same data with the source, as we expected, meaning the translation (using the SMMU) and the DMA transaction worked.

With the tests we did, we also noticed that if we give other addresses that were not mapped as a physical and virtual address entry in the translation table (context bank), the DMA transaction would normally happen with no translation at all (as if all addresses that were given were physical).

\section{Transaction from PL with single-page entry}
\setlength{\parindent}{3ex}

The experiments we did for this module, as expected from the description above (Section \ref{Transaction-from-the-PL}), at first included the initialization of the SMMU and a context bank. For this, we followed the same steps that were described in Section \ref{Transaction-from-the-PS}. Then, we used again the utility programs of the Laboratory in order to open part of the /dev/mem and write to physical addresses of the fields of the IP, the destination address, the data to be sent to the destination, the cacheability bits and the AXI ID of the transaction. When we were ready to trigger the transaction, we had
\vfill
\footnoterule
\pagebreak
to write the value of one (1) to another field of the IP block, that was triggering the transaction from the PL.
\setlength{\parindent}{3ex}

Since we knew that the IP block in the PL was connected to the PS via the HPC0 port, the corresponding StreamID would be the \textbf{0x200} or the 000 00\textbf{00 10}00 0000 - in bold we see the four Master Device (port) bits for the HPC0 port - and not the default StreamID.

The change we did to a device that was already (by default) an SMMU master (in our example the first channel of the FPD-DMA) was in its StreamID field, in order to give our own StreamID for HPC0 port and AXI ID 0x0, which is translated to the StreamID "0x200". As we already mentioned in Section \ref{Transaction-from-the-PL}, in the "smmu" field of the source of the device-tree, the first channel of the FPD-DMA has by default this entry: "0x29 0x14e8" -- so, what we did, in order to pass our StreamID, is that we changed the "0x14e8" to "0x200". We also changed the StreamID field that the device (the first FPD-DMA channel) had in its description in the source of the device-tree, as we mentioned in the Section \ref{Transaction-from-the-PL}.

The assumption that was proven correct by changing the StreamID of a device, that was already a SMMU master, in the device tree, was that when it comes to the SMMU, all the transactions that will eventually go through it, will be "checked" if they are a "match" with any of the StreamIDs that exist in the Stream Mapping Table, no matter the device they are coming from.

By doing these changes, in a "hacking" way, we managed to add our StreamID to the Stream Mapping Table (Figure \ref{fig:4.19}), so each transaction that would come through the HPC0 port with AXI ID 0 (since the six least significant bits were zero), would be a match for the SMMU and in our case, it would also "request a translation". And so it did.

We also tested the case where the IP block in the PL, that works like a DMA, is not connected with the PS via HPC0 port, but via HPC1 port and we had the same results -- that is not a surprise, since the HPC1 port is also coherent and almost identical to HPC0 port.

Below there is an example that we tested, where the IP block in the PL that was connected to PS via HPC0 port has as the base physical address: 0xb0000000. These were the addresses of its fields:
\begin{itemize}
	\item Destination Address (least significant bits) : 0xb000{\_}0030
	\item Destination address (least significant bits): 0xb000{\_}0034
	\item Destination data : 0xb000{\_}0038
	\item Cacheability bits : 0xb000{\_}003c
	\item AXI ID : 0xb000{\_}0040
	\item Trigger : 0xb000{\_}0044
\end{itemize}

In order to initialize the SMMU, we have to insert (and initialize) the SMMU kernel module, that except from the initialization, it also adds one (physical address, virtual address) entry of the destination of our transaction from the PL. In the test we did, we had the 0x60002000 as the physical address and the 0x70002000 as the virtual address. 

\vfill
\footnoterule
\pagebreak

\setlength{\parindent}{3ex}	
In our example the StreamID is the one we mentioned above: 0x200, meaning the expected AXI ID is 0x0. The cacheability bits, it is suggested to have as a value the 0xf, so that is what we used in our test.

The following lines are the lines we wrote using a terminal that triggered the transaction from the PL that indeed used the SMMU to translate the virtual address before the transfer.

\begin{lstlisting}
# Insert the SMMU module, that initializes the SMMU 
# and adds the 0x60002000 0x70002000 mapping
insmod test_iommu.ko

# Check the original values of the destination address
./read_physical32 0x60002000		

# Set the IP block in the PL
./write_physical32 b0000030 0x70002000
./write_physical32 b0000034 0x00000000
./write_physical32 b0000038 0xcafebeef
./write_physical32 b000003c 0xf
./write_physical32 b0000040 0x0
./write_physical32 b0000044 0x1

# To make sure the destination has the same data
# with the data the IP block had
./read_physical32 0x60002000    

\end{lstlisting}

\section{Transaction from PL using process page table}
\setlength{\parindent}{3ex}	
For the testing of the exanest{\_}virt module, we did not change anything in terms of the parameters -- the only changes that we had are the changes mentioned in the description of the Section \ref{Process-Page-Table} and the Appendix \ref{AppendixD}.

Let's suppose that the source code of the user program described in Section \ref{Process-Page-Table} is called \textbf{run{\_}exanest{\_}virt.c} and we already have an executable of it, with the name \textbf{run{\_}exanest{\_}virt}. The following lines are the lines we wrote to a terminal in order to test this module.

\begin{lstlisting}
	insmod exanest_virt.ko
	./run_exanest_virt 
\end{lstlisting}


As mentioned in the description of the module, it works since it’s overwriting the user buffer, but it seems that the data are not coherent.

\noindent

\setlength{\parindent}{3ex}

\lstset{style=mystyle,
	basicstyle=\ttfamily\color{white}}

\vfill
\footnoterule

\chapter{Conclusion} 

\label{Chapter6} 


\setlength{\parindent}{3ex}


All the SMMU modules worked, meaning in all the environments (transactions from the Processing System and from the Programmable Logic), that were described in detail previously in Chapter \ref{Chapter4}, we managed to have the translation, from a virtual address to a physical address.

There are still many areas of the IOMMU (SMMU) to be discovered, since it is still difficult to extract information from the existing documentations, that is related to the implementation or the use of the  IOMMU in Linux.
For instance, when it comes to the StreamID, it is still not clear how it works - we had an assumption for the 10 least significant bits, that, as mentioned, we used and tested in our implementation, but the whole picture with the StreamID remains unclear.
Also, in our case, we only used one context bank (translation table) and although it might seem easy after the analysis in the previous chapters, an implementation of using many context banks, allocating many domains etc is important and should happen in the future.
In the last module with the page table of a process, although we managed to make it work, as we mentioned, we noticed lack of coherence, so in order to make it work properly, we flushed the cache lines (only four cache lines, for each one of the levels of the translation table) and changed the snooping option in the FSBL - this change, although it looked optimistic at the beginning, did not fix the coherence problem completely, meaning that sometimes we still had non-coherent data. Another approach was to allocate another much bigger buffer, that we used before triggering the transaction from the PL and by doing that, we ensured that our first buffer had the last/updated data. In the future it should be investigated if there are any options that solve the coherence problem in a more clear way (for instance: by making the process page table write-through, using the proper cacheability flags in the IOMMU settings etc).

To sum up, the greater goal triggering transactions that use the global virtual address space in a remote level 
did not happen yet, since our implementations were not tested for an environment like the environment the Unimem describes, where many remote coherent islands (nodes) are trying to communicate. As we mentioned at the beginning, this thesis was aiming to help and support some of the big goals of the Unimem (Section \ref{unimem}), by providing information and tested ideas, which we hope that we achieved. By expanding the ideas, approaches and perhaps the modules that were described in this thesis, it should be easier for someone to reach Unimem's greater goal.


\vfill
\footnoterule

	\bibliographystyle{unsrt}
	\nocite{*}
	\bibliography{main}
	
	\appendix 
	
	

\chapter{Appendix} 

\label{AppendixA} 

\lstset{
	language=bash,
	basicstyle=\footnotesize\color{white}
}

\section{Booting from the SD Card}
\setlength{\parindent}{3ex} 

In this Appendix, we will write all steps we followed in order to boot Linux using an SD card. Before that, we should mention that in the Laboratory we used a specific format of the SD Card - the "u-boot" file "knew" from what partitions it could find all files that were needed. That means that files should have specific names (e.g.: the "Image" file should have the name "Image" and not "Image2"). The first partition should have all the needed files except the "rootfs", that belongs to the second partition.

\setlength{\parindent}{0ex}

In the first partition, in order for our system to boot, we needed those files:

\begin{enumerate}
	\item BOOT.bin
	\item Image (kernel)
	\item system.dtb (device-tree)
\end{enumerate}

\subsection{BOOT.bin}
\label{bootbin}
There are specific steps someone should follow to create the image of the BOOT.bin file. Below we write all the files that are needed for BOOT.bin file.

\begin{enumerate}
	\item bitstream (.bit file)
	\item FSBL (fsbl.elf)
	\item u-boot (u-boot.elf)
	\item bl31.elf
\end{enumerate}
\setlength{\parindent}{3ex} 
\noindent

The bitstream is generated from a Xilinx Vivado project. The "u-boot" file, as we mentioned before, is already created so it is like something fixed in this Appendix. Also the bl31.elf is fixed for the purposes of this Appendix - in general, it can be generated easily with the help of Petalinux.

The \textbf{FSBL} file is generated by the .hdf file, that can also be exported from a Xilinx Vivado project.
Below (next page) we explain how.
\label{fsbl_generation}
\vfill
\footnoterule
\pagebreak
\setlength{\parindent}{0ex} 
\textbf{Step 1}: ~We open Xilinx SDK \\
\textbf{Step 2}: We give a path and a name for our workspace (for instance here it is: {\/}workspace{\_}Dec19)

\begin{figure}[h!]
	
	\centering
	\captionbox[Text]{Step 2 \label{fig:A.1}}{%
		\includegraphics[width=0.8\textwidth]{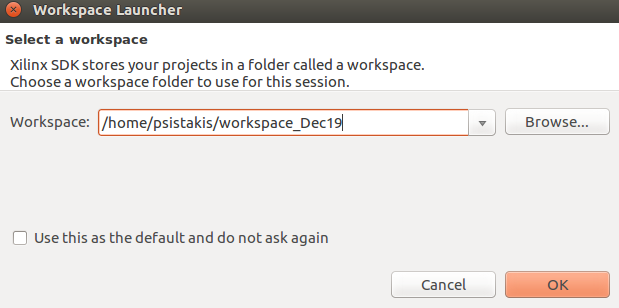}  
	}
	
\end{figure}

\textbf{Step 3}: File $\rightarrow$ New $\rightarrow$ Application Project

\begin{figure}[h!]
	
	\centering
	\captionbox[Text]{Step 3 \label{fig:A.4}}{%
		\includegraphics[width=0.8\textwidth]{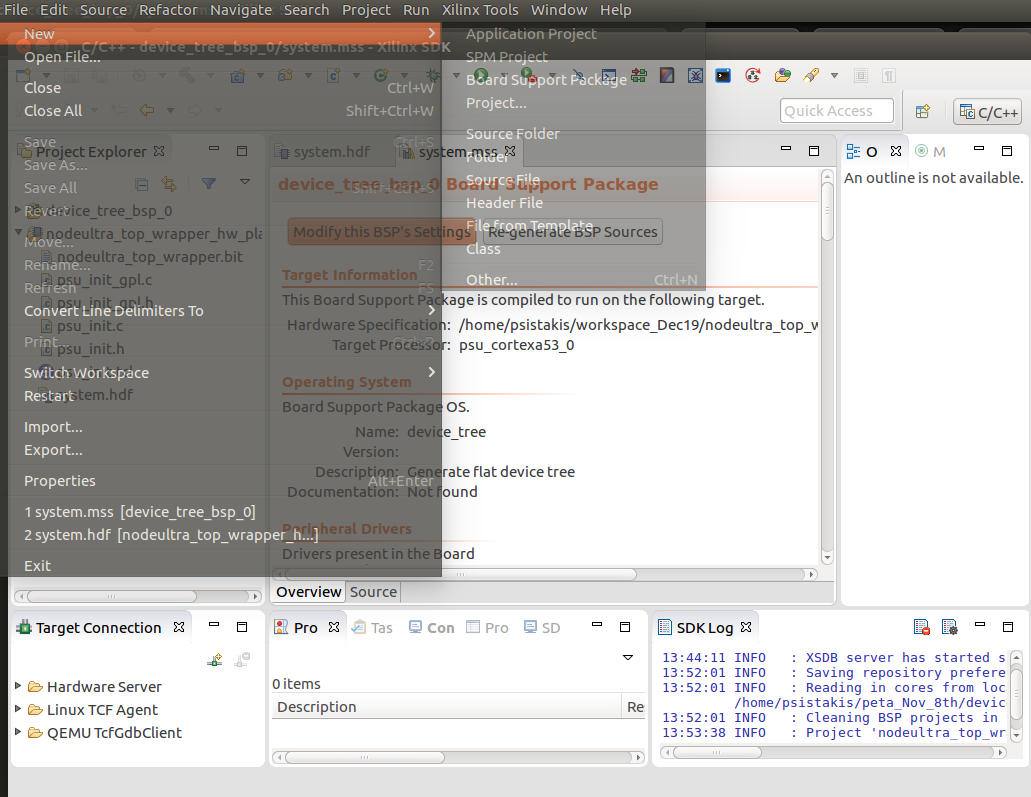}  
	}
	
\end{figure}

\vfill
\footnoterule
\pagebreak

\textbf{Step 4}: Hardware Platform $\rightarrow$ New
\begin{figure}[h!]
	
	\centering

	\captionbox[Text]{Step 4 \label{fig:A.2}}{%
		\includegraphics[width=0.7\textwidth]{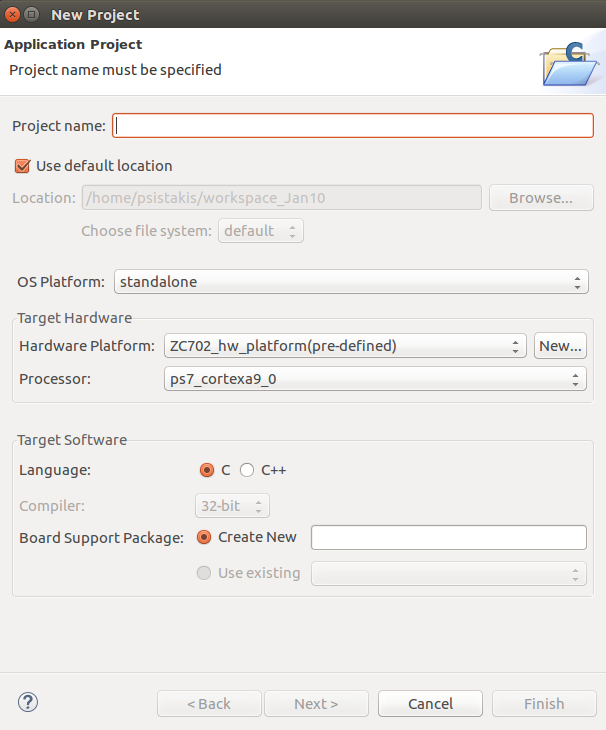}  
	}

\end{figure}

%
%
%
%

\textbf{Step 5}: Specify and browse the path of the .hdf file $\rightarrow$ The project name will appear.

\begin{figure}[h!]
	
	\centering
	
	\captionbox[Text]{Step 5 \label{fig:A.3}}{%
		\includegraphics[width=0.7\textwidth]{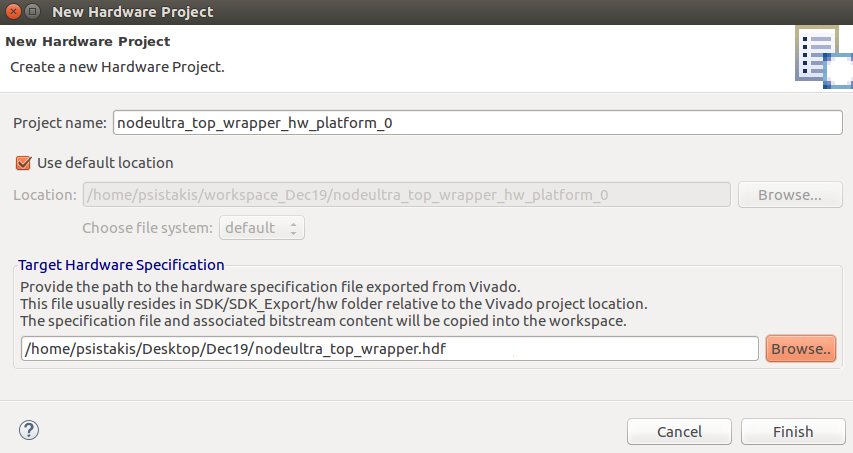}  
	}
	
\end{figure}

\vfill
\footnoterule
\pagebreak

\textbf{Step 6}: In "Project name" field write: fsbl
\begin{figure}[h!]
	
	\centering
	\captionbox[Text]{Step 6 \label{fig:A.5}}{%
		\includegraphics[width=0.5\textwidth]{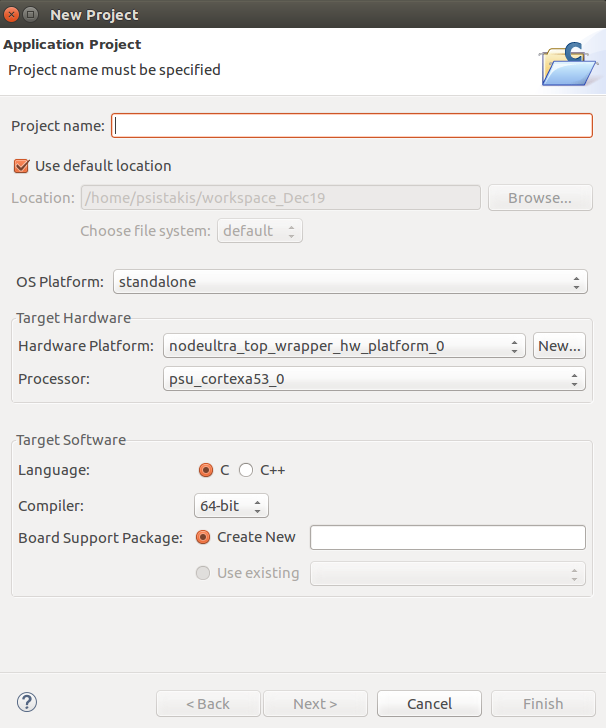}  
	}
	
\end{figure}

\textbf{Step 7}: Click "Next"
\begin{figure}[h!]
	
	\centering
	\captionbox[Text]{Step 7 \label{fig:A.6}}{%
		\includegraphics[width=0.5\textwidth]{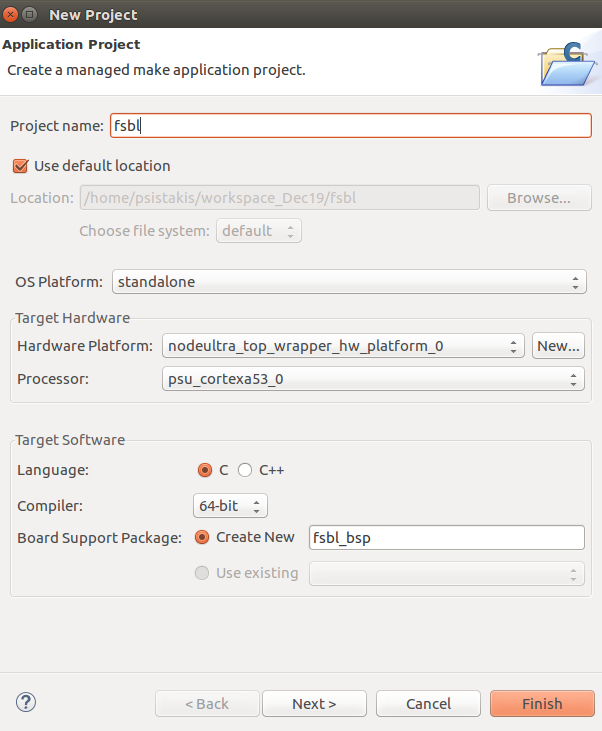}  
	}
	
\end{figure}

\vfill
\footnoterule
\pagebreak

\textbf{Step 8}: Choose "Zynq MP FSBL" -> Click "Finish".
\begin{figure}[h!]
	
	\centering
	\captionbox[Text]{Step 8 \label{fig:A.7}}{%
		\includegraphics[width=0.5\textwidth]{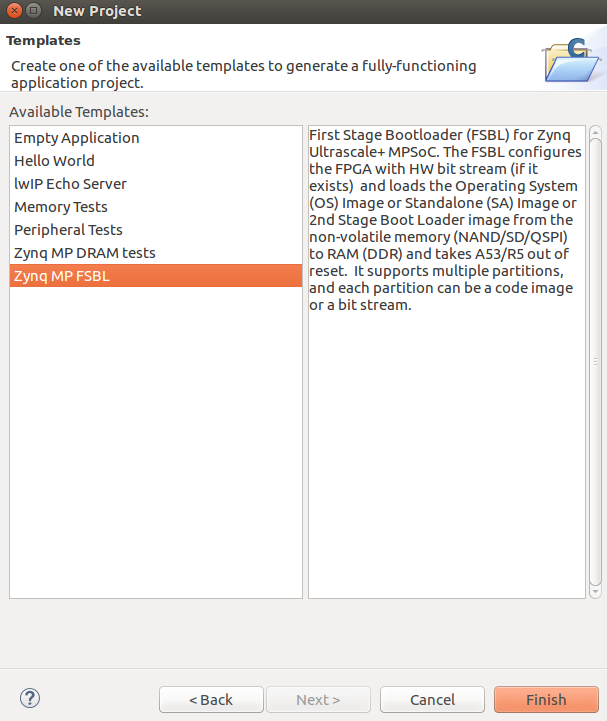}  
	}
	
\end{figure}

\textbf{Step 9}: We go at the workspace folder and see some folders/files like in the image below. The "fsbl" is the folder we care about.

\begin{figure}[h!]
	
	\centering
	\captionbox[Text]{Step 9 \label{fig:A.8}}{%
		\includegraphics[width=0.5\textwidth]{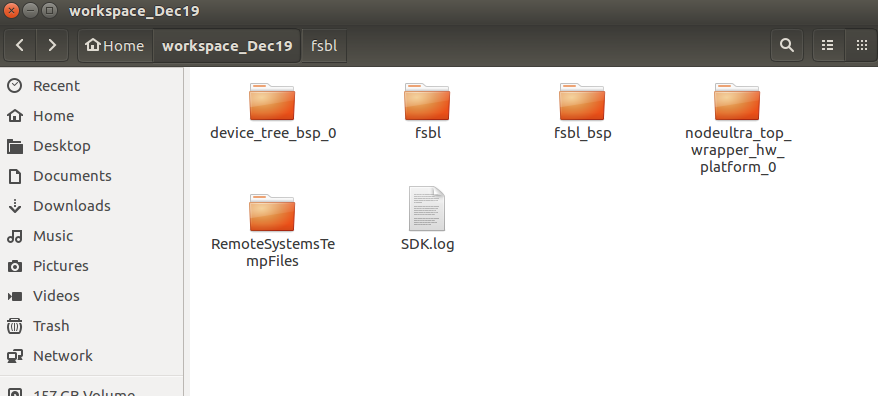}  
	}
	
\end{figure}

\textbf{Step 10}: Inside the "Debug" folder there is the fsbl.elf that was generated.

\begin{figure}[h!]
	
	\centering
	\captionbox[Text]{Step 10 \label{fig:A.9}}{%
		\includegraphics[width=0.6\textwidth]{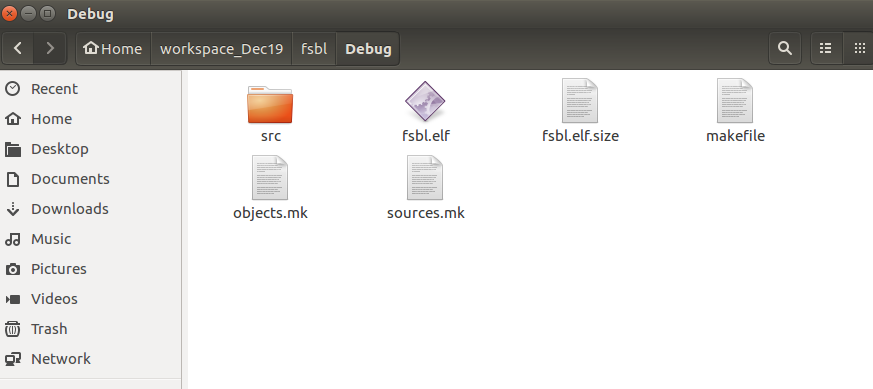}  
	}
	
\end{figure}

\vfill
\footnoterule
\pagebreak

\setlength{\parindent}{3ex} 
\noindent

Except from all the needed files that we mentioned for the BOOT.bin file, we will also need a .bif that mentions them.

\setlength{\parindent}{0ex} 

An example of the sd{\_}boot.bif we used:

\lstset{
	language=C,
	basicstyle=\footnotesize\color{white}
}

\begin{lstlisting}
//arch = zynqmp; split = false; format = BIN
the_ROM_image:
{
	[fsbl_config]a53_x64
	[bootloader]fsbl.elf
	[destination_cpu = a53-0]bl31.elf
	[destination_cpu = a53-0]u-boot.elf
	[offset = 0x100000, destination_device = pl]nodeultra_top_wrapper.bit
}
\end{lstlisting}

\lstset{
	language=bash,
	basicstyle=\footnotesize\color{white}
}

Now we have the .bif, here is the command we run to a terminal, in order to create the BOOT.bin file:
\begin{lstlisting}
	bootgen -arch zynqmp -image sd_boot.bif -o BOOT.bin -log info
\end{lstlisting}

\subsection{Image}
\setlength{\parindent}{3ex}

Having our own configuration file (.config) in the root file of our Linux sources, we use a cross compiler in order to build the kernel in the same environment (Linux version) we want our boards to have.
To do that, we run the following commands:
\begin{lstlisting}
export ARCH=arm64
export CROSS_COMPILE=<path of the cross compiler sources>/bin/aarch64-linux-gnu-
\end{lstlisting}

\setlength{\parindent}{3ex} 
\noindent

By running the Makefile (with the command "make") in the root file of our Linux sources (kernel), we build/create the "Image" file that will be located in this path: <Linux 4.4.0 sources>/arch/arm64/boot/Image.

\subsection{system.dtb}

Appendix (\ref{AppendixC}) is dedicated to the generation of the Device-Tree.

\vfill
\footnoterule
\lstset{style=mystyle,
	basicstyle=\ttfamily\color{white}}

\chapter{Appendix} 

\label{AppendixB} 

\section{Xilinx Zynq UltraScale+ DMA in the PS}

Although in \cite{Reference5}, the DMA in the Programming System (PS) is mentioned as one unit, the general purpose DMA (ZDMA), we actually see it as two DMAs: one is located in the LPD, the low-power domain DMA (LPD-DMA) and another is located in the FPD, the full-power domain DMA (FPD-DMA).

The basic differences between those two are:
\setlength{\parindent}{3ex}

\begin{enumerate}
	\item The LPD-DMA can be optionally configured as I/O coherent, like all of the PS masters, including the RPU but excluding the full-power DMA controller (FPD-DMA).
	
	The FPD-DMA does not have a physical path through the CCI and does not support I/O coherency. The LPD-DMA does have an alternate path to the DDR controller through the CCI, which allows it to be marked as I/O coherent.
	\item 
	\begin{itemize}
		\item FPD-DMA is an 8-channel DMA with a 128-bit AXI bus width.
		\item LPD-DMA is an 8-channel DMA with a 64-bit AXI bus width.
	\end{itemize}
	
\end{enumerate}
The addresses of the channels of the \textbf{FPD-DMA} :\\

	\setlength{\parindent}{3ex}
	0xFD500000 (channel 0)
	\setlength{\parindent}{3ex}
	\noindent
	
	0xFD510000 (channel 1)
	\setlength{\parindent}{3ex}
	\noindent
	
	0xFD520000 (channel 2)
	\setlength{\parindent}{3ex}
	\noindent
	
	0xFD530000 (channel 3)
	\setlength{\parindent}{3ex}
	\noindent
	
	0xFD540000 (channel 4)
	\setlength{\parindent}{3ex}
	\noindent
	
	0xFD550000 (channel 5)
	\setlength{\parindent}{3ex}
	\noindent
	
	0xFD560000 (channel 6)
	\setlength{\parindent}{3ex}
	\noindent
	
	0xFD570000 (channel 7)\\
	\setlength{\parindent}{3ex}
	\noindent
	
\setlength{\parindent}{0ex}
\noindent
\setlength{\parindent}{3ex} The addresses of the channels of the \textbf{LPD-DMA}: \\

	0xFFA80000 (channel 0)
	\setlength{\parindent}{3ex}
	\noindent
	
	0xFFA90000 (channel 1)
	\setlength{\parindent}{3ex}
	\noindent
	
	0xFFAA0000 (channel 2)
	\setlength{\parindent}{3ex}
	\noindent
	
	0xFFAB0000 (channel 3)
	\setlength{\parindent}{3ex}
	\noindent
	
	\vfill
	\footnoterule
	\pagebreak
	
	\setlength{\parindent}{3ex}
	\noindent
	
	0xFFAC0000 (channel 4)
	\setlength{\parindent}{3ex}
	\noindent
	
	0xFFAD0000 (channel 5)
	\setlength{\parindent}{3ex}
	\noindent
	
	0xFFAE0000 (channel 6)
	\setlength{\parindent}{3ex}
	\noindent
	
	0xFFAF0000 (channel 7)
	\setlength{\parindent}{3ex}
	\noindent
	
\vfill
\footnoterule

\chapter{Appendix} 

\label{AppendixC} 

\lstset{
	language=C,
	basicstyle=\footnotesize\color{white}
}

\section{The Device-Tree}

\subsection{Generating the Device-Tree}
\setlength{\parindent}{3ex}

In general, the \textbf{device tree} is a data structure for describing hardware, which originated from Open Firmware. The data structure can hold any kind of data as internally it is a tree of named nodes and properties. Nodes contain properties and child nodes, while properties are name–value pairs.

In order to generate the device-tree we want to use for an Operating System (Linux), we only need the .hdf (hardware file) that someone can easily export from Xilinx Vivado, with a specific design for a specific board.

\setlength{\parindent}{0ex}

\textbf{Step 1}: We open Xilinx SDK\\
\textbf{Step 2}: We give a path and a name for our workspace (for instance here it is: {\/}workspace{\_}Dec19) 

\begin{figure}[h!]
	
	\centering
	
	\captionbox[Text]{ Step 2 \label{fig:C.1}}{%
		\includegraphics[width=0.5\textwidth]{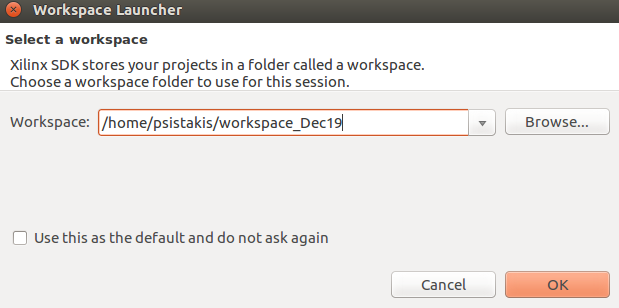}  
	}
	
\end{figure}
\textbf{Step 3}: Xilinx Tools $\rightarrow$ Repositories

\begin{figure}[h!]
	
	\centering
	
	\captionbox[Text]{ Step 3 \label{fig:C.2}}{%
		\includegraphics[width=0.4\textwidth]{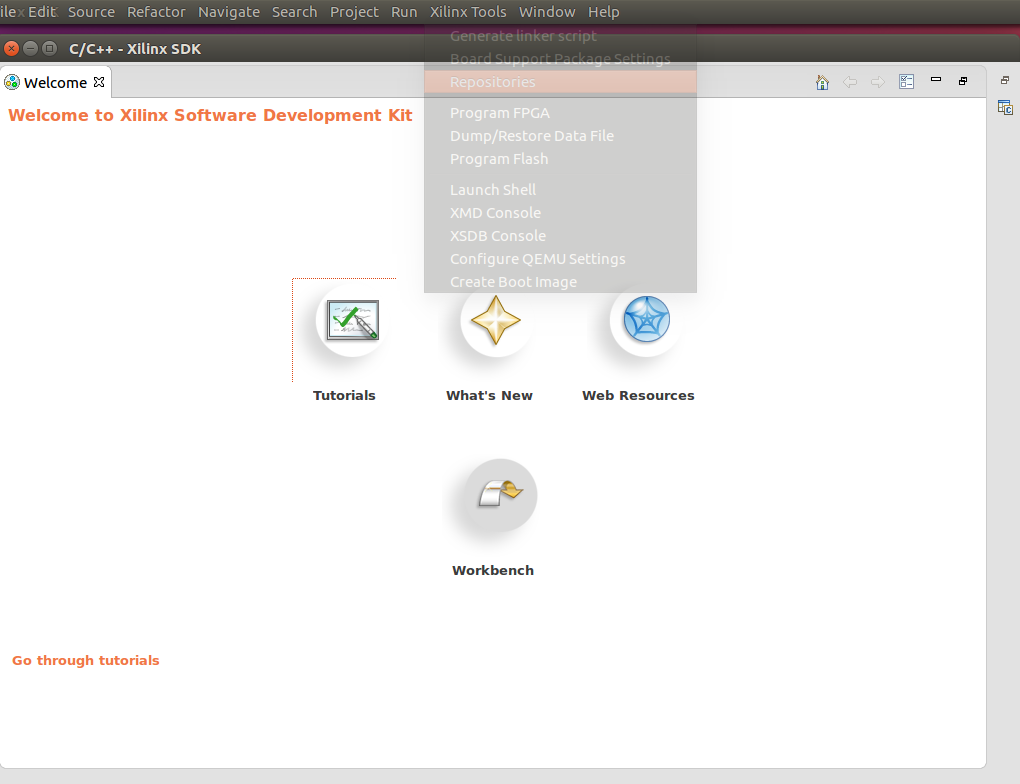}  
	}
	
\end{figure}

\vfill
\footnoterule
\pagebreak

\textbf{Step 4}: Click "New" (in Local Repositories)

\begin{figure}[h!]
	
	\centering
	
	\captionbox[Text]{ Step 4 \label{fig:C.3}}{%
		\includegraphics[width=0.6\textwidth]{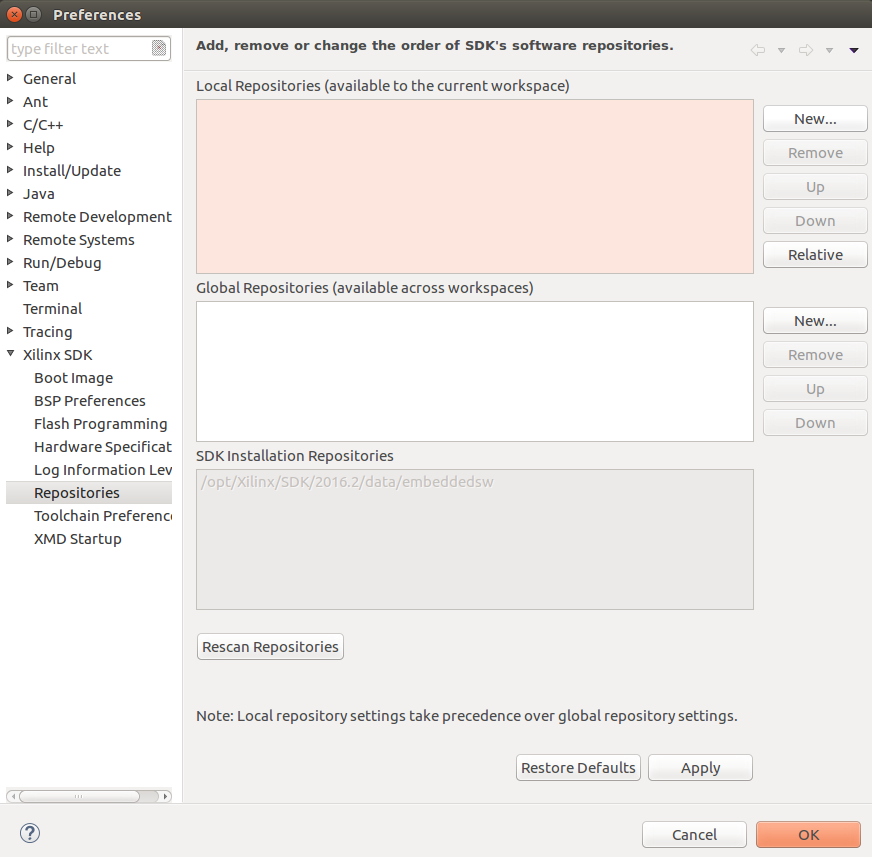}  
	}
	
\end{figure}
%
%
%
%

\textbf{Step 5}: Add specific repository (the device tree source files), provided, in our case, by the \href{https://github.com/xilinx}{Xilinx Git}. For instance, we were working with Xilinx Vivado and SDK 2016.2 version, so we also downloaded the 2016.2 version of the device tree sources.

\begin{figure}[h!]
	
	\centering
	
	\captionbox[Text]{ Step 5 \label{fig:C.4}}{%
		\includegraphics[width=0.6\textwidth]{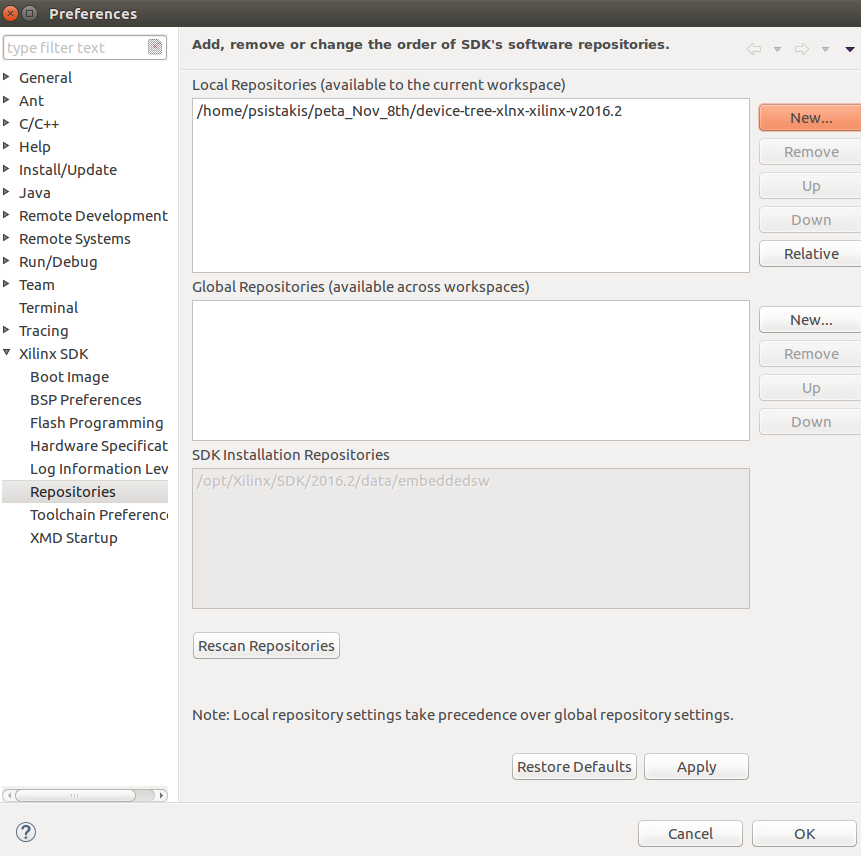}  
	}
	
\end{figure}

\vfill
\footnoterule
\pagebreak

\textbf{Step 6}: File $\rightarrow$ New $\rightarrow$ Board Support Package

\begin{figure}[h!]
	
	\centering
	
	\captionbox[Text]{ Step 6 \label{fig:C.5}}{%
		\includegraphics[width=0.6\textwidth]{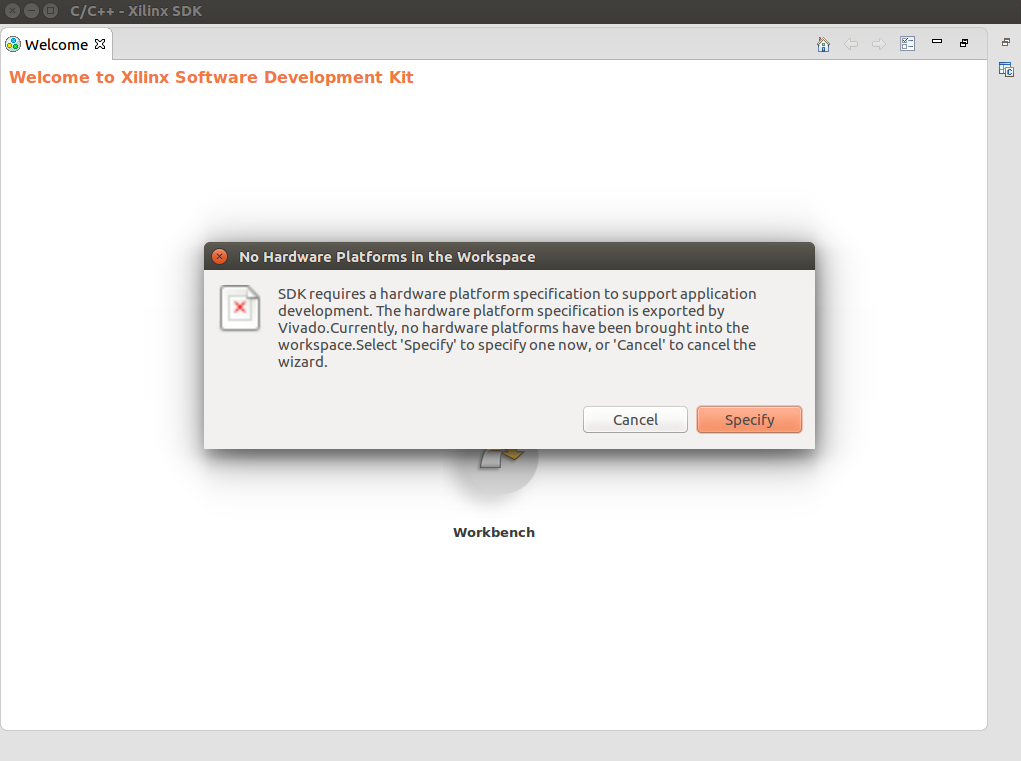}  
	}
	
\end{figure}

\textbf{Step 7}: Specify and browse the path of the .hdf file $\rightarrow$ The project name will appear.

\begin{figure}[h!]
	
	\centering
	
	\captionbox[Text]{ Step 7 \label{fig:C.6}}{%
		\includegraphics[width=0.6\textwidth]{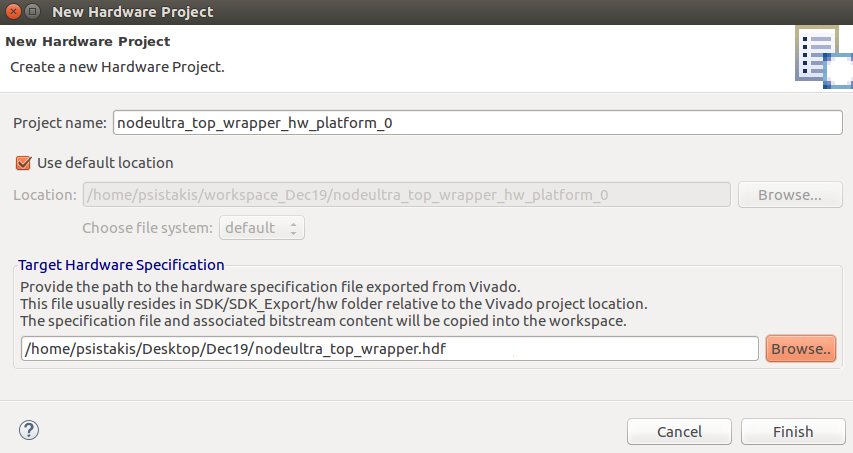}
	}
	
\end{figure}

\textbf{Step 8}: Choose the "device{\_}tree" $\rightarrow$ Click "Finish".

\begin{figure}[h!]
	
	\centering
	
	\captionbox[Text]{ Step 8 \label{fig:C.7}}{%
		\includegraphics[width=0.5\textwidth]{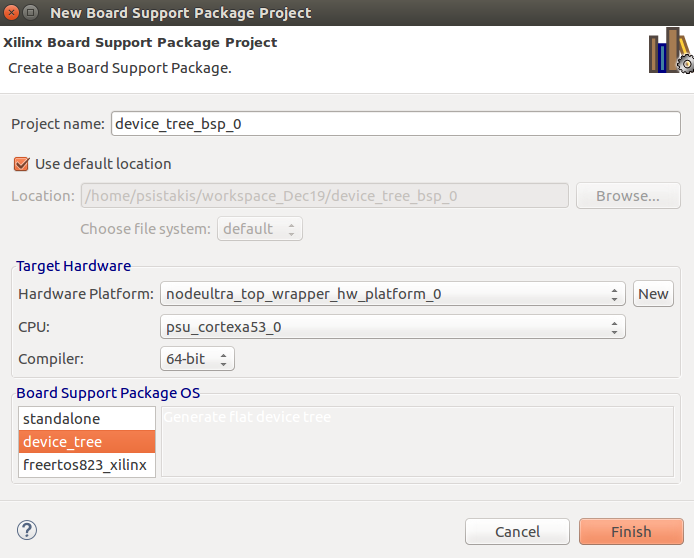}  
	}
	
\end{figure}

\vfill
\footnoterule
\pagebreak

\textbf{Step 9}: At this step we do not need to change anything.

\begin{figure}[h!]
	
	\centering
	
	\captionbox[Text]{ Step 9 \label{fig:C.8}}{%
		\includegraphics[width=0.6\textwidth]{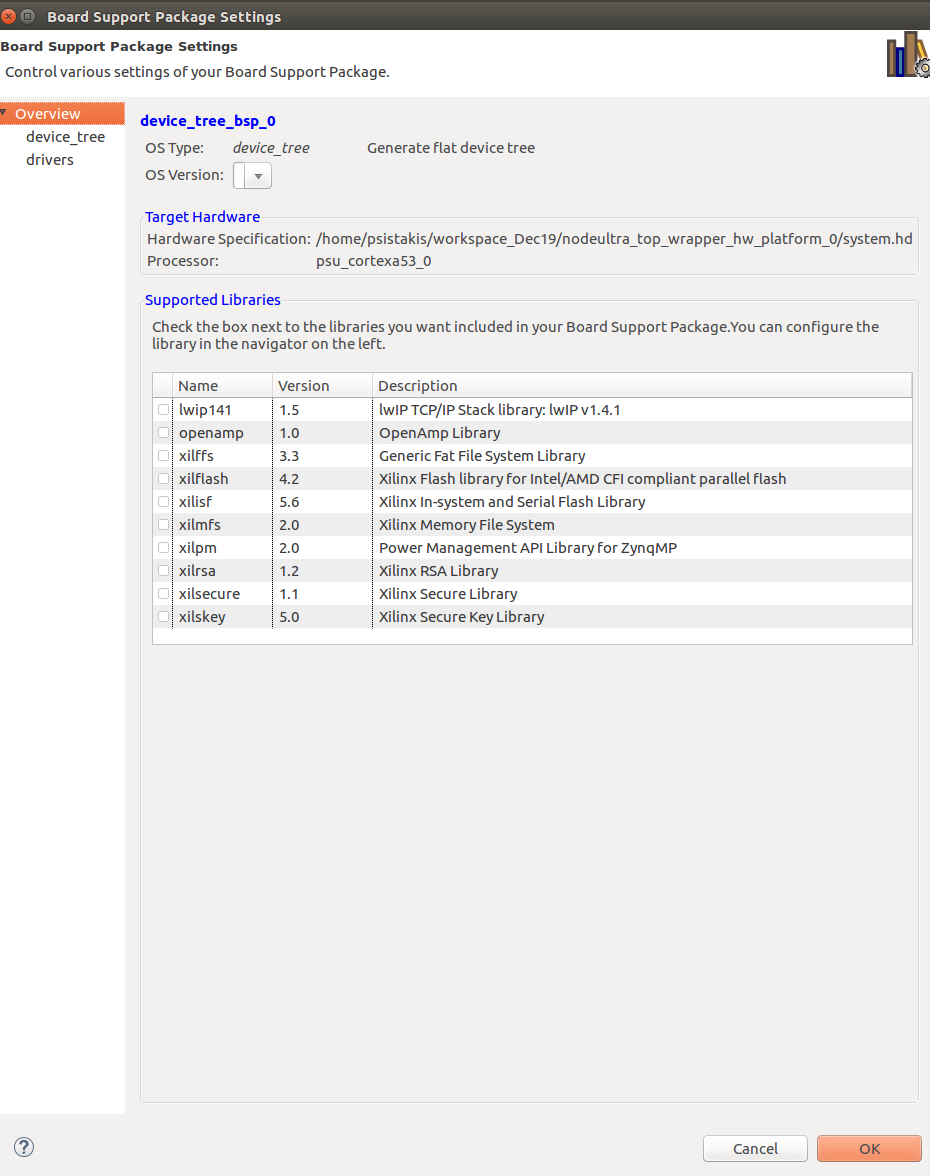}  
	}
	
\end{figure}

\textbf{Step 10}: We go at the workspace folder and see some folders/files like in the image below. The "device{\_}tree{\_}bsp{\_}0" is the folder we care about.

\begin{figure}[h!]
	
	\centering
	
	\captionbox[Text]{ Step 10 \label{fig:C.9}}{%
		\includegraphics[width=0.8\textwidth]{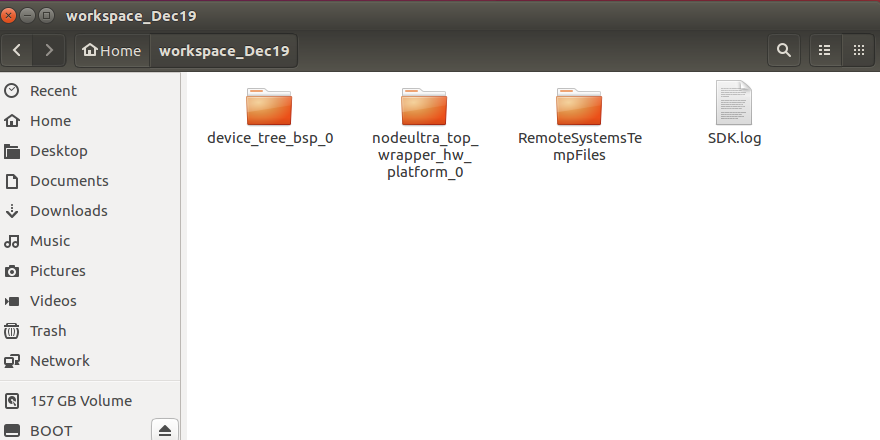}  
	}
	
\end{figure}

\vfill
\footnoterule
\pagebreak

\textbf{Step 11}: Inside the "device{\_}tree{\_}bsp{\_}0" we see the files like in the image below.

\begin{figure}[h!]
	
	\centering
	
	\captionbox[Text]{ Step 11 \label{fig:C.10}}{%
		\includegraphics[width=0.8\textwidth]{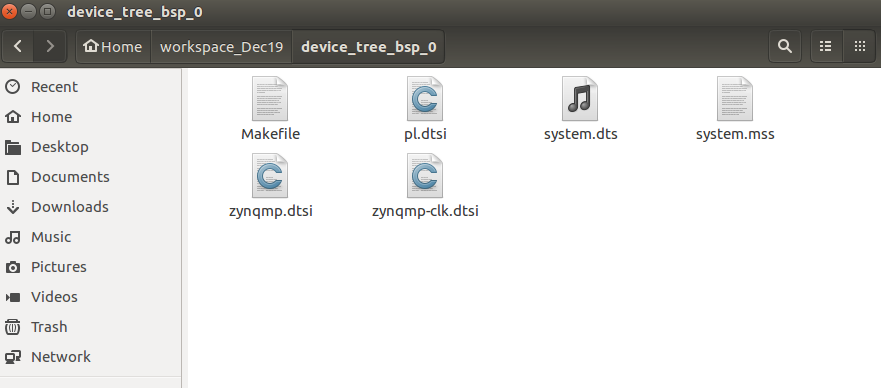}  
	}
	
\end{figure}
\label{Step 11}

\textbf{Step 12}: We open a terminal and write this command: 
\textbf{dtc -I dts -O dtb -o system.dtb ./system.dts}. This command will create the Device-Tree Blob (.dtb) the kernel needs. The "system.dts" file "includes" all the other files, so now with the .dtb we created, we have them all in one file.

\begin{figure}[h!]
	
	\centering
	
	\captionbox[Text]{ Step 12 \label{fig:C.11}}{%
		\includegraphics[width=0.8\textwidth]{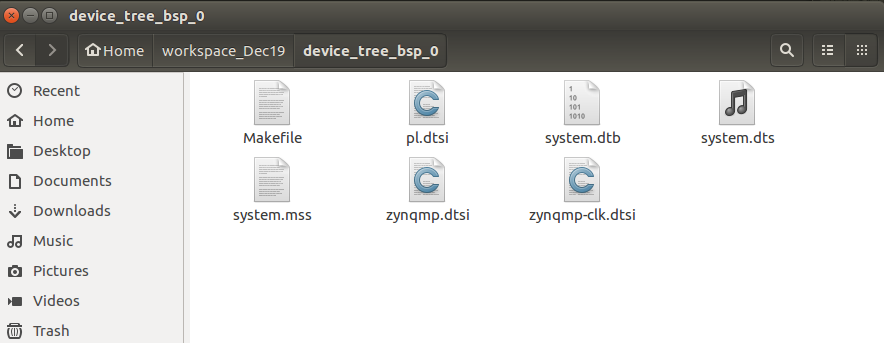}  
	}
	
\end{figure}

\textbf{Step 13}: If we want to change the source of the device-tree, we open a terminal and run this command: 
\textbf{dtc -I dtb -O dts -o system{\_}NEW.dts ./system.dtb}. Now in the "system{\_}NEW.dts" file we have all the sources of the device-tree (the files we saw in Figure \ref{fig:C.11}) in one file.

\begin{figure}[h!]
	
	\centering
	
	\captionbox[Text]{ Step 13 \label{fig:C.12}}{%
		\includegraphics[width=0.8\textwidth]{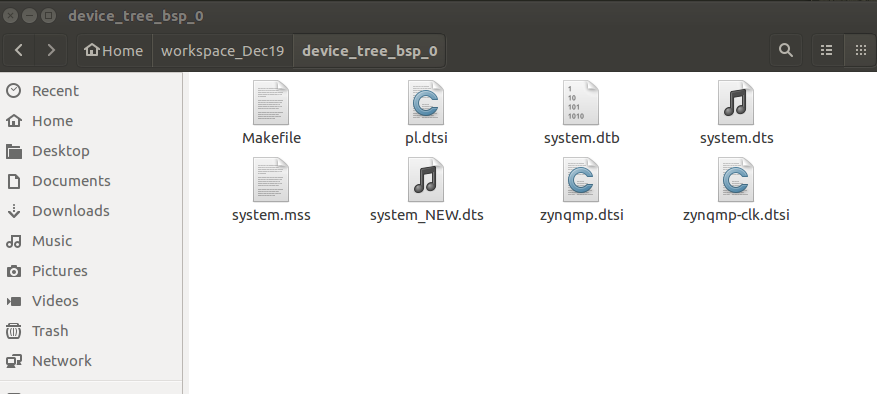}  
	}
	
\end{figure}

\vfill
\footnoterule
\pagebreak

\subsection{Enabling the LPD-DMA}
\lstset{language=C}
\setlength{\parindent}{3ex}

By default, all the LPD-DMA channels are disabled in the generated device-tree. In order to fix that, in the source of the device-tree (in our case: the "system{\_}NEW.dts" file) we add the "\textbf{clock-names = "clk{\_}main", "clk{\_}apb";}" and the "\textbf{clocks = <0x7 0x3>;}" fields, in all LPD-DMA channels, like the example below.

%
%

\begin{lstlisting}
dma@ffa80000 {
	status = "okay";
	compatible = "xlnx,zynqmp-dma-1.0";
	reg = <0x0 0xffa80000 0x1000>;
	interrupt-parent = <0x1>;
	interrupts = <0x0 0x4d 0x4>;
	~\textbf{clock-names = "clk{\_}main", "clk{\_}apb";}~
	xlnx,id = <0x0>;
	xlnx,bus-width = <0x40>;
	#stream-id-cells = <0x1>;
	iommus = <0x5 0x868>;
	power-domains = <0x8>;
	~\textbf{clocks = <0x7 0x3>;}~
	linux,phandle = <0x21>;
	phandle = <0x21>;
};
\end{lstlisting}

\subsection{SD card}
\setlength{\parindent}{0ex}

If we want to boot from the SD card, we also need to do the following changes in the source of the device tree.

\begin{enumerate}
	
	\item Change the bootargs field, in order to be able to detect the root and some other things that apparently are needed. Instead of this: \textbf{bootargs = "console=ttyPS0,115200";}, write this: \textbf{bootargs = "console=ttyPS0,115200 root=/dev/mmcblk0p2 rw rootwait earlyprintk bootmem{\_}debug=1 memblock{\_}debug=1";}. 
	
	Below we see the part of the source code of the device-tree, that we changed:
	%
	%
	
	\begin{lstlisting}
	chosen {
		~\textbf{bootargs = "console=ttyPS0,115200";}~	
	};
	aliases {
		ethernet0 = "/amba/ethernet@ff0e0000";
		serial0 = "/amba/serial@ff000000";
		spi0 = "/amba/spi@ff0f0000";
	};	
	memory {
		device_type = "memory";
		reg = <0x0 0x0 0x0 0x80000000>;
	};
	\end{lstlisting}

	\vfill
	\footnoterule
	\pagebreak
	\item If you have many boards, you have to change the MAC address according to the board/box you are currently working on. This is a "mini-hack" in order to avoid many boards having the same MAC address - we achieve that by simply changing the second field of zeros (from the end) to the "cx", where x is the number of the board (for instance for board 7: c7). In the example below, instead of this: \textbf{"local-mac-address = [00 0a 35 00 00 00];"}, we write this: \textbf{"local-mac-address = [00 0a 35 00 c7 00];"}.
	%
	%
	
	\begin{lstlisting}
	ethernet@ff0e0000 {
		compatible = "cdns,zynqmp-gem";
		status = "okay";
		interrupt-parent = <0x1>;
		interrupts = <0x0 0x3f 0x4 0x0 0x3f 0x4>;
		reg = <0x0 0xff0e0000 0x1000>;
		clock-names = "pclk", "hclk", "tx_clk";
		#address-cells = <0x1>;
		#size-cells = <0x0>;
		#stream-id-cells = <0x1>;
		iommus = <0x5 0x877>;
		power-domains = <0xe>;
		clocks = <0xb 0xb 0xb>;
		~\textbf{local-mac-address = [00 0a 35 00 c7 00]; }~
		phy-mode = "rgmii-id";
		xlnx,ptp-enet-clock = <0x0>;
		linux,phandle = <0x1d>;
		phandle = <0x1d>;
	};
	\end{lstlisting}

	\item Add this: \textbf{"no-1-8-v"} at the end of the sd-node that is activated. In our case, it was the "sdhci@ff170000" sd-node. 
	
	Below we see the part of the source code of the device-tree, that we changed:
	%
	%
	
	\begin{lstlisting}
	sdhci@ff170000 {
		compatible = "arasan,sdhci-8.9a";
		status = "okay";
		interrupt-parent = <0x1>;
		interrupts = <0x0 0x31 0x4>;
		reg = <0x0 0xff170000 0x1000>;
		clock-names = "clk_xin", "clk_ahb";
		broken-tuning;
		#stream-id-cells = <0x1>;
		iommus = <0x5 0x871>;
		power-domains = <0x19>;
		clocks = <0x18 0x18>;
		clock-frequency = <0xa37c42e>;
		linux,phandle = <0x32>;
		phandle = <0x32>;
		~\textbf{no-1-8-v;}~ /* New field */
	};
	\end{lstlisting}
\end{enumerate}

\setlength{\parindent}{0ex} 

\textbf{Note}: After all the changes in the source code, build again the device tree, by creating the Device-Tree blob (Step 12).

\vfill
\footnoterule
\lstset{style=mystyle,
	basicstyle=\ttfamily\color{white}}

\chapter{Appendix} 

\label{AppendixD} 

\lstset{
	language=C,
	basicstyle=\footnotesize\color{white}
}

\section{Changes to the Linux kernel}

\setlength{\parindent}{3ex}

In order to make the fourth module of this thesis work (Section \ref{Process-Page-Table}), we had to make some changes to some Linux drivers, like the arm-smmu.c driver, which is the driver of the ARM SMMU, the iommu.c driver, which is the IOMMU driver that implements the IOMMU API and the iommu.h library. Below we see all dominant changes for each file:

\subsection{iommu.h}

\setlength{\parindent}{0ex}

In the iommu.h, we just added an extra field in the iommu{\_}domain structure.
\begin{lstlisting}
	struct iommu_domain {
		unsigned type;
		const struct iommu_ops *ops;
		iommu_fault_handler_t handler;
		void *handler_token;
		struct iommu_domain_geometry geometry;
		void *iova_cookie;
		~\textbf{u64 ttbr;}~ /* new field */
	};
\end{lstlisting}

\subsection{iommu.c}
\setlength{\parindent}{0ex}

In the iommu.c driver, we only edited the \textbf{{\_\_}iommu{\_}domain{\_}alloc()} function, which was called each time we had an allocation of an IOMMU domain. More specifically, we initialized the field "ttbr", that we added in the iommu.h.

\begin{lstlisting}
	static struct iommu_domain *__iommu_domain_alloc(struct bus_type *bus,
	unsigned type){
		pr_info("%s\n", __func__);
		struct iommu_domain *domain;
		if (bus == NULL || bus->iommu_ops == NULL)
			return NULL;
		domain = bus->iommu_ops->domain_alloc(type);
		if (!domain)
			return NULL;
		domain->ops  = bus->iommu_ops;
		domain->type = type;
		~\textbf{domain->ttbr = 0;}~ /* new field */
		return domain;
	}
\end{lstlisting}

\vfill
\footnoterule
\pagebreak

\subsection{arm-smmu.c}

\setlength{\parindent}{0ex}

The idea is that when we allocate an IOMMU domain in our exanest{\_}virt module, we would want to "pass" to the arm-smmu driver the pointer of the page table of a process of the user. So, when we allocate the IOMMU domain, we also "pass" this pointer (as we mentioned before, the pgd, which is not equal to zero) to the new field of the structure (the ttbr). Then, the arm-smmu driver, after the changes we did (we can see them below), will understand that this value of the ttbr is different from the expected (default) value, which is zero, and will take the ttbr and "pass" it to its own context bank. After that, the context bank we initialized will have as a translation table, not the translation table that was created at the initialization of the context bank, but the translation table of the user process.

In the function \textbf{arm{\_}smmu{\_}init{\_}context{\_}bank()}:

\begin{lstlisting}
	if(smmu_domain->domain.ttbr!=0){ /* new field */
		printk(KERN_ALERT "BYPASSED TTBR0\n");
		reg64 = smmu_domain->domain.ttbr;
		printk("smmu_domain->domain.ttbr %lu \n", smmu_domain->domain.ttbr);
	}
	else{
		printk(KERN_ALERT "DEFAULT TTBR0\n");
		reg64 = pgtbl_cfg->arm_lpae_s1_cfg.ttbr[0];
	}
\end{lstlisting}

\setlength{\parindent}{3ex} 
\noindent

In order to fix the issues about the size of the addresses, that were mentioned in Section \ref{Process-Page-Table}, we did some changes to the same function:
\begin{lstlisting}
	/* TTBCR */
	if (stage1) { 
		reg = pgtbl_cfg->arm_lpae_s1_cfg.tcr;
		if ( smmu_domain->domain.ttbr != 0 ) {
			// Use processor given size from TCR_EL1
			// VIRTUAL ADDRESS SIZE
			// Overides bit calculation based on cfg->ias from 
			// arm_64_lpae_alloc_pgtable_s1
			reg = (reg & 0xffffffc0) | 0x19;
		}
		writel_relaxed(reg, cb_base + ARM_SMMU_CB_TTBCR);
		pr_info("TTBCR write 0x%x\n",reg);
		if (smmu->version > ARM_SMMU_V1) {
			reg = pgtbl_cfg->arm_lpae_s1_cfg.tcr >> 32;
			if ( smmu_domain->domain.ttbr != 0 ) {
				// Use processor given size from TCR_EL1
				// PHYSICAL ADDRESS SIZE
				// Overides bit calculation based on cfg->oas from 
				// arm_64_lpae_alloc_pgtable_s1
				reg = (reg & 0xfffffff8) | 0x2;
			}
			reg |= TTBCR2_SEP_UPSTREAM; 
			writel_relaxed(reg, cb_base + ARM_SMMU_CB_TTBCR2);
			pr_info("TTBCR2 write 0x%x\n",reg);
		}
	} else {
		reg = pgtbl_cfg->arm_lpae_s2_cfg.vtcr;
		writel_relaxed(reg, cb_base + ARM_SMMU_CB_TTBCR);
	}
\end{lstlisting}

\vfill
\footnoterule
\lstset{style=mystyle,
basicstyle=\ttfamily\color{white}}

	
\end{document}